\documentclass[AMA,STIX1COL]{WileyNJD-v2}
\usepackage{tikz}
\newcommand\copyrighttext{%
  \footnotesize This work has been submitted to the International Journal of Network Management (IJNM) and accepted for publication. Copyright may be transferred without notice, after which this version may no longer be accessible.}
\newcommand\copyrightnotice{%
\begin{tikzpicture}[remember picture,overlay]
\node[anchor=south,yshift=10pt] at (current page.south) {\fbox{\parbox{\dimexpr\textwidth-\fboxsep-\fboxrule\relax}{\copyrighttext}}};
\end{tikzpicture}%
}

\usepackage{enumitem}
\usepackage{multirow}
\usepackage{bigstrut}
\usepackage{mathtools}
\usepackage{wrapfig}
\usepackage{comment}
\usepackage{textgreek}
\usepackage{setspace}
\usepackage{subfigure}

\usepackage[utf8]{inputenc}
\usepackage{algorithm} 
\usepackage{algpseudocode} 
\usepackage[subfigure]{tocloft}
\usepackage{amssymb}
\usepackage{pifont}

\let\oldnl\nl
\newcommand{\nonl}{\renewcommand{\nl}{\let\nl\oldnl}}
\usepackage{setspace}
\newcommand{\sepp}{\hspace{8mm}}

\definecolor{color1}{RGB}{198, 90, 103}
\definecolor{color2}{RGB}{175, 42, 48}
\definecolor{color3}{RGB}{191, 6, 29}
\definecolor{color4}{RGB}{179, 43, 59}

\definecolor{darkgreen}{rgb}{0.0, 0.5, 0.0}

\newcommand{\cpa}[1]{{\textcolor{black}{#1}}}

\newcommand{\hsep}{\hspace{6 mm}}

\articletype{Article Type}

\received{26 April 2016}
\revised{6 June 2016}
\accepted{6 June 2016}

\begin{document}

\title{Dissemination Control in Dynamic Data Clustering For Dense IIoT Against False Data Injection Attack}

\author[1]{Carlos Pedroso*}



\author[1,2]{Aldri Santos}

\authormark{PEDROSO \textsc{et al}}

\address[1]{\orgdiv{Department of Informatics}, \orgname{Federal University of Paraná}, \orgaddress{\state{Paraná}, \country{Brazil}}}

\address[2]{\orgdiv{Department of Computer Science}, \orgname{Federal University of Minas Gerais}, \orgaddress{\state{Minas Gerais}, \country{Brazil}}}

\corres{Aldri Santos, Department of Informatics, Federal University of Paraná, Paraná, Brazil. Email: aldri@inf.ufpr.br}


\abstract[Summary]{
The IoT has made possible the development of increasingly driven services, like industrial IIoT services, that often deal with massive amounts of data. Meantime, as IIoT networks grow, the threats are even greater, and false data injection attacks (FDI) stand out as being one of the most aggressive. The majority of current solutions to handle this attack do not take into account the data validation, especially on the data clustering service. Aiming to advance on the issue, this work introduces CONFINIT, an intrusion detection system for mitigating FDI attacks on the data dissemination service performing in dense IIoT networks. CONFINIT combines watchdog surveillance and collaborative consensus strategies for assertively excluding various FDI attacks. The simulations showed that CONFINIT compared to DDFC increased by up to 35\% - 40\% the number of clusters without attackers in a gas pressure IIoT environment.  CONFINIT achieved attack detection rates of 99\%, accuracy of 90 and F1 score of 0.81  in multiple IIoT scenarios, with only up to 3.2\% and 3.6\% of false negatives and positives rates, respectively. Moreover, under two variants of FDI attacks, called Churn and Sensitive attacks, CONFINIT achieved detection rates of 100\%, accuracy of 99 and F1 of 0.93 with less than 2\% of false positives and negatives rates.}

\keywords{IIoT, Data Dissemination, Secure Clustering, Consensus, Watchdog, False Data Injection Attacks}
\jnlcitation{\cname{
\author{C. Pedroso}, and 
\author{A. Santos}} (\cyear{2021}), 
\ctitle{Dissemination Control in Dynamic Data Clustering For Dense IIoT Against False Data Injection Attack}}

\maketitle

\copyrightnotice

\section{Introduction}
\label{sec:intro}

The Internet of Things (IoT) enables the connection of different types of physical objects through technologies such as wireless sensor networks (WSN), RFID, GPS and NFC, among others. The objects that make up the IoT hold characteristics like identity, physical attributes, heterogeneity, and many times different interfaces to establish communication with each other and ways of mobility~\cite{gubbi2013internet,pal2020security}. Besides the IoT environments many times support mobile and fixed devices, whose infrastructure varies depending on the interaction among the objects. Further, many devices own limited resources, i.e. low energy, low processing, and storage capacity, in addition to suffering losses in connection links~\cite{borgia2014internet,qiu2018can}. IoT is embedded in dense and complex domains, such as industrial processes, logistics, public security and smart cities. The Smart Industry has been challenging for researchers due to the variety, heterogeneity, and integration of the sub-domains~\cite{ageed2021survey}. Particularly the Industrial Internet of Things (IIoT) paradigm has stood out for its easy integration between industrial devices, allowing them to act in a synchronized and organized manner~\cite{mumtaz2017massive, chihana2018iot}. Furthermore,  as Cyber-Physical Systems (CPS) allow the integration between the physical and digital environment, IIoT envision as part of Industry 4.0 or 4th industrial revolution~\citep{lee2014service,pivoto2021cyber,qin2020recent}. For instance, fully connected factories controlled by autonomous machines; production managed by sensors capable of estimating the failure of a machine; and inventory management and logistics integrated with the production line to provide information to the control center~\cite{qin2020recent}. The availability of such information correctly allows us to reduce the production time and  provide real-time diagnostic, improving overall performance. In this way, IIoT is essential to collect, disseminate and handle the volume of data required by different applications in their decision making~\cite{minoli2017iot, akpakwu2018survey, souza2019digital}.

IIoT services commonly yield a huge volume of data, which are collected by numerous devices, disseminated, and made available safely and efficiently. Hence providing security to the data dissemination service on IIoT  plays an essential role in the evolution of industrial application and trustworthiness~\cite{kouicem2018internet,akpakwu2018survey, khan2020industrial}. These applications require data to support their decision-making, so when the data face some inconsistencies or manipulations, the fully industrial system can show up results different from those expected. Thus, handling and disseminating the large volume of data due to interactions among multiple devices expose IIoT to various~\cite{bodkhe2020secure} security issues.  This exposure makes IIoT target of numerous threats that violate  the integrity, authenticity, and availability of services~\cite{yaqoob2017rise, kouicem2018internet, borhani2020secure}; and hence they damage the functioning of various end-applications~\cite{miorandi2012internet, yang2017survey}. Thus, providing data availability and authenticity is essential to achieve services free from threats that interfere in the  decision making~\cite{hamad2020realizing}.

The attack of false data injection (FDI) stands out among the internal threats to the data dissemination service in IIoT, being one of the most harmful intrusion attacks on data-driven networks due to the data inconsistency and the attacker unpredictability~\cite{sen2017risk,yang2017robust, aboelwafa2020machine}. Due to the FDI attack complexity, its detection becomes a non-trivial task as malicious devices  authenticated in the networks perform their standard data collection and dissemination functions~\cite{deng2016false, zhang2020false}. Further, attacks can emerge at different moments and with different times, disrupting the network~\cite{hug2012vulnerability,liao2020evaluating}. The FDI attack takes place when an external attacker captures the device and manipulates its data or when the device itself presents misconduct and manipulates its own data. Hence, those behaviors make the attack difficult to identify and increases network malfunction times. Besides, a variety of other attacks, like camouflage, recommendation, and deduction, act in a similar way to the  FDI attack, each one of them with its peculiarities~\cite{bhusal2022coordinated}. Therefore, the fast identification and isolation of malicious devices in IIoT reduce long periods of network malfunctions, that can affect the data dissemination service, by mitigating that those misbehaving devices generate inconsistent sensed data. 

When applied in dense IIoT various approaches effective against FDI attacks in WSN~\cite{lu2012becan, kumar2018deterministic, yi2021route},~\textit{Smart Grids}~\cite{li2017distributed,tahsien2020machine} or IoT~\cite{yang2017robust,aboelwafa2020machine} usually face difficulties because either they have failed or they are not suitable by requiring high resource consumption, do not verifying data, and having a centralized decision making. Among all approaches, the most commonly used are En-route Filtering Schemes, Collaborative Detection, Machine Learning (ML), and Intrusion Detection Systems (IDS).Schemes of En-route filtering are usually applied in WSN and take the report of packets verification in intermediate  devices between source and destination in order to discard ones with any inconsistency.Although the reports are endorsed to guarantee authenticity, they disregard changes in the sensed data. Collaborative detection approaches have been employed in Smart Grids as an alternative to deal with FDI attacks, where each device also performs as a detection agent. Machine learning (ML) approaches in general are applied in the training machines to identify multiple attacks at an early stage and provide defensive policies corresponding to each attacker. Lastly, IDSs deal with attacks in different contexts, being employed on devices or networks to monitor all types of systems; and many of them can generate high resource consumption and new vulnerability gaps. Therefore, the IIoT evolution demands solutions that can detect and isolate assertively against the arising of threats, ensuring greater robustness for the data collection and dissemination services.

This work presents the CONFINIT (\emph{\textbf{CON}sensus Based Data \textbf{FI}lteri\textbf{N}g for I\textbf{I}o\textbf{T}}) system to detect and isolate FDI attacks acting on data dissemination services in dense IIoT networks.  CONFINIT focuses on fixed IIoT networks with a high volume of continuously collected and disseminated data. CONFINIT meets clustering by data similarity to handle the density of devices~on the network, combines a watchdog strategy between devices for data self-monitoring and a collaborative consensus for detecting and excluding FDI attacks. We have made a evaluation by simulation between the performance of CONFINIT and DDFC (Dynamic Data-aware Firefly-based Clustering)~\cite{gielow2015data} in a gas pressure IIoT environment. The results showed that CONFINIT increased by up to 35\% - 40\% the number of clusters without FDI attackers compared to DDFC. It also achieved attack detection rates of around 99\% with accuracy of 90 and F1 score of 0.81 in the multiple IIoT scenarios, and only up to 3.2\% of false negatives and 3.6\% of false positives. Further, under Churn and Sensitive FDI attacks, CONFINIT achieved a detection rate of around 100\%, with accuracy of 99 and F1 score of 0.93,  and less than 2\% of false positive and negative rates.

The paper is organized as follows: Section~\ref{sec:related} presents a background and main related works. Section~\ref{sec:proposal} describes the IIoT network, FDI attack, data dissemination models, as well as assumptions taken by CONFINIT, its components and operation. Section~\ref{sec:eva} shows a performance evaluation between CONFINIT and DDFC and discusses the results obtained. Section~\ref{sec:concl} presents conclusions and future work.

\section{Background and Related Work}
\label{sec:related}

The evolution of computing devices capable of collecting and disseminating information has enabled the emergence of personalized and unique services. For instance, services for measuring temperature, locating objects, and monitoring equipment conditions, among others, can be offered quickly and efficiently~\cite{diene2020data,liao2020evaluating}. Thus, the devices are initialized in different locations, starting to work in an organized and dynamic way~\cite{peng2018boat}. Though, certain issues  can compromise the availability of existing services such as the heterogeneity of devices, the power outages, failure of devices, and the presence of attackers~\cite{kumar2020integration,khanam2020survey,bhushan2021unification}. Furthermore, the wireless communication medium is normally exposed to interference, noise, and collisions, which can damage the quality and availability of the service provided. In this scenario, an attack, like False Data Injection (FDI), can take advantage of these issues. In addition, FDI compromises the authenticity of the disseminated data by tampering, thus damaging the availability of data in the network~\cite{gopalakrishnan2020security,ahmed2020false}.

\begin{figure}[ht]
    \centering    
    \includegraphics[width=0.85\linewidth]{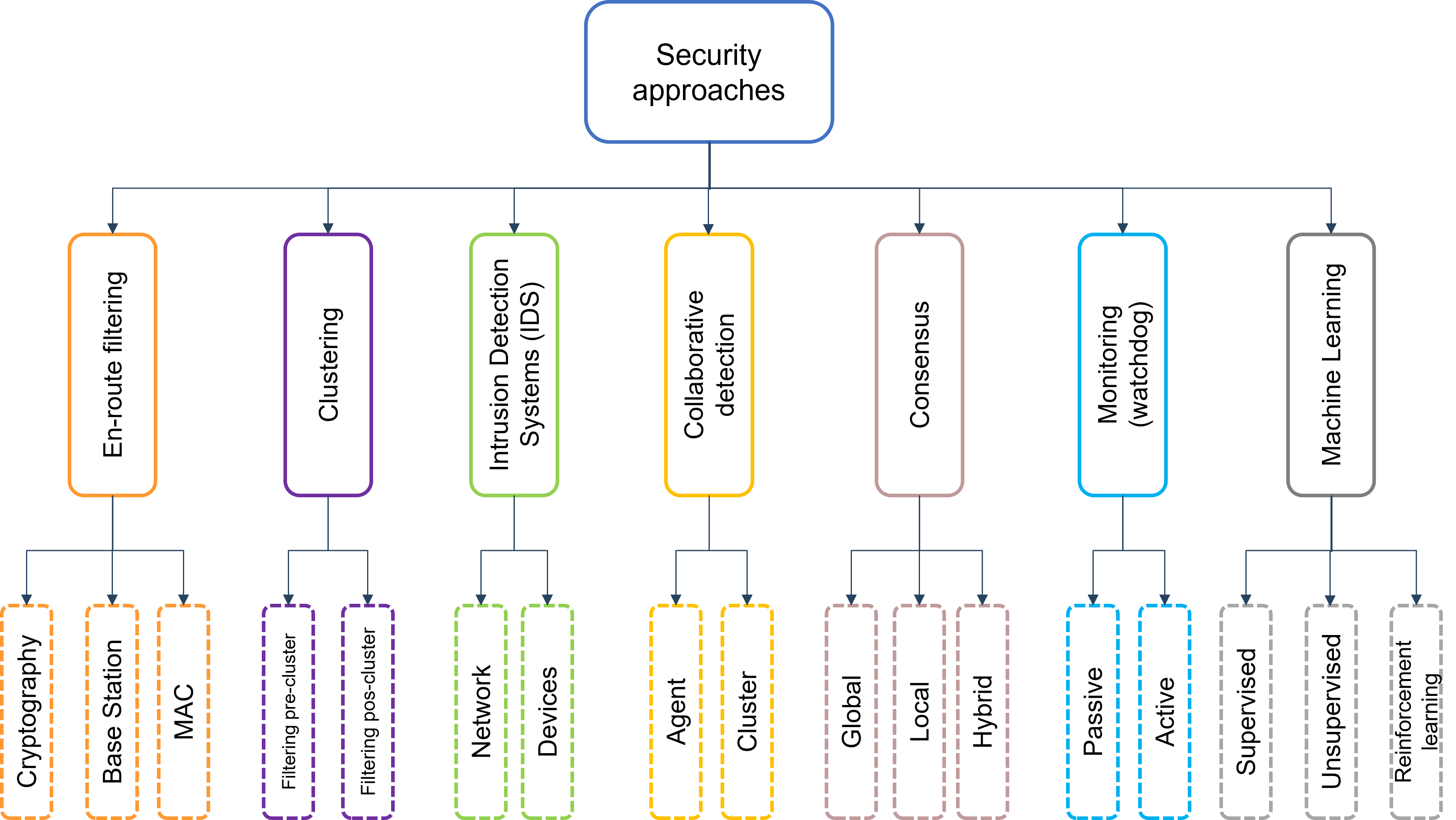}
      \caption{Taxonomy of the approaches against FDI attacks}
      \label{Fig:tax}
\end{figure}

\cpa{To understand better the potentials approaches against the FDI attacks on the dissemination and routing services, we divided them into categories:} {\bf En-route filtering schemes}~\citep{lu2012becan}, {\bf Clustering}~\citep{wang2014defending}, Intrusion {\bf Detection Systems (IDS)}~\citep{yang2017robust}, {\bf Collaborative Detection}~\cite{li2017distributed}, {\bf Collaborative Consensus}~\cite{toulouse2015consensus}, {\bf Monitoring (watchdog)}~\cite{santos2019clustering}, and {\bf Machine Learning (ML)}~\cite{tahsien2020machine, aboelwafa2020machine} and summarize in the taxonomy shown in Figure~\ref{Fig:tax}. The en-route filtering scheme takes into account packet filtering between the source and the destination by assigning the marking and discarding of packets by the intermediate nodes and the packet aggregates Base Station (BS). This scheme employs techniques such as Message Authentication Code (MAC), cryptography and packet classification at the BS to deal with data falsification and network failures. However, they disregard validation of collected data, dynamics of devices, and detection of the source of the attack;~\cpa{hence decreasing} the effectiveness of the solutions. 

Clusters-based approaches take into account the elimination of false data in two ways, pre-cluster, and post-cluster formation. In pre-cluster, attack detection acts so that only honest nodes integrate the clusters, preventing attackers from integrating a cluster. In post-cluster, attack detection occurs after cluster establishment, being performed by the In both forms, detection of the attackers can be carried out by a central entity or by devices present on the network. Collaborative detection techniques are alternative means to deal with FDI attacks in~\textit{Smart Grids}. In this technique, each device is responsible for performing two functions. The first pattern refers to data collection and dissemination, and the second as a collaborative agent for attack detection. Thus, network security is maintained only by the devices, without the need for an external entity.

Intrusion Detection Systems (IDS) are comprehensive solutions, which often support different resources and aim to improve the security of computer systems or networks in the presence of threats~\cite{shamshirband2020computational, sicato2020comprehensive}. The IDS operates as a second layer of defense and generally deals with internal threats,~\cpa{which bypass anomaly prevention measures or services such as access control.} While detecting and monitoring events, they search for anomalies among different types of systems. IDSs are classified in machine-based (\textit{host}) and network-based. The first one monitors, detect, and respond to attack activity on a given \textit{host}. The second one uses a machine or software to monitor packages in real-time for verifying possible anomalies. They are usually configured at specific points on the network to manage all traffic looking for anomalies.

There are other ways to address collaboratively and dynamically the FDI attacks. The combination of monitoring~\textit{watchdog} and collaborative consensus brings up several benefits to protect the network, like collaboration between participants, continuous monitoring, real-time detection, and precision in decision making.  Meantime, those techniques require organization and standardization to avoid yielding vulnerability gaps in the network. Various systems have carried out \textit{watchdog} techniques to monitor and detect anomalies, acting either passively or actively. In a passive way, everyone monitors everyone searching for anomalies, when detecting an anomaly the participants inform a central entity that acts to contain it. In an active way, participants monitor and act against anomalies, without requiring a central entity. On the other hand, a manner to achieve collaborative detection in dynamic systems is through the adoption of consensus among devices on a network. The consensus strategy is usually applied to dynamic systems with interaction between all participants. In addition, the majority of consensus approaches have been classified into three categories: global, local, and hybrid~\citep{zhan2017cluster, colistra2014task}. The global strategies have been applied to systems where all participants make decisions about any activity or change of state. The local strategies have been adopted in partitioned systems or divided into extensions, where each group is responsible for managing an area and obtaining a decision. In this model, one decision does not interfere with another. Finally, the hybrid strategies have been employed in systems divided into several local subgroups in order to establish a large global consensus and make a decision about something in the system.

Machine Learning (ML) approaches train machines using different algorithms and aid devices to learn from their experience rather than explicitly programming them~\cite{tahsien2020machine}. ML does not require human assistance, owns simple classification mathematical equations, and can work in dynamic networks. In the past few years, ML techniques have been advanced remarkably for IoT security purposes~\cite{zikria2021next,haji2021attack}. Further, solutions that take into account the environment characteristics can be provided using different ML algorithms for performing in IoT devices with limited resources. ML techniques can commonly be divided into supervised, unsupervised, and reinforcement learning. In supervised learning techniques, there is no output data for given input variables. Besides most of the data are unlabeled where the system tries to find out the similarities among this dataset. Based on that, it classifies them into different groups as clusters. Supervised learning techniques are the most common methods among classifiers, its output is classified based on the input employing a pre-trained dataset to obtain the result. As most of the data are unlabeled where the system tries to find out the similarities among this data set, it classifies them into different groups as clusters. In unsupervised learning techniques, the system will cluster unsorted data according to similarities and differences even though there are no categories provided; and some data cannot be labeled when it tries to discover the similarities between the data set. The reinforcement learning techniques enable machines to learn through interactions with the environment as new interactions and new changes enable new learning. In this way, that learning can be applied to detect intelligent attacks on IoT devices and establish a defense policy in protecting the network.

The demand for secure data dissemination services in dense networks in the presence of various means of intrusion attacks has been the focus of different studies~\cite{lu2012becan, kumar2016security, yang2017robust, santos2019clustering, bostami2019false, tahsien2020machine}. Particularly, FDI attacks on the data dissemination service in dense IIoT become harmful to the network due to the large volume of data generated by the devices and their importance to decision-making by applications. Thus, the proposition of new systems capable of mitigating and isolating such threats as FDI attacks is essential to enable the advancement of specific applications as well as diverse services on the IIoT. Next, we present the main works found in the literature to deal with FDI attacks in several contexts and networks. Besides, we discussed their advantages and disadvantages in the domain of dense IIoT.

\subsection{Related Work}
En-route filtering schemes have been an approach usually applied against FDI attacks in WSN~\cite{lu2012becan, yu2010dynamic, wang2014defending, yi2021route}. In~\cite{lu2012becan}, the authors propose a cooperative authentication scheme to filter false data in Wireless Sensor Network (WSN), that requires  that all nodes have a fixed number of neighbors to manage the authentication in a distributed  across data routing to the base station.  The scheme adopts the compressed bit technique for not overloading the channel, making it suitable for filtering false data injected since authentication takes place point-to-point. Furthermore, the usage of a routing protocol prevents malicious devices from compromising other devices. However, the proposal does not consider data manipulation, in addition to not identifying the compromised devices. In~\cite{yu2010dynamic}, the authors present a dynamic En-route filtering scheme to deal with false data injection and Denial of Service (DoS) attacks on WSN. In this scheme, each node owns a set of authentication keys used to endorse reports. The authentication service is guaranteed by a group of nodes chosen previously. Each node offers its key to the forwarding nodes that must disclose their keys, allowing them to check all reports. Meantime, the solution ignores the issue of data manipulation and demands an additional cost due to the constant exchange of keys between devices. In~\cite{yi2021route} the authors proposed a new route filtering scheme called EMAS, aiming to improve the probability and efficiency of route filtering through MAC verification. EMAS checks out endorsement nodes identifier and position, the packet pre-verification, and the monitoring and reporting mechanism in order to know the manipulated packets. Before generating an event report to the base station, EMAS first checks the endorsements provided by node that detected the manipulated packets and then discards them. In addition, it applies a forward reporting strategy to resist selective forwarding attacks for balancing the residual energy of nodes and prolonging the network life. However, the authors ignore the manipulation of sensed data and add extra cost to devices due to constant key exchange, in addition to not identifying compromised devices.

In~\cite{yang2017robust}, the authors presented an IDS based on the detection of anomalies by using monitoring~\textit{watchdog} to mitigate the false data injection attack. They attempt to predict natural events by using data collected from IoT environmental surveillance through Bayesian Spatial-Temporal (HBT) monitoring. Next, they apply a statistical decision strategy in a sequential probability test to identify malicious activities. However, the HBT model presents high-energy consumption, and although the probabilistic test is effective, it overloads the network and does not apply any data validation. In~\cite{santos2019clustering}, the authors present an IDS for attack mitigation~\textit{sinkhole} and~\textit{selective forwarding} in the routing of dense IoT networks, where the network is organized in~\textit{clusters} and whose nodes are classified  as leaders, associates and members. A tiered~\textit{watchdog} strategy monitors the relationship between forwarding and received data, defining which nodes  exhibit malicious behavior. Despite this strategy to get an effective detection, computing the trust between participant nodes based on the amount of data received and transmitted, it does not exclude the chance of the node falsifying the collected data. In\cite{krishna2016f} the authors presented a framework for detecting false data injection attacks in Smart Grid.  They take into account a theft detector based on Kullback-Leibler (KL) divergence to detect cleverly-crafted electricity theft attacks that circumvent detectors. Even though the solution is successful against FDI attack, when applied to IIoT, it should also take into account factors such as dynamicity and low processing power. In~\cite{martins2022privacy} the authors highlighted a framework to preserve data privacy on IoT devices in a smart home. They have developed an end-to-end data flow model that autonomously extracts data and classifies it on the anomaly detection server. After the extraction, they analyze and model the patterns in the activities in the physical channel to train the proposed model with data from a real environment and noise-generating devices. However, there is a limitation since the required training time and centralized decision-making do not meet the dynamics and density of the IIoT.

One way to achieve collaborative detection is through the application of a~\textit{consensus} technique among devices on a network~\cite{colistra2014task}. In~\cite{toulouse2015consensus}, the authors proposed a distributed system for detecting anomalies based on a medium consensus protocol among participants. The system aims at identifying anomalies that generate a Distributed Denial of Service (DDoS) attacks. Thus, it performs a pattern analysis at each data collection point using a Bayes classifier. In additional such analyses occur redundantly, i.e. parallel to the level of each data collection point, in order to avoid the single point of failure. The adopt of consensus in a distributed way facilitates collaborative decision making. On the other hand, the computational cost of communication between the participants tends to overload the network and decrease the effectiveness of the system. In~\cite{kailkhura2016data}, the authors develop a robust consensus algorithm based on distributed weighted average. The algorithm aims to allow an adaptation to the local rules stipulated by the network, where a learning technique was developed to estimate operational parameters or the weight of each node. Thus, it becomes possible to automate local mergers or update rules to mitigate attacks. Although, it performances in a distributed network, the authors by ignoring dynamism as an impact factor, offer a breach to malicious actions. In~\cite{hao2020consensus} the authors apply a distributed consensus Kalman filtering algorithm to detect the FDI attack in a WSN. The Kalman filtering aids to estimate the partial and full states of the system in relation to possible attackers in the network. The algorithm adopts a threshold determined by the estimation performance in order to detect the attack and reduce its effects. Further, it  reduces the error generated by communication and measurement. Despite the solution is successful against FDI attack, they ignore the dynamism network as an impact factor and the data readings falsification. In~\cite{bhattacharjee2018towards} the authors proposed a scheme based on consensus correlation to detect data falsification in the Smart Meters (SM). The solution is based in a  semi-supervised learning-based trust scoring model to detect falsification of power consumption data and then identified the compromised meters. Despite this strategy to get an effective detection, computing the trust scoring by participant nodes does not exclude the node's chance of falsifying the collected data from the SM and does not eliminate the source of the attack.

In~\cite{aboelwafa2020machine} the authors introduced a new method for FDI attack detection using  machine learning autoencoders (AEs). The method exploits the sensor data correlation in time and space to identify the falsified data in an industrial IoT.  Despite the method obtaining an effective detection, the required training time and the centralized decision making do not meet the requirements expected for a dynamic environment such as IIoT. In~\cite{saba2021intrusion} the authors built an IDS based on machine learning to detect various attacks in IoT networks, including FDI. The IDS takes a genetic algorithm (GN) jointly with classifiers vector machine (SVM), decision tree, and ensemble to identify intruders in the network. Thus, it can select the attacks  characteristics and apply them to recognize each type of attack. However, the time spent in training together with the choice of classifier affects the IDS performance under real-time networks like IIoT. 
In~\cite{toshpulatov2021anomaly} the authors designed a hierarchical Self-Organizing Map (SOM) for anomaly detection on Smart Meters. They applied an unsupervised learning-based approach to identify additive, deductive and camouflage attacks on smart meters using only three features. For that, they used temporal information by exploiting timestamps and a shift register with tap-delay lines. Even though effective, the required training time and the centralized decision-making do not meet the requirements expected for a dynamic environment such as IIoT. In \cite{marvi2021augmented} the authors proposed an unsupervised machine learning (ML)-based approach for the detection of different types of DDoS attacks and improved the performance of the K-means clustering algorithm. They employed a hybrid method for extracting encoded features, which can better separate the clusters of benign and malicious network flows. The technique improves the selection and extraction features by sequentially combining an Integrated Feature Selection (IFS) algorithm and a Deep Autoencoder (DAE). Even though the model effectively creates threat-free clusters, the time spent on training and decision-making does not meet the requirements expected for a dynamic environment such as IIoT. In~\cite{white2021unsupervised} the authors proposed a machine learning method for detecting anomalies on home IoT devices. They considered an unsupervised one-class learning to monitor the network devices, being flexible to handle a variety of attack patterns exhibited by different devices on the network. The proposal can create a profile of home IoT based on their characteristics to help identify attacks such as DoS. However, only the identification of device characteristics does not eliminate the network attack; and this way, when considering the density and dynamics of the IIoT, the model becomes limited.

\begin{table}[ht]
\centering
\scriptsize
\caption{Security Solutions Requirements for IIoT}
\label{tab:tab1}
\begin{tabular}{c|c c c c c c}
\hline
\multirow{2}{*}{\textbf{Works}} & \multicolumn{5}{c}{\textbf{Requirements}}                              \\  \\ \cline{7-6} 
                                & \textbf{Collaborative} & \textbf{Exclusion} & \textbf{Identification} & \textbf{Distributed} & \textbf{Scalability} \\ \hline 
Lu et al.,  2012                 &             &         &              & \textcolor{blue}{\ding{51}}         & \textcolor{blue}{\ding{51}}         \\ \hline
Yu and Gun,  2010                &            &         &             &          &  \textcolor{blue}{\ding{51}}       \\ \hline
Yi C.  2021          &            &        & \textcolor{blue}{\ding{51}}            & \textcolor{blue}{\ding{51}}         & \textcolor{blue}{\ding{51}}       \\ \hline
Yang et al.,  2017               & \textcolor{blue}{\ding{51}}            &\textcolor{blue}{\ding{51}}       & \textcolor{blue}{\ding{51}}            &         &           \\ \hline
Cevantes et al.,   2019           & \textcolor{blue}{\ding{51}}            & \textcolor{blue}{\ding{51}}        & \textcolor{blue}{\ding{51}}            & \textcolor{blue}{\ding{51}}          &           \\ \hline
Toulouse et al.,   2015           &            &         & \textcolor{blue}{\ding{51}}             & \textcolor{blue}{\ding{51}}          &           \\ \hline
Kailkhura et al.,  2016           & \textcolor{blue}{\ding{51}}           &         & \textcolor{blue}{\ding{51}}            & \textcolor{blue}{\ding{51}}          &          \\ \hline
Hao J et al.,   2020           & \textcolor{blue}{\ding{51}}           &         & \textcolor{blue}{\ding{51}}           & \textcolor{blue}{\ding{51}}        &          \\ \hline
Aboelwaf et al.,   2020           &             &         & \textcolor{blue}{\ding{51}}          & \textcolor{blue}{\ding{51}}        &          \\ \hline
Saba T et al.,  2021           & \textcolor{blue}{\ding{51}}           &         & \textcolor{blue}{\ding{51}}           & \textcolor{blue}{\ding{51}}         &          \\ \hline
Krishna  et al.,   2016           &         & \textcolor{blue}{\ding{51}}          &  \textcolor{blue}{\ding{51}}            &       &        \\ \hline
Martins P  et al.,   2022           &           &    \textcolor{blue}{\ding{51}}     & \textcolor{blue}{\ding{51}}            &          &           \\ \hline
Bhattacharjee S et al., 2018           &          &         & \textcolor{blue}{\ding{51}}          &        & \textcolor{blue}{\ding{51}}          \\ \hline
Toshpulatov M, et al.,   2021           &            &         &   \textcolor{blue}{\ding{51}}           &          &  \textcolor{blue}{\ding{51}}         \\ \hline
Marvi M,  et al.,   2021           &  \textcolor{blue}{\ding{51}}             &       & \textcolor{blue}{\ding{51}}               &      & \textcolor{blue}{\ding{51}}              \\ \hline
White J  et al.,   2021           &         & \textcolor{blue}{\ding{51}}        &  \textcolor{blue}{\ding{51}}        &         &           \\ \hline 
\end{tabular}
\end{table}

\begin{table}[ht]
\renewcommand{\arraystretch}{0.8}
\centering
\scriptsize
\caption{Countermeasures against attacks on different types of services}
\label{tab:tab2}
\begin{tabular}{c| c c c c c c}
\hline
\multirow{2}{*}{\textbf{Works}} & \multicolumn{5}{c}{\textbf{Characteristic}}                                                                                                                     \\ \\ \cline{7-6} 
                                & \textbf{Network }                                               & \textbf{Service }      & \textbf{Dissemination} & \textbf{Attack}                                                      & \textbf{Approach} \\ \hline \hline
Lu et al.,\\ 2012                 & \begin{tabular}[c]{@{}c@{}}WSN \\ massive\end{tabular} & Routing       & Individual    & FDI                                                         & \textcolor{blue}{En-route}        \\ \hline
Yu and Gun,\\ 2014                & \begin{tabular}[c]{@{}c@{}}WSN \\ massive\end{tabular} & Routing       & Clustering    & FDI                                                         & \textcolor{blue}{En-route}          \\ \hline 
Yi C.,\\ 2021                & \begin{tabular}[c]{@{}c@{}}WSN \\ massive\end{tabular} & Routing       & Clustering    & FDI                                                         & \textcolor{blue}{En-route}            \\ \hline \hline
Yang et al.,\\ 2017               & \begin{tabular}[c]{@{}c@{}}IoT\\ small\end{tabular}    & Dissemination & Clustering    & FDI                                                         & \textcolor{blue}{IDS  and  Watchdog}  \\ \hline
Cevantes et al.,\\ 2019          & \begin{tabular}[c]{@{}c@{}}IoT\\ Massive\end{tabular}  & Routing       & Individual    & \begin{tabular}[c]{@{}c@{}}Sinkhole\\ Foreword\end{tabular} & \textcolor{blue}{IDS  and  Watchdog}  \\ \hline 
Krishna  et al.,  \\ 2016          & \begin{tabular}[c]{@{}c@{}}Smart\\Grid\end{tabular}  & Application       & Individual    & \begin{tabular}[c]{@{}c@{}}Theft\end{tabular} & \textcolor{blue}{IDS and KL}  \\ \hline
Martins P  et al.,\\ 2022          & \begin{tabular}[c]{@{}c@{}}IoT\\Small\end{tabular}  & Application       & Individual    & \begin{tabular}[c]{@{}c@{}}DoS\end{tabular} & \textcolor{blue}{IDS}  \\ \hline
\hline
Toulouse et al., \\ 2015           & Network                                                & Application   & Individual    & DDoS                                                        & \textcolor{blue}{Consensus}        \\ \hline

Kailkhura et al., \\ 2016           & \begin{tabular}[c]{@{}c@{}}Smart\\Grid\end{tabular}                                                & Application   & Individual    & DDoS                                                        & \textcolor{blue}{Consensus and ML}      \\ \hline
Hao J et al., \\ 2020           & \begin{tabular}[c]{@{}c@{}} WSN \\massive\end{tabular}                                                & Application   & Individual    & FDI                                                        & \textcolor{blue}{Consensus and Kalman} \\ \hline 
Bhattacharjee S et al.,\\ 2018          & \begin{tabular}[c]{@{}c@{}}Smart\\Grid\end{tabular}  & Application       & Individual    & \begin{tabular}[c]{@{}c@{}}FDI\end{tabular} & \textcolor{blue}{Consensus and ML} \\ \hline \hline
Aboelwaf et al.,\\ 2020         & \begin{tabular}[c]{@{}c@{}}IIoT\\\end{tabular}  & Routing       & Individual    & \begin{tabular}[c]{@{}c@{}}FDI\\\end{tabular} & \textcolor{blue}{ML}  \\ \hline
Saba T et al.,\\ 2021          & \begin{tabular}[c]{@{}c@{}}IoT\\Small\end{tabular}  & Routing       & Individual    & \begin{tabular}[c]{@{}c@{}}Sinkhole\\DDos\\ Sybil\end{tabular} & \textcolor{blue}{IDS and ML}  \\ \hline
Toshpulatov M, et al.,\\ 2021          & \begin{tabular}[c]{@{}c@{}}Smart\\Grid\end{tabular}  & Application       & Individual    & \begin{tabular}[c]{@{}c@{}}FDI\end{tabular} & \textcolor{blue}{IDS and ML}  \\ \hline
Marvi M,  et al.,\\ 2021          & \begin{tabular}[c]{@{}c@{}}Network\end{tabular}  & Application       & Clustering    & \begin{tabular}[c]{@{}c@{}}DDoS\end{tabular} & \textcolor{blue}{IDS and ML}  \\ \hline
White J  et al.,\\ 2021          & \begin{tabular}[c]{@{}c@{}}IoT\\Small\end{tabular}  & Application       & Individual    & \begin{tabular}[c]{@{}c@{}}DDoS\end{tabular} & \textcolor{blue}{IDS and ML}  \\ \hline
\end{tabular}
\end{table}

Table~\ref{tab:tab1} summarizes the requirements that must be taken into account by a secure data dissemination \textit{middleware} for dense IIoT networks to preserve the authenticity of the  collected data. A solution should commonly preserve the availability of devices, not interfere with their functioning, and include the least overhead. Further, it should respect security requirements, such as availability and authenticity, and  be tailored to the characteristics of a dense IIoT. Table~\ref{tab:tab2} presents the main countermeasures already applied against FDI attacks in different network infrastructures. Thus, we observe that most of the existing solutions do not adopt the verification of sensing data, do not exclude the origin of the attack and centralize the decision making. In addition, machine learning solutions, while effective, require a central unit and training and learning time which, when applied to dynamic IIoT, allows attacks to act for a while until they are detected. To overcome these drawbacks, the solution must act dynamically in IIoT networks, considering the source of the attack, authenticating the sensed data and decentralizing decision-making through network devices. 

\section{A System for Mitigating FDI Attacks in IIoT Networks}
\label{sec:proposal}

This section introduces the CONFINIT (\emph {\textbf{CON}sensus Based Data \textbf{FI}lteri\textbf{N}g for I\textbf{I}o\textbf{T}}) system for mitigating false data injection (FDI) attacks on dense IIoT networks over the data dissemination service.  CONFINIT acts as \textit{middleware} increasing the security of the data dissemination service against the rise of FDI attacks, intending to deteriorate the functioning of the dissemination service in the network. It works on fixed IIoT networks with a high volume of data continuously collected and disseminated. CONFINIT applies surveillance by  {\it watchdog} monitoring among participants to detect FDI attacks on the IIoT network and makes use of a \textbf{Collaborative Consensus} technique for decision-making. The adoption of both approaches aims to create accurate and dynamic collaborative filtering. CONFINIT supports the authenticity and availability of the data collected and disseminated on the IIoT network to assist the application decision making, and act as a third layer of defense. In this work, we assume that all nodes in the network went through some access control mechanism, ensuring the authenticity and legitimacy of the node's actions in the network. Initially, we describe the models of IIoT network, clustering, and data dissemination in the dense environment where CONFINIT plays, as well as the FDI attack behavior acting in the data dissemination service. Next, we present details about the architecture and CONFINIT algorithms and its operation manner.

\subsection{Network and Dissemination Model}
\label{sec:rede}
\begin{figure}[ht]
    \centering    
    \includegraphics[width=0.80\linewidth]{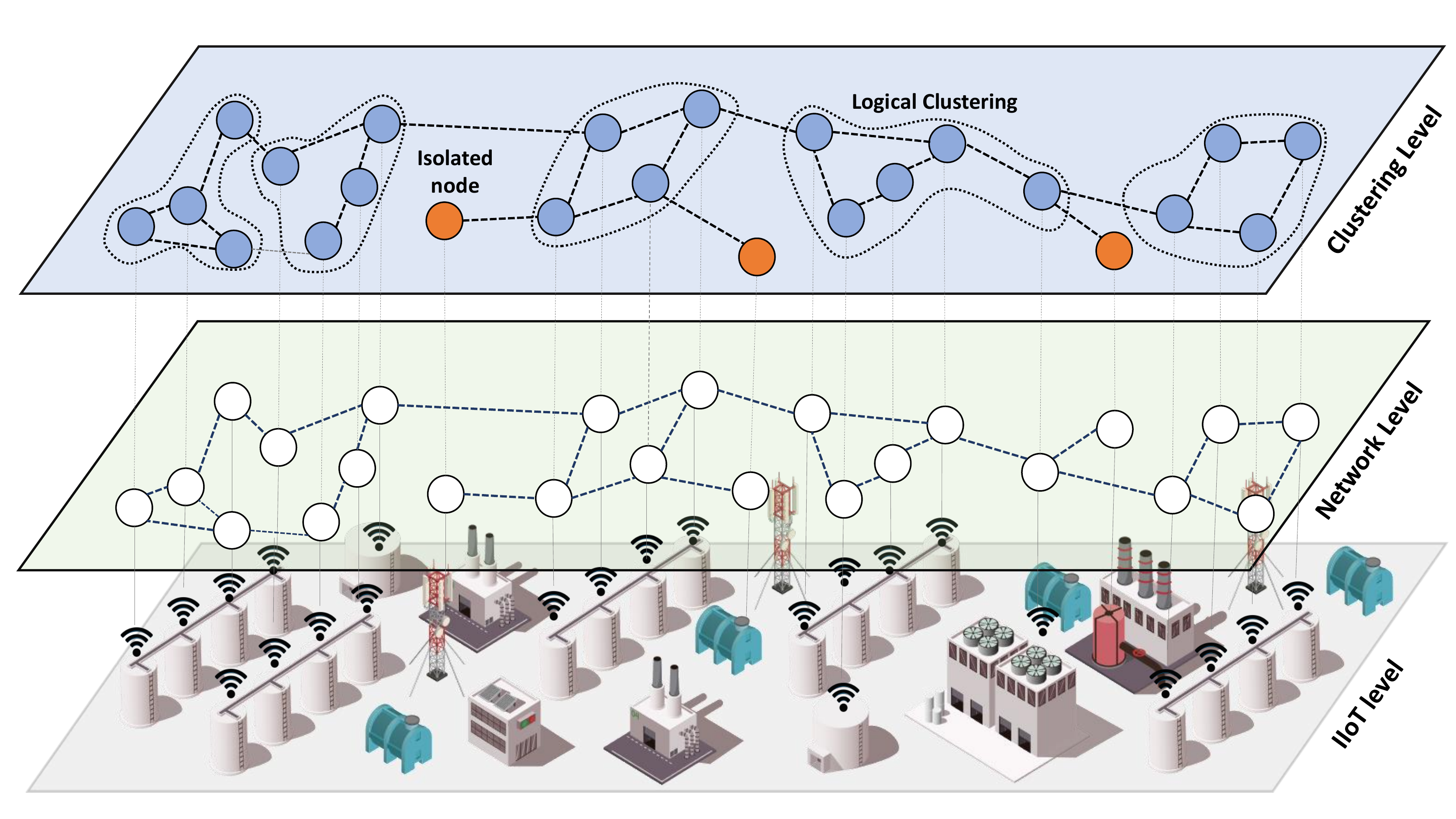}
      \caption{The IIoT network model}
      \label{Fig:network}
\end{figure}

We consider a dense IIoT network composed by a set of nodes denoted by $N = \{n_1, n_2, n_3, ..., n_n\}$, where $n_i \in N$. Each node $n_i$ owns a unique identity address ($Id$)
to identify it in the network. Moreover, all nodes 
in the IIoT perform their functions of collecting and disseminating data according to the IIoT application domain in which they are inserted.~\cpa{Each environment establishes the type and frequency of the data transmission.} Further, we consider all statics nodes, without energy restrictions, because they belong to an industrial environment, such as the petrochemical industry. We also adopt the cluster configuration to improve network organization, save resources, optimize communication between participants, and increase the lifetime of the network operation. Figure~\ref{Fig:network} presents the network model, defined in a three-level structure, and the first level comprises the IIoT nodes, the second level performs the communication between the nodes and the third level coordinates the formation of clusters.

Clusters are established according to the similarity between the data collected by the nodes in the network and form the subset ${A_{\alpha}}\subseteq N$ of nodes; $\forall$~$n_{j} \in {N}$. For the cluster establishment, we regard greatness associated with space and time among the nodes, i.e physically close and similar data sharing, respectively.  In this way, the clusters are formed considering the data similarity relation between the nodes, taking into account the time and physical proximity, that is, within the transmission radius, which enables direct communication between them. All nodes start its operation alone, looking for neighbors to establish clusters and hence spreading its data. Each node can only integrate one cluster per time. The similarity technique works to ensure that only nodes with similar readings integrate the clusters. This technique is an adaptation of the bio-inspired clustering model developed by~\cite{gielow2015data}. We also label all nodes as honest, suspicious, and attackers. Honest nodes integrating a cluster can play as a common node or leader.  Common nodes have the function of collecting and disseminating their data to the cluster leaders. Leaders are responsible for receiving data from honest nodes and making it available to the application. Suspicious nodes cannot integrate any cluster because they can present a crash failure or misbehaving, i.e. damaging the data dissemination service. Attacker's nodes operate by disseminating deliberately false data on the network and exhibit this behavior more than once.

We represent the IIoT network structure over time by a simple graph $G = (V, E)$, where the set of vertices means the nodes, and the set of edges means the connections among those nodes. That representation allows us to identify variations of communication interaction among nodes in the network~\cite{sizemore2017dynamic}. On that IIoT graph (network level), we set up temporal clustering graphs (clustering level), named as $H = (V, E, T)$ that represents the data-sharing relationship on space and time by the nodes, searching for cluster establishment. Time brings up the influence on readings variations between nodes, and space points out communication ties among nodes.  In this way,  the occurrence of contact among nodes is represented by a triple ($i, j, t$) indicating an existing edge between nodes $i$ and $j$ (or from node $i$ to node $j$ in the directed case) at time $t$. Those interactions over time establish  graphs  $H_0, H_1, ..., H_n$, being one for reach moment $T = 0, 1, ...,T$, throughout the network functioning.

\begin{figure}[ht]
    \centering 
    \includegraphics[width=0.75\linewidth]{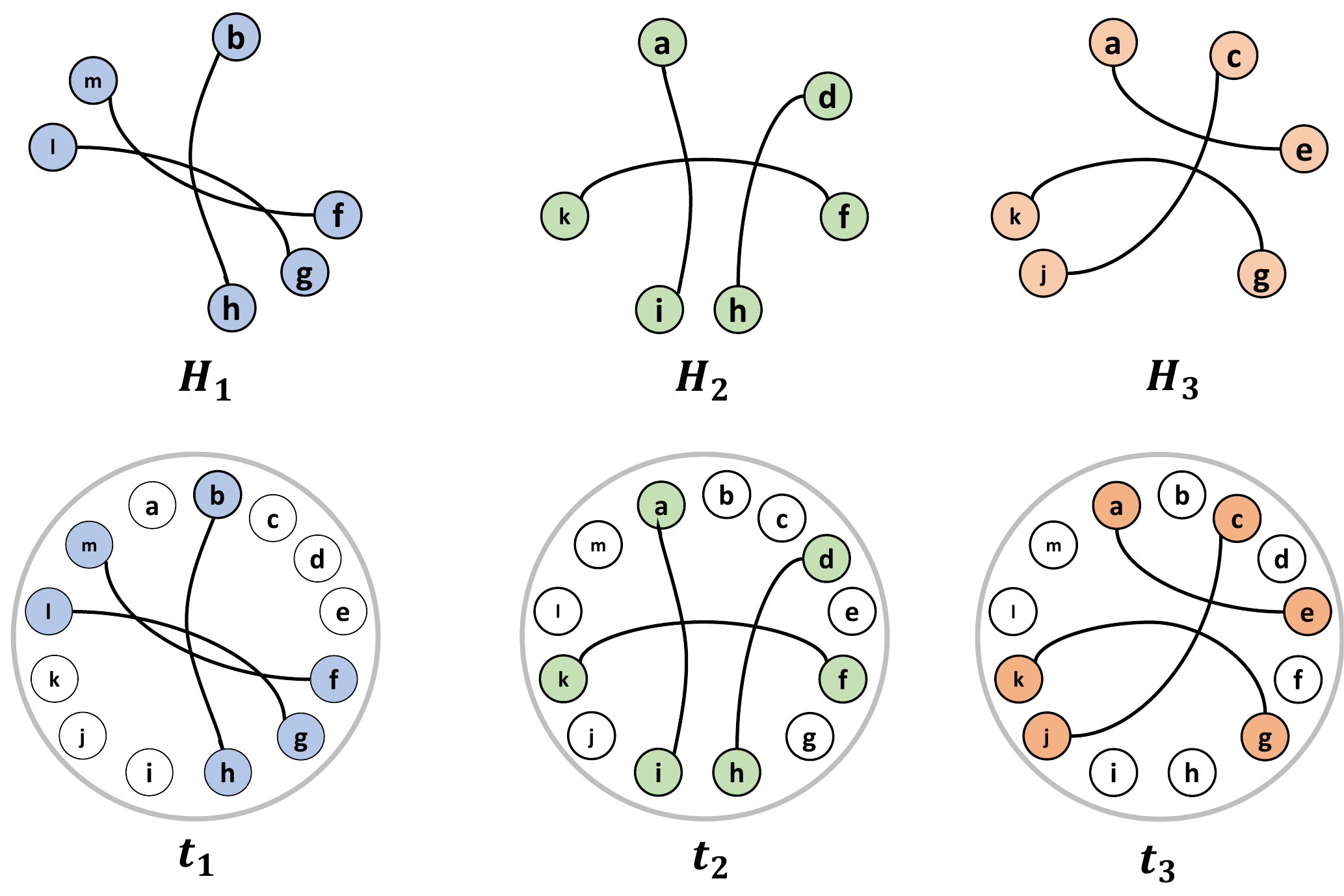}
      \caption{Clusters connections variations over time in network}
      \label{Fig:grafte1}
\end{figure}

Figure~\ref{Fig:grafte1} illustrates the IIoT nodes interactions over time represented by graph  $H = (V, E, T)$  with $V = \{a, b, c, d, e, f, g, h, i, j, k\}$, and whose $E$ varies over time, and for each instant $t_1, t_2, t_3$, emerges a new instance of graph $H$ with active connections,  where  $E_1 (H_1) = \{\{b, h \},\{l, g\}, \{m, f\} \}$, $E_2(H_2) = \{\{a, i\}, \{k, f\}, \{d, h\}\}$; and  $E_3(H_3) = \{\{j, c\}, \{a, e\}, \{k, g\}\}$. Thus, the nodes keep dynamic links over time so that the data collection among them determines when the cluster should be continued or interrupted.

The dissemination service requires that incoming nodes initially disseminate data looking for neighbors to form clusters and by cluster leader making their data available to the application.  In addition, the type of data collected is numeric, with a continuous and dynamic flow that impacts how the data is made available. Due to the data's large volume generated from IIoT nodes, the delivery of the disseminated data follows in a cluster way, being the data similarity the criterion for setting up the clusters.  Therefore, only the nodes that are part of the clusters make their data available to the application. Thus, isolated attackers and nodes are unable to disseminate their data to the application. Also, we consider standards aiming to preserve the security attributes of authenticity and availability of the collected data by the nodes. By ensuring these requirements, we provide safe and threat-free dissemination, in addition to meeting the demands imposed by applications.

\subsection{False data Injection Attack Model}
\label{sec:ataque}
Among a variety of attacks that act similarly to data injection attack (FDI), we highlight camouflage, recommendation, and deduction, each one with its peculiarities~\cite{bhusal2022coordinated}. For simplicity, we adopting FDI as the standard because it is more insistent against data-oriented networks like IIoT. We apply a FDI attack model based on~\cite {deng2016false}, we adapt the way the attack acts on the data to change only one data at a time, instead of the data array as in the original. Thus, attacking the model of data collected and disseminated in our IIoT becomes adequate. The FDI attack takes place in two ways to target the disturbance in the network. In a first way, the attacker captures an authenticated node ${N_{a}}$ on the network and manipulates its data, whether changing or falsifying them. In a second way, ${N_{a}}$ itself is the attacker, who makes the same actions to defraud the network. Both FDI models employ the same principle to generate distinct types of data inconsistencies, overload and degrade the network. Therefore, to simplify the explanation of the model,  we apply the first form as the main one. Thus, honest node ${N_{c}}$ after having the manipulated data by ${N_{a}}$, spreads this false data to damage the process of forming network clusters.

\begin{figure}[ht]
    \centering
    \includegraphics[width=0.65\linewidth]{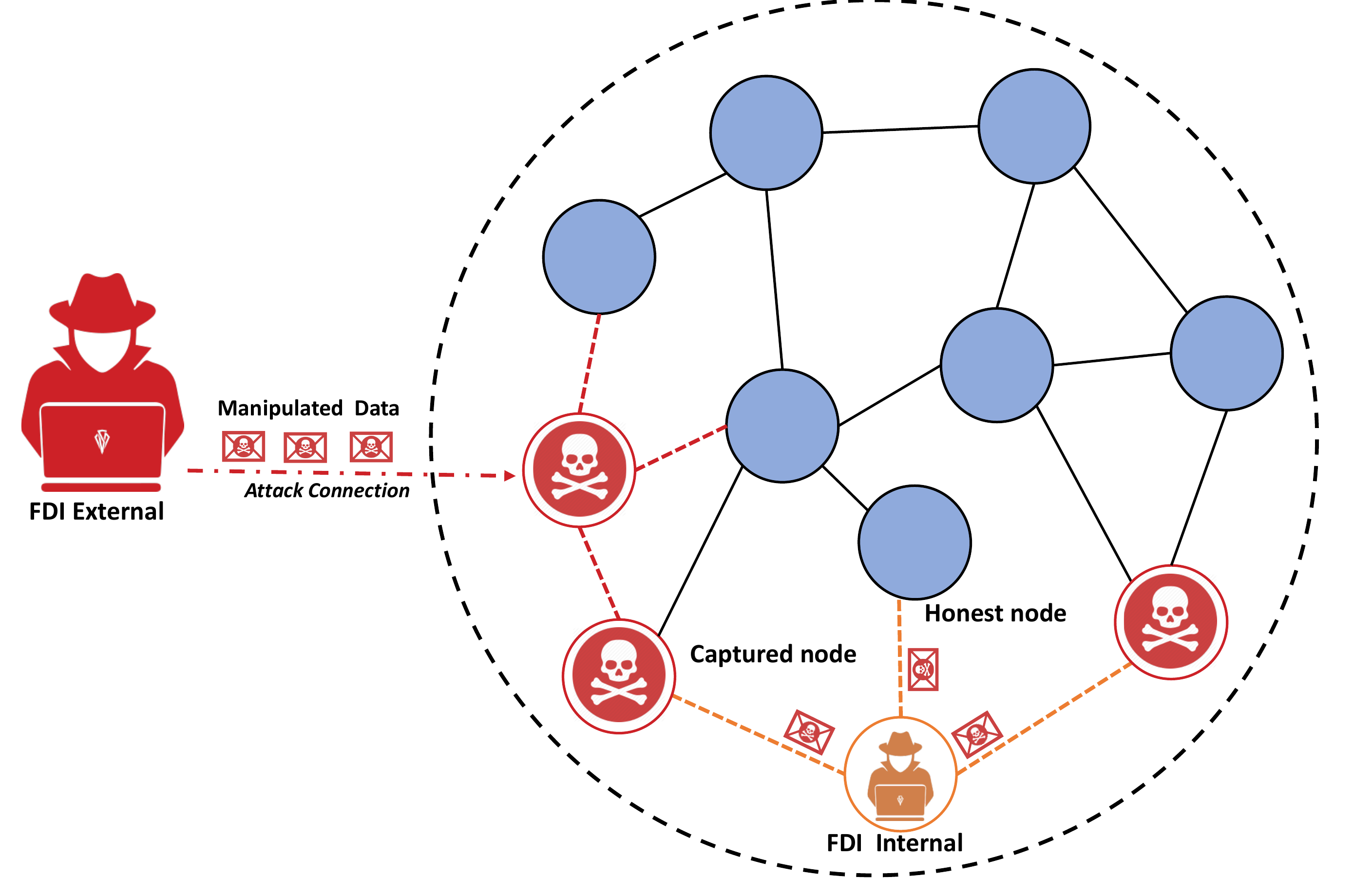}
      \caption{Behavior of FDI attacks in the IIoT network}
      \label{Fig:ata_f}
\end{figure}

Figure~\ref{Fig:ata_f} exhibits the two manners of FDI attacks acting on the IIoT network. In the first one, the external attacker bypasses the access control of the network in order to capture a node and makes it propagate false data. In the second one, the internal attacker circumvents access control and initiates the propagation of false data on the network. We consider that FDI attacks occur by the exploitation of vulnerabilities arising from previous attacks against the networks like Sybil, black hole, among others which target finding loopholes and paving the way for attacks like FDI. The attackers usually target to analyze the network for understanding its operation, and, thus,  carry out a more effective attack.  Besides, we consider that to violate the IIoT network, the attacker must have full knowledge about the network operation and the type of data traffic. This type of behavior makes it hard to identify the source of the FDI attack, increases the time the network malfunctions, and creates inconsistency on the data made available. Therefore, understanding this behavior becomes a challenge for the mechanisms mitigating the FDI attack from the IIoT. Besides, the faster and more efficient the identification of the attacker, the better the network recovery and works normally.

\subsection{CONFINIT architecture}
\label{sec:mec}

\begin{figure}[ht]
    \centering
    \includegraphics[width=0.90\linewidth]{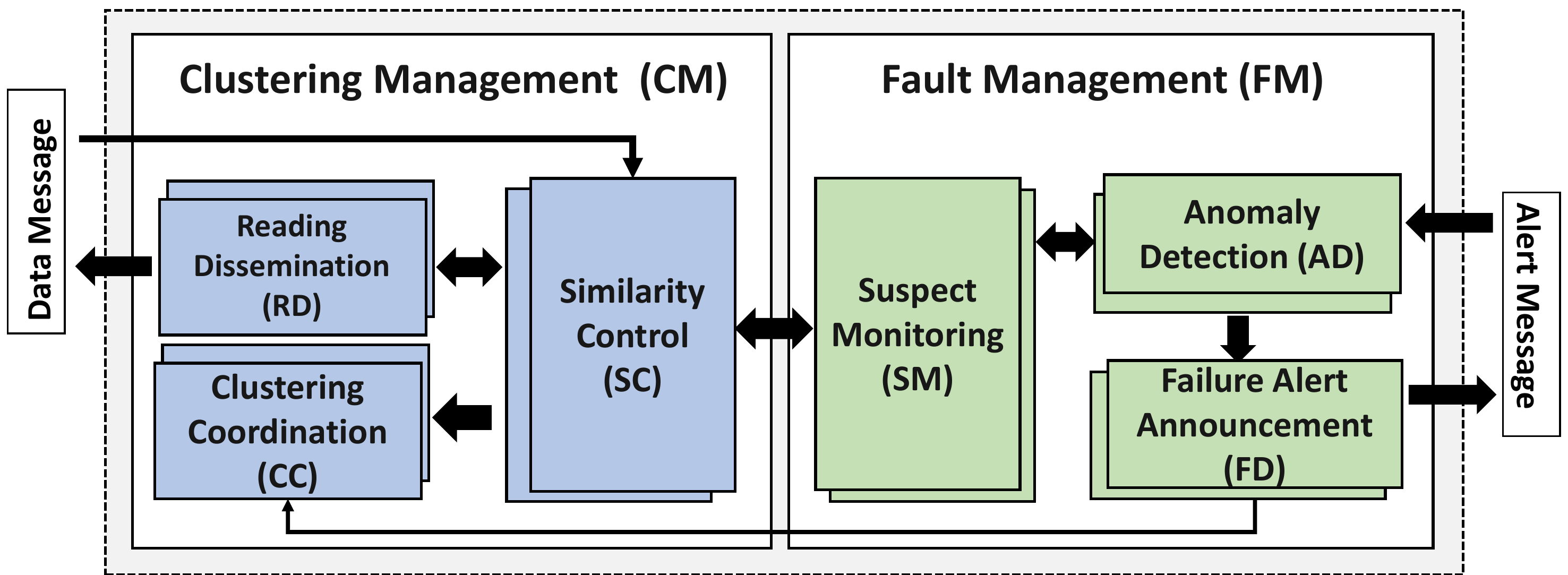}
    \caption{The CONFINIT architecture}
    \label{Fig:arquii}
\end{figure}

 The CONFINIT architecture is composed by the ~\textbf{Clustering Management} and \textbf{Fault Management} modules, as shown in Figure~\ref{Fig:arquii}. They work jointly to provide infrastructure n and security for the data disseminated in the IIoT network; so that the first module organizes the network in~\textit{clusters} and the second module deals with monitoring and detecting and isolating malicious nodes acting as FDI attackers.

\textbf{Clustering Management Module (CM)} sets up clusters over time based on a similarity threshold of readings data from the nodes physically close to determine when they can integrate or not a cluster. To carry out its role, CM owns the~\textit{Similarity Control (SC)},~\textit{Clustering Coordination (CC)} and~\textit{Reading Dissemination (RD)} components. Thus, SC coordinates the receipt and interpretation of data messages exchanged by nodes by applying a similarity threshold. CC acts to establish and maintain the clustering based on the similarity threshold from the nodes data readings, as well as it controls the leader election. RD works propagating the nodes readings, the number of readings, and the amount of neighbors. Thus, all nodes that receive the data message will know when they are part of a cluster. The clustering process runs locally at each node, using readings that respect the similarity threshold between the participants. So each node maintains only a partial view of the network due to not overload its functions.

\textbf{Fault Management Module (FM)} ensures the security of data dissemination service among nodes, allowing only authentic readings are disseminated over the network. For that, it takes the \textit{Suspect Monitoring (SM)}, \textit{Anomaly Detection (AD)}, and~\textit{Failure Alert Announcement (FD)} components. SM is responsible for monitoring nodes that transpassing the similarity threshold. AD deals with the collaborative consensus to detect FDI nodes, that acts as an agreement and uniformity of opinions established by nodes through data reading exchanges among the cluster neighbors. Lastly, the FD component coordinates the isolation of FDI nodes and informs the cluster leaders about the threat. The fault management process happens every time a node is detected as an attacker so that security is maintained by the network participants themselves without the help of external entities

\subsection{Clustering Management Module (CM)}

The \textbf{CM} module organizes the IIoT network in clusters based on the leaders in order to create a topology to organize and facilitate the data exchange among nodes and hence its delivery to the application. Initially,  nodes begin their operations isolated, transmitting and collecting data messages from neighboring nodes for the composition of clusters. The nodes propagate data messages in broadcast to be known by the physically close neighbors. The data messages exchanged among nodes to form the cluster contain five fields $<DM,Id,{L_{ind}},{N_{viz}},{L_{agr}}>$: the type of message ($DM$), the identifier of the emitter node ($Id$), individual reading (${L_{ind}}$), number of neighbors of the node (${N_{viz}}$) and aggregate reading (${L_{agr}}$). If any field is empty, the message is automatically discarded by the receiving node. Otherwise,  the component \textit{Similarity Control (CS)} analyzes the information of the fields and calculates the similarity of readings between nodes considering the values of individual readings, number of neighbors, and aggregated readings of these neighbors.

\begin{algorithm}
\caption{Clustering establishment}
\label{alg:clus}
\begin{algorithmic}[1]

\Procedure{SendDataMessage()}{$a,b$}\\
\sepp $Send(Id, {L_{ind}}, {L_{agr}},|{N {viz}}|)$ \\
\sepp $WaitIntervalnextmsg$ \\
\sepp $RControlTimerExpire()$
\EndProcedure 
\Procedure{ReceiveDataMessage}{($Org,iR,aR,nR$)} \\
\hsep ${N_{viz}}[Org] \leftarrow \{iR, aR, nR\}$ \\
\sepp $localRead \leftarrow {L_{agr}}()$ \\
\sepp \textbf{if} ${(|iR - localRead| < Threshold)}$ \textbf{and} ${(|{L _{ind}}() - aR| < Threshold)}$ \textbf{then} \\
\sepp \sepp ${N_{viz}} \leftarrow S{N_{viz}} \bigcup \{org\}$\\
\sepp \textbf{else if} \\
\sepp \sepp ${N _{viz}} \leftarrow {N _{viz}} \bigcap \{org\}$\\
\sepp \textbf{and if}
\EndProcedure 
\end{algorithmic}
\end{algorithm}


Algorithm~\ref{alg:clus} exhibits how \textbf{CM} works to establish cluster in the IIoT. Periodically, each node sends a data message (DM) in~\textit{broadcast}, stating the device identifier ($Id$), the current reading (${L_{ind}}()$), the average aggregated reading of the neighbor nodes (${L_{agr}}()$) and the number of neighbors (${N_{viz}}()$). Further, the DM sending takes into account a predefined time interval, in our case it was established  in \textit{2ms}, to avoid simultaneous transmissions among all nodes (\textit {l}.1-\textit{l}.5). Upon receiving a DM (\textit{l}.6), the node will know the origin ($org$), the individual reading ($iR$) of the emitter node, the aggregate reading ($aR$) of its neighborhood, and the number of neighbors  ($nR$). The receiver node then updates ($N_{viz}$) with all those information ($iR$, $aR$ and $nR$) (\textit{l}.7). The average aggregate reading forwarded by the emitter node collaborates to know whether the reading of the receiver node satisfies or not the similarity threshold (\textit {l}.8-9) With the similarity successfully, the neighborhood structure (${N_{viz}}$) is updated to add the emitter node ($org$) (\textit{l}.10-11). Otherwise, the emitter node ($org$) cannot participate in the neighborhood (${N_{viz}}$) (\textit {l}.12). The clustering process takes place dynamically at each node in the IIoT network, ensuring that everyone can keep their neighborhood structure updated.

We compute the similarity relationship between two nodes based on the values of the reading data, in addition to considering the threshold value between them. Equation~\ref{eq:firefc} verifies the similarity between two readings, following the model proposed by~\cite{gielow2015data}. In its composition, the number of neighbors and the aggregate readings of these neighbors are attributed to the node readings, to determine when there is a similarity between two nodes. In the equation, \textit{X} means the current reading of the node, and \textit{Y} the reading to be compared whether the similarity threshold is being respected or not.

\vspace{0.3cm}
\begin{equation}
\label{eq:firefc}
	\left | Y -
	\frac{X + \sum_{v \in S{N_{viz}}} ({N_{viz}}[v].aR * {N _{viz}}R[v].nR)}
	{1 + \sum_{v \in S{N _{viz}}} ({N_{viz}}[v].nR)}
 \right |< CThresh
\end{equation}
\vspace{0.3cm}

During the formation of clusters, the election of leaders happens by considering the number of neighbors as a criterion for defining the leaders. Further, clusters can contain more than one leader in reason of the large number of nodes in the network. In this way, all cluster nodes will be able to identify the cluster leaders. As new interactions between nodes occur, the election process also takes place dynamically.  The leader definition establishes a hierarchy among the nodes and organizes how the data dissemination occurs between leaders and common nodes. Thus, all data collected by the common nodes must be forwarded to the leader, it makes them available for the application. Clusters carry out maintenance under failures or exit of the nodes, redoing the process whenever necessary. In addition, all nodes maintain an updated neighbor list as nodes enter or leave the cluster, and the neighborhood structure updating occurs in conjunction with the DM exchanges.

\subsection{Fault Management Module (FM)}

The \textbf{FM} module takes care to analyze threats preventing FDI attackers from spreading false data in the IIoT network. Therefore, When detecting an FDI attack, the detector nodes propagate an alert message so that the cluster leader can disseminate it to other cluster leaders in the IIoT network. The data into the alert message consists of four fields $<AM,Id_{int},Id_{ataq},L_{ataq}>$: the type of message ($AM$), the node identifier of the attacker (${Id_{int}}$), attacker's identifier ($Id_{ataq}$), and individual reading ($L_{ataq}$). Algorithm~\ref{alg:dect} details the FM operation against FDI attacks. The detection action takes place effectively after the first DM exchange, due to nodes require other messages to compare it with ones received previously. Firstly, FM classifies  as suspects nodes those whose forwarded values do not respect the similarity threshold of readings over the clustering formation, and adds them in the suspect list of each node integrating of the cluster (\textit{l}.1-\textit{l}.7). The suspects list supports the failure detection as nodes can sometimes only exhibit crash behavior, which does not mean an attack. Case the node  does not belong to the suspect list but its readings are suspicious, FM will insert it on the suspect list (\textit{l}.8-\textit{l}.16). Otherwise, FM will be removed of the suspect list when its readings meet the threshold (\textit{l}.17-\textit{l}.20). Thus, after the detection confirmation, FM performs the exclusion of the FDI attacker to prevent it from degrading the IIoT. After being identified, mitigated, and excluded, through AM the attacker has its identifier and its readings disseminated by the cluster leader to the other leaders.

\begin{algorithm}
\caption{Attacker detection}
\label{alg:dect}
\begin{algorithmic}[1]
\Procedure{CheckSuspicious}  $(Id, ConsensusParticipant)$ \\
\sepp \textbf{if}$(ID$ $\in$  $SuspectList$ $\And$ $ConsensusParticpant$ == $False$)\\
\sepp \sepp $return 1$ \\
\sepp \textbf{else if}$(ID$ $\in$  $SuspectList$ $\And$ $ConsensusParticpant$ == $True$)\\
\sepp \sepp $return 2$ \\
\sepp \textbf{end if}
\EndProcedure 

\Procedure{CheckAttack}$(Id, Read)$\\
\sepp $Valid \longleftarrow checkSuspicious  (Id, ConsensusParticpant)$ \\
\sepp \textbf{if}$(Read \leq Thresholdconsensus$)\\
\sepp \sepp $Switch \longleftarrow Valid$\\
\sepp \sepp \sepp \textbf{\textit{Case 1}}\\
\sepp \sepp \sepp $Atklist \leftarrow  Attacklist \bigcup \{Id, Read\}$ \\
\sepp \sepp \sepp \textbf{\textit{Case 2}}\\
\sepp \sepp \sepp $Suspectlist \leftarrow Suspectlist \bigcap\{Id, Read\}$ \\
\sepp \sepp  $Switch \longleftarrow Valid$\\
\sepp \textbf{else}\\
\sepp \sepp $Suspectlist \leftarrow Suspectlist \bigcup\{Id, Read\}$ \\
\sepp \textbf{end if}
\EndProcedure 
\end{algorithmic}
\end{algorithm}


The Consensus Equation~\ref{eq:conse} computes the value of the deviation standard to be applied in two steps in the collaborative filtering to optimize the attack detection. The collaborative consensus is the agreement and uniformity of opinions established by nodes through the data exchange among the cluster neighbors. The usage of standard deviation aims to determine how many discrepant the readings are in a given moment. At first, it takes into account the data from the set $M_{c}$ = $({{c _{i}},{c _{i+1}}},...,{{c_{n}}})$ comes from  the neighborhood information of the receiving node. Thus,~\textit{SD} is the value from the standard deviation, obtained by $\sum_{i=1}^{n}$ which, in turn instead, it adds all values of the set~\textit{$M_{c}$}, from the first position to the last one. The value of $X_{i}$ means  the position~\textit{i} in the data set $M_{c}$ to be compared with $M_{A}$; and $M_{A}$ corresponds to the arithmetic average of the data. The value of $N$ means  
the amount of data in~\textit{$M_{c}$}. In the next step,  seen in Figure~\ref{Fig:filtra}, the equation employs a new of set, called $M_{b}$ = $({{x_{i}}, {x_{i + 1}}},...,{{x_{n}})}$, obtained by the consensus region, so that $\sum_{i = 1}^{n}$ adds all values from the set $M_{b}$ from the first position \textit{(i = 1)} to the position~\textit{n} of the total nodes belonging to the consensus.  $X_{i}$ represents the $i_{th}$ value in the data set $M_{b}$,  $M_{A}$ means the arithmetic average of the data. $N_{ma}$ determines the amount of data to be evaluated in the equation. Lastly,~\textit{ThresholdConsensus} sets up the threshold for comparing values, adjusting it according to the data variation and the end application requirement.

\vspace{0.3cm}
\begin{equation}
\label{eq:conse}
	 SD =
	\sqrt{\frac{\sum_{i=1}^{n}\left (X_{i} - M_{A}  \right )^{2}}{N}}
    \leq ThresholdConsensus
\end{equation}
\vspace{0.3cm}

\subsection{Collaborative Filtering}

\begin{figure}[ht]
    \centering
    \includegraphics[width=0.60\linewidth]{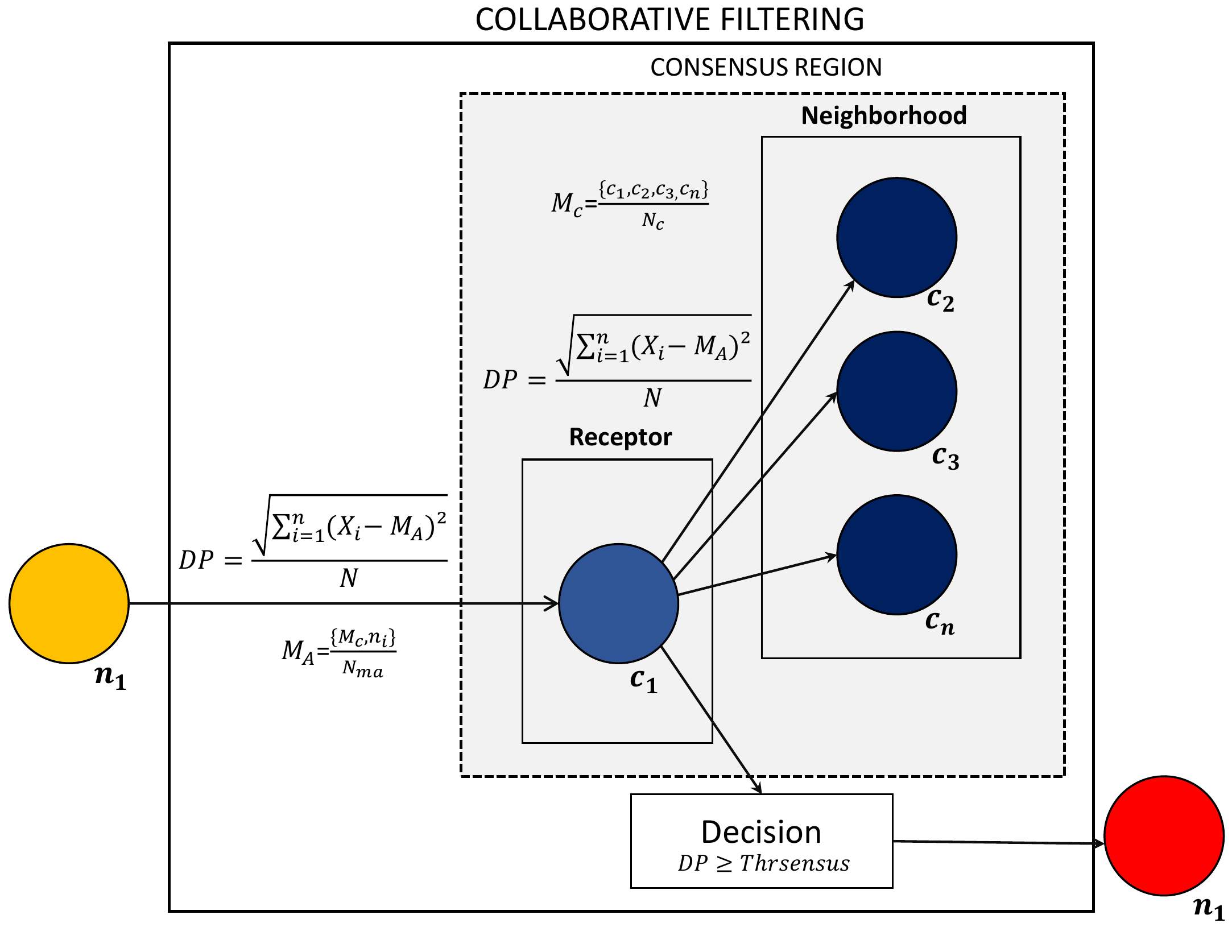}
    \caption{Steps of the Collaborative Filtering}
    \label{Fig:filtra}
\end{figure}

We apply the collaborative filtering in two steps to appoint out attacker nodes that aim to damage the data dissemination service during the establishment of the clusters process. For that, we employ a dynamic and collaborative filtering scheme to classify nodes like honest or attacker by assigning a predefined \textit{consensus threshold}. Thus, the filtering  acts as an additional layer in the IIoT protection, enabling accurate detection while minimizing incorrect attacker identification by CONFINIT. Figure~\ref{Fig:filtra} presents the stages of the collaborative filtering adopted by CONFINIT and how Equation~\ref{eq:conse} is applied over this phase. When receiving a data message from $n_1$ (suspicious node), $C_1$ will analyze the received reading value in order to determine or not  $n_1$ as an attacker. Firstly, $C_1$ applies Equation~\ref{eq:conse} with its reading value and neighbor reading values to get the medium reading value on its consensus region. Next, $C_1$ reapplies the equation taking into account the valor of consensus region and the reading value of $n_1$. 
Lasty, $C_1$ gets to know whether $n_1$ respects the \textit{ThresholdConsensus},  described as Thrsensus in the figure, and hence if $n_1$ is or not an FDI attacker. The collaborative filtering scheme addresses to meet the devices density and data volume generated in the IIoT network, allowing the decision-making process about the presence of FDI attackers to be precise and distributed among nodes, hence eliminating the use of external entities.

\begin{figure}[ht]
    \centering    
    \includegraphics[width=0.85\linewidth]{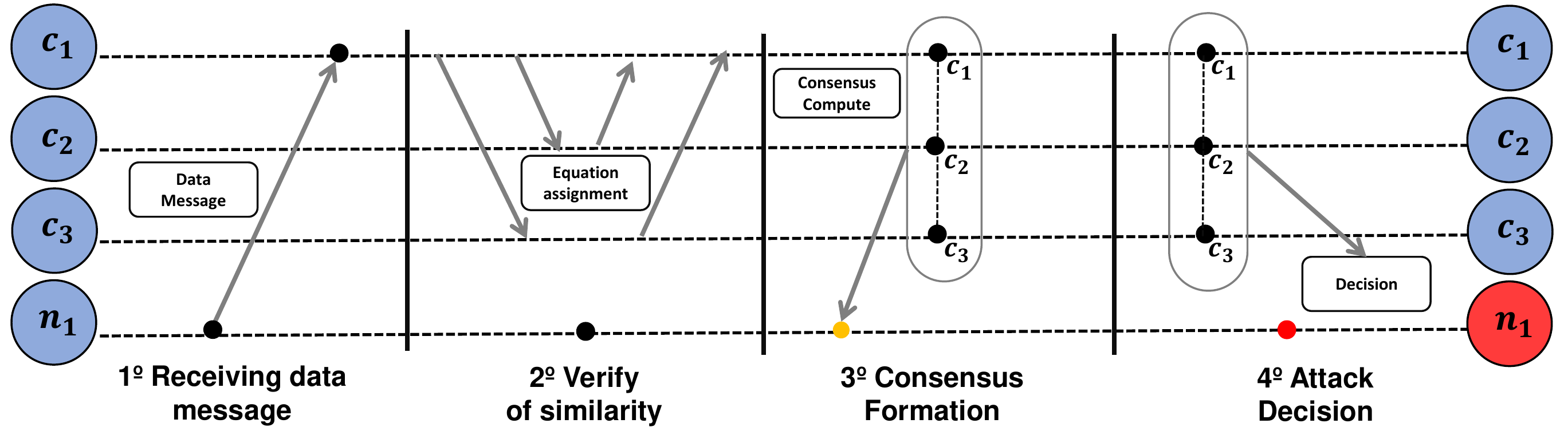}
     \caption{Interactions among entities during CONFINIT phases}
      \label{Fig:relac}
\end{figure}

We describe the relationships between the different phases from the formation of clusters and detection of FDI attackers by CONFINIT in the IIoT network. One phase interferes with another since, from the first data message exchange, all nodes are being monitored, creating a collaboration between all IIoT network nodes. Besides, one phase interferes with the other, as when a malicious node is authenticated as honest, it can participate in the cluster and spread its false data across the network. These actions are highlighted by each component of the mechanism, creating synergy between all. Figure~\ref{Fig:relac} illustrates the CONFINIT operation starting with the exchange of data messages and the similarity verification for the formation of clusters, followed by the formation of consensus and collaborative filtering for detecting attacks. In summary, CONFINIT consists of four phases related to the performance of nodes in the IIoT network. Each phase establishes a partial view of the process of organizing the network and providing data authenticity, determining how these relationships collaborate for the correct functioning of IIoT.

\subsection{CONFINIT Operation}
\label{sec:funci}

This section presents an example of CONFINIT  operation in the petrochemical industry, starting with the formation of clusters and continuing until the detection and exclusion of the FDI attacker. Suppose a petrochemical industry with several gas storage tanks where IIoT objects are marked monitoring the reservoir level that continuously changes its values due to the large volume of gas produced. In this way, CONFINIT works starting the first exchange of messages among these tanks for forming and maintaining clusters. The clustering establishment with CONFINIT takes place dynamically for each node of the IIoT network.

\begin{figure}[ht]
    \centering    
    \includegraphics[width=0.70\linewidth]{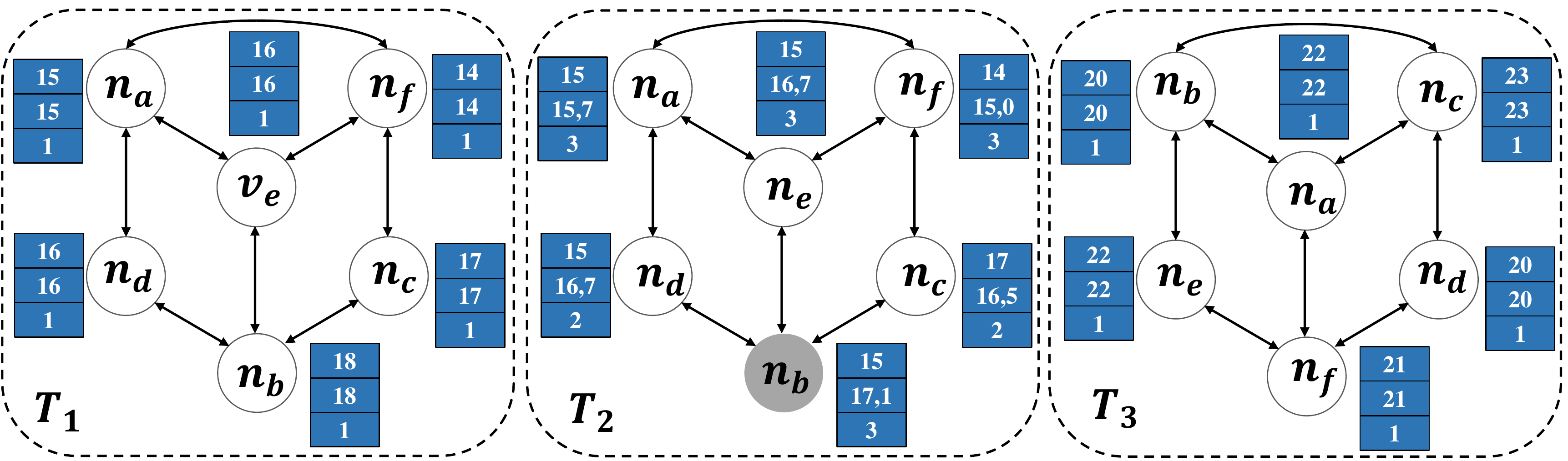}
    \caption{Clustering formation and leader election over the time}
    \label{Fig:funci}
\end{figure}

Figure~\ref{Fig:funci} illustrates how CONFINIT acts in the cluster formation and leader election. Solid edges indicate nodes with transmission range overlap and that can exchange data messages (DM) among them. The boxes next to each node mean, from top to bottom, the individual reading of the node, the neighbor's node's aggregate reading, and the value of aggregate readings. Further, we run with the $CThresh$ equal to three for cluster formation, and each instant~\textit{T} corresponds to a DM exchange round among nodes to establish clusters and elect leaders. In the figure, CONFINIT makes use of a ($CThresh$) =~\textit{3}, and in $T_1$ all nodes start exchanging data messages to update the neighbors list and aggregate readings by Equation~\ref{eq:firefc}. They begin the clusters forming and  and electing leaders based on the readings obtained at the early instant~$Ra_{Tn-1}$. In $T_2$, It is achieved the similarity calculation among the nodes, 
$Ra_{T2}(n_a) = \frac{15+1*16}{1+1}$,
$Ra_{T2}(n_b) = \frac{16+1*15+1*18}{1+1+1}$,
$Ra_{T2}(n_c) = \frac{18+1*16+1*16+1*16+1*17}{1+1+1+1}$,
$Ra_{T2}(n_d) = \frac{17+1*16+1*18}{1+1+1}$,
$Ra_{T2}(n_e) = \frac{16+1*17+1*18}{1+1+1}$. 
After that, since node~\textbf{$n_b$} owns the largest number of neighbors, it is elected as the cluster leader. As the clustering works dynamically on each node, in $T_3$, it is again carried out due to a new data message exchanged among the nodes, and thus the new similarity calculation is applied to maintain the cluster formation,
$Ra_{T3}(n_a) = \frac{20+1*22+1*21+1*23}{1+1+1+1}$,
$Ra_{T3}(n_b) = \frac{22+1*20+1*23}{1+1+1}$,
$Ra_{T3}(n_c) = \frac{23+1*22+1*20}{1+1+1}$,
$Ra_{T3}(n_d) = \frac{21+1*24+1*20}{1+1+1}$,
$Ra_{T3}(n_e) = \frac{24+1*21}{1+1}$.
Cluster configuration enables better organization and scalability for the IIoT network.

\begin{figure}[ht]
    \centering 
    \includegraphics[width=0.70\linewidth]{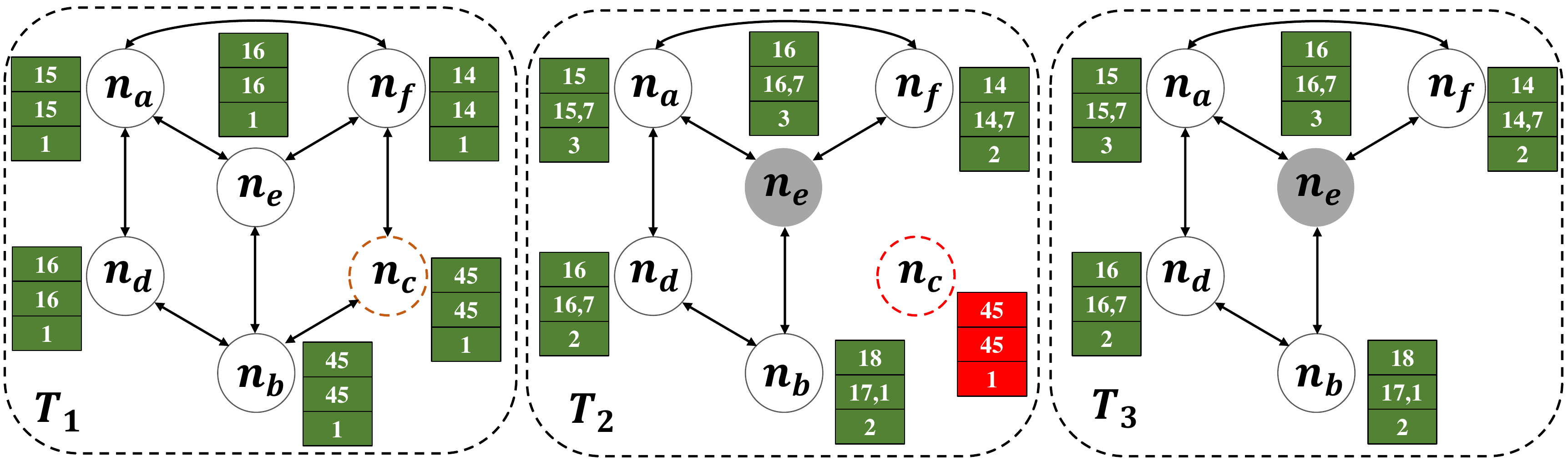}
    \caption{Detection and isolation of an FDI Attack}
    \label{Fig:ata}
\end{figure}

The attackers' detection works simultaneously with the formation of clusters. Thus, nodes whose readings do not respect the similarity threshold ($CThresh$ =~\textit{3}) are first added to the suspect list and later removed from the network. Figure~\ref{Fig:ata} illustrates the FDI attack detection, where CONFINIT makes use of consensus threshold, with $ThresholdConsensus$=~\textit{5}, and following the clustering instants and $CThresh$ value above described,  at the instant $T_1$ nodes~\textbf {$(n_a, n_b, n_d, n_d, n_f)$} own reading values varying from $14$ to $18$, respecting the $CThresh$ among them.  Meantime, ~\textbf{$n_c$} has a reading value of $45$, which deviates greatly from its spatial neighbors, not respecting the $CThresh$. Hence, it cannot integrate the cluster at that time. At $T_2$, \textbf{$n_c$} again, tries to join the cluster, but its $Id$ is already on the suspect list. Thus, nodes apply Equation~\ref{eq:conse} to check out the $ThresholdConsensus$ based on consensus participates readings and compared with ~\textbf{$n_c$}. Next, they  conclude~\textbf{$n_c$} is an attacker and cannot integrate any cluster. At $T_3$, \textbf{$n_c$} is excluded from the network, and the detector $n_b$ has sent an alert message (AM) to the leader $n_e$ with the \textbf{$n_c$} $Id$. After receiving AM, $n_e$ propagates AM to other leaders in the IIoT network in order to avoid that $n_c$ integrates another clustering.

\section{Performance Evaluation}
\label{sec:eva} 

We implement the CONFINIT and  DDFC  ({Dynamic Data-aware Firefly based Clustering})~\cite{gielow2015data} systems in NS3-simulator, version 3.28, in order to analyze them and compare their performance in a IIoT network.The DDFC code was modified to work on the version of the simulator. Besides, we implemented the FDI attack model seen in~\cite{deng2016false} against a smart grid network, in which the attacker targeted to replace the data from an energy reading matrix by inserting either small or full changes on the collected data. We adopted the attacker with the second misbehavior way to achieve a better attack performance in our scenario. In addition, some devices randomly act as attackers in the network and they are fully aware of the data type transmitted in the IIoT, easing its interaction with the target nodes. Additionally, devices may malfunction due to crash failures, which do not mean an attack.

\begin{figure}[ht]
    \centering   
    \includegraphics[width=0.60\linewidth]{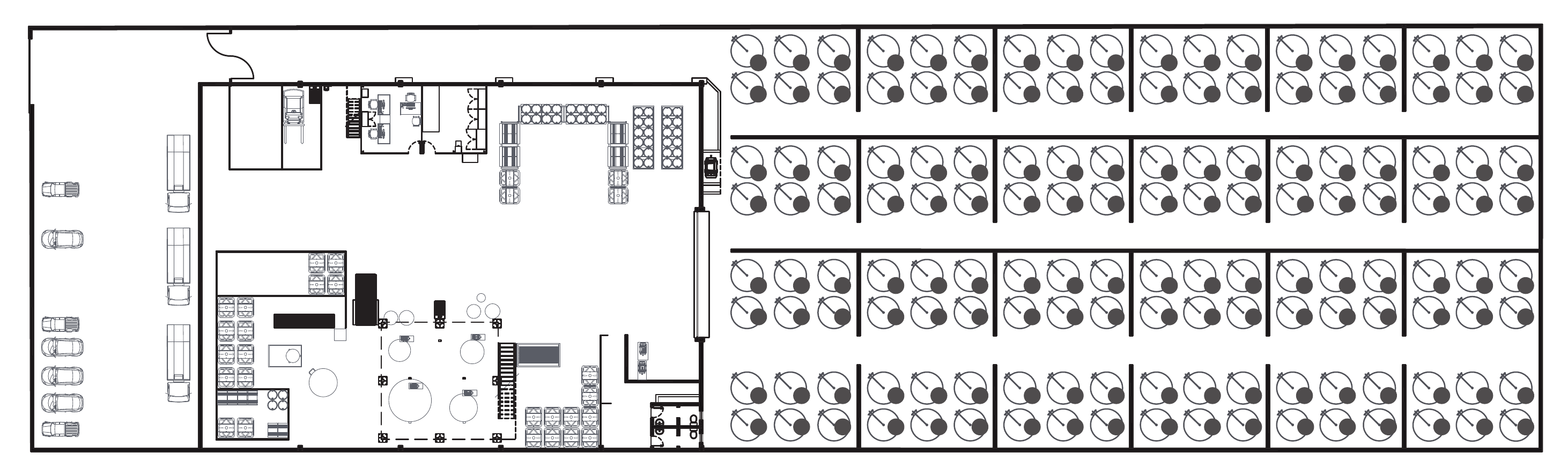}
    \caption{Simulation scenario}
    \label{Fig:cena_fina}
\end{figure}

We have made all simulations taking into account the same IIoT environment (\textit{an  smart industry}) and  with various workloads aiming to analyze the behavior and performance of 
both systems. Particularly, we define and variate the workload according to the number of devices and the percentage of FDI attacks in the IIoT network. For simplicity, we assume that all devices as authenticated and validated by a previous access control system. The IIoT environment comprehends devices embedded in industrial machinery of petrochemical, in which ones collect and monitor gas pressure data. For applying real data reading, we use data collected from gas pressure sensors pick up in 2008, available by the Machine Learning Repository (UCI)~\cite{UCI} laboratory. We have applied this \textit{dataset} as it is the most suitable among the ones that represent the chosen IIoT model. Furthermore, it provides various types of information, such as a high number of devices, continuous data flow, and different variations in reading values. These properties fulfill our necessity to create the IIoT environment of the petrochemical industry. Thus, with this~\textit{dataset}, we attempted to reach an IIoT dense environment closer to the real one, as illustrated in Figure~\ref{Fig:cena_fina}\footnote[1] {We took this from the bibliocad - https://bit.ly/2Z0dUJa -  website and adapted to suit the dense IIoT industrial model proposed in this work.}, whose several tanks and gas valves create an interface between physical and digital environments.

The \textit{Smart Industry} corresponds to a region of 200m x 200m  based in~\cite{santos2019clustering, kumar2018deterministic}, meaning  a dense industrial IIoT. All devices adopt the IEEE 802.15.4 standard (6LoWPAN), being the most suitable for low capacity devices since the header encapsulation and compression mechanism allows the forwarding of IPV6 packets with 127~\textit{bytes} messages; they also employ the User Datagram Protocol (UDP). We  consider that devices do not present communication failures and, in order to avoid to packet loss due to interference and bottleneck, a delay of \textit{2ms} was added for messages exchanged among devices. We set up for each simulation a number of gas pressure sensor devices 50, 75, 100 and 120. They operate for~\textit{1200s}, and in each simulation round, the amount of FDI attacks ranges among 2\%, 5\%, 10\%, 15\% and 20\%. We also set up the transmission radius in~\textit{100m} taking into account the proportion of the defined coverage area to not compromise the results obtained, as a greater range could interfere with the formation of clusters and consensus. After a careful analysis of the distribution of data in the \textit{dataset} and based on the low sensitivity of variations in the data values, we assigned the thresholds values. Thus, we apply the value of three for the similarity threshold and five for the consensus threshold. These values took into account the procedures of clusters formation and attacks detection, choosing different values that could overlap clusters and affect how the filtering of honest nodes is classified. This difference between the thresholds allows us to have a more significant assertiveness in the attack detection. Furthermore, as the nodes are physically close, they present a low variation of the collected data. Therefore, if the node reading exceeds the threshold of similarity and consensus then means that the system is under attack. The results obtained from the simulations correspond to the average of 35 simulations carried out for each scenario, with a 95\% confidence interval. A comparative analysis with the DDFC was only the clustering service face FDI attacks. Since DDFC does not take into the  data authenticity and hence FDI attacks can be successful, see more details in~\cite{gielow2015data}, the detection metrics were not analyzed. About other works in the literature, the closest model is ~\cite{yang2017robust}, whose the clustering principle is different, and it does not take into account the network density.

We have selected the metrics based in~\cite{gielow2015data,yang2017robust,ferdowsi2019generative} aiming to validate the efficiency and effectiveness  of detection, exclusion, and availability and authenticity of data disseminated by CONFINIT on the IIoT network.  The efficiency metric consists only of the amount of~\textbf{\textit{Established Clusters}}, which corresponds to how many clusters are formed over the simulation time relates to the number of nodes available on the network. The effectiveness metrics against FDI attacks are described below:

\textbf{\textit{FDI attack detection rate (DR):}} calculates the percentage of the FDI attacks correctly identified through interactions between participants. The value of $({T_{det}})$ understands the ratio between the total detection $({det_{ni}})$, and the amount of attacks entered $({A_{ins}})$.

\vspace{0.3cm}
\begin{equation}
{T _{det}} =\frac{\sum {det_{ni}}}{{A_{ins}}}
\label{eq:td_}
\end{equation}
\vspace{0.3cm}

\textbf{\textbf{Accuracy} (AC):} evaluates the precision in detecting attacks,  corresponding to the total of attacks detected from the $({det_{ni}})$ and the correct identification of legitimate network nodes $({det_{nl}})$, divided by total attacks detected on the network $({T_{det}})$ and $({det_{nl}})$, $({T_{fp}})$ and $({T_{fn}})$ (Eq~\ref{eq:Ac}). The $({R_{a}})$ also results in discrete values between 0 and 100, the closer to 100 the more accuracy is. 

\vspace{0.3cm}
\begin{equation}
{R _{a}} =\frac{\sum {det_{ni}} + \sum {det_{nl}}}{{T_{det}} + {det_{ni}} + {T_{fn}} + {T_{fp}}}
\label{eq:Ac}
\end{equation}
\vspace{0.3cm}

\textbf{\textit{False positive rate (FPR):}}  defines the number of times 
whose correct devices were identified as FDI attacker. It is calculated by dividing the $({det_{ni}})$ detection amount by the total interactions made to the $({T_{int}})$ network.

\vspace{0.3cm}
\begin{equation}
{T_{fp}} =\frac{\sum {det_{ni}}}{{T_{int}}}
\label{eq:td}
\end{equation}
\vspace{0.3cm}

\textbf{\textit{False negative rate (FNR):}} calculates the percentage of devices misidentified as non-intruders. The value  means the difference between the total interactions~\textit{X} , i.e. sending control messages, and $({T_{det}})$ the attack detection rate.

\vspace{0.3cm}
\begin{equation}
{T_{fn}} = |X| - {T_{det}}
\label{eq:tfn}
\end{equation}
\vspace{0.3cm}

\textbf{\textit{F1 score (FC):}} is the harmonic mean of precision and recall. Precision is the proportion of data classified as successful as an attack to all data classified as an attack, as in Equation~\ref{eq:pre}. Recall is the proportion of data classified as an attack for all attack data, as in  Equation~\ref{eq:recll} The highest possible value of an F-score is 1.0, indicating perfect precision and recall, and the lowest possible value is 0, if either the precision or the recall is zero.

\begin{equation}
precision = \frac{t_p}{t_p+f_p}
\label{eq:pre}
\end{equation}
\begin{equation}
recall = \frac{t_p}{t_p + f_n}
\label{eq:recll}
\end{equation}
\begin{equation}
{F_{C}} = \frac{ 2 * (precision * recall)}{(precision * recall)}
\label{eq:tfn}
\end{equation}

\subsection{Availability of clusters and Attack detection}
\label{sec:valid}

\begin{figure}[ht]
    \centering
    \includegraphics[width=65mm]{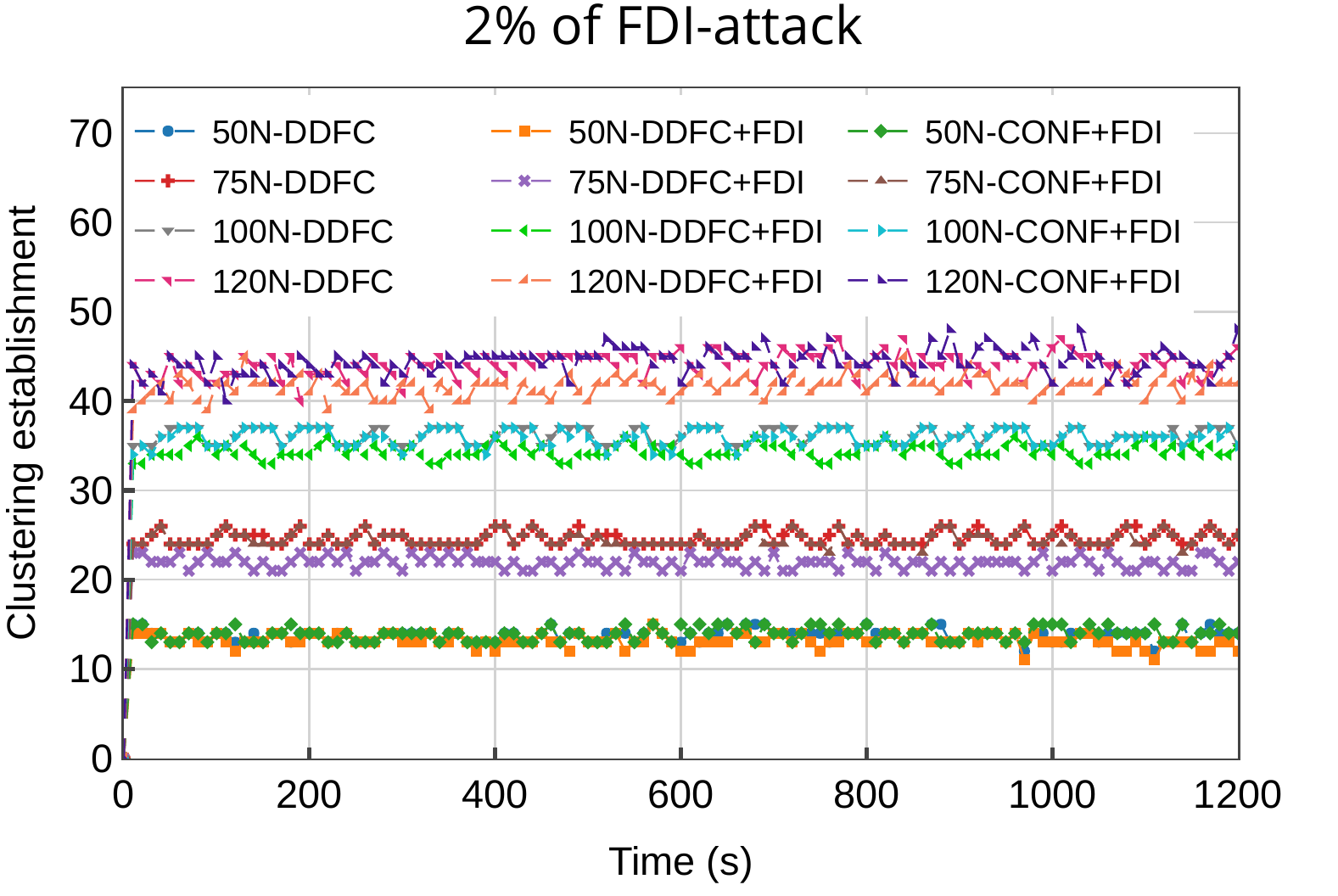}
    \includegraphics[width=65mm]{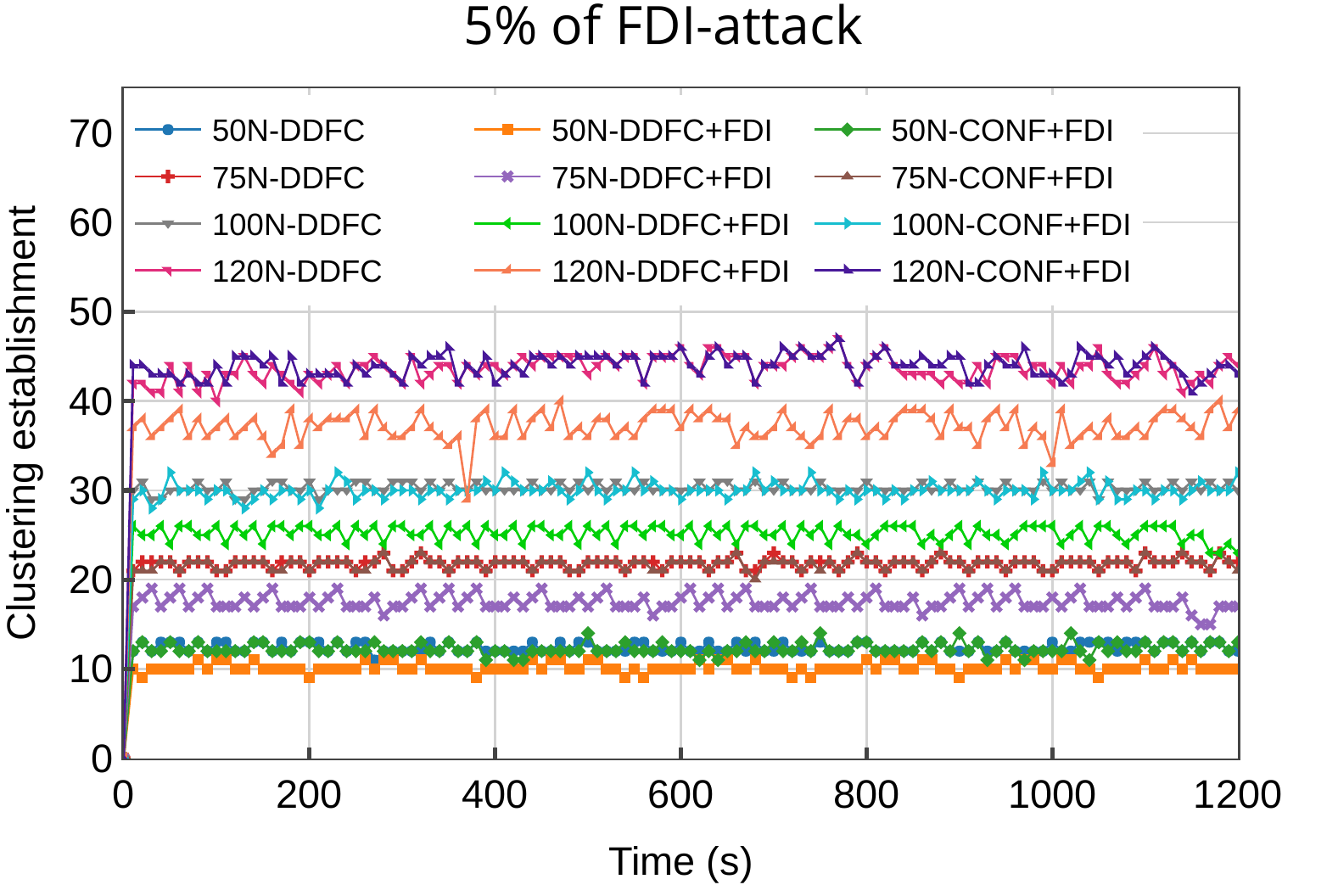}
    \includegraphics[width=65mm]{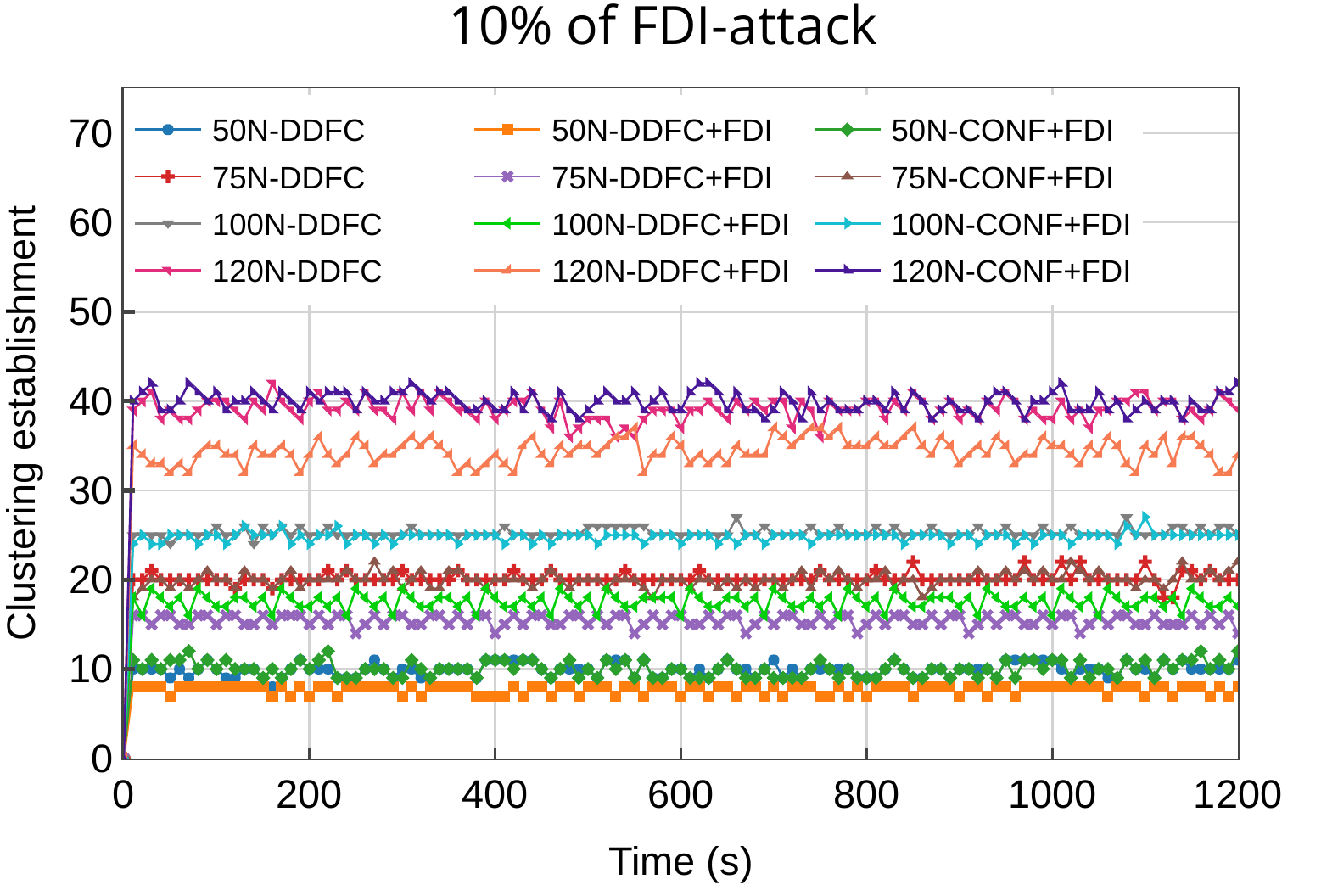}
    \includegraphics[width=65mm]{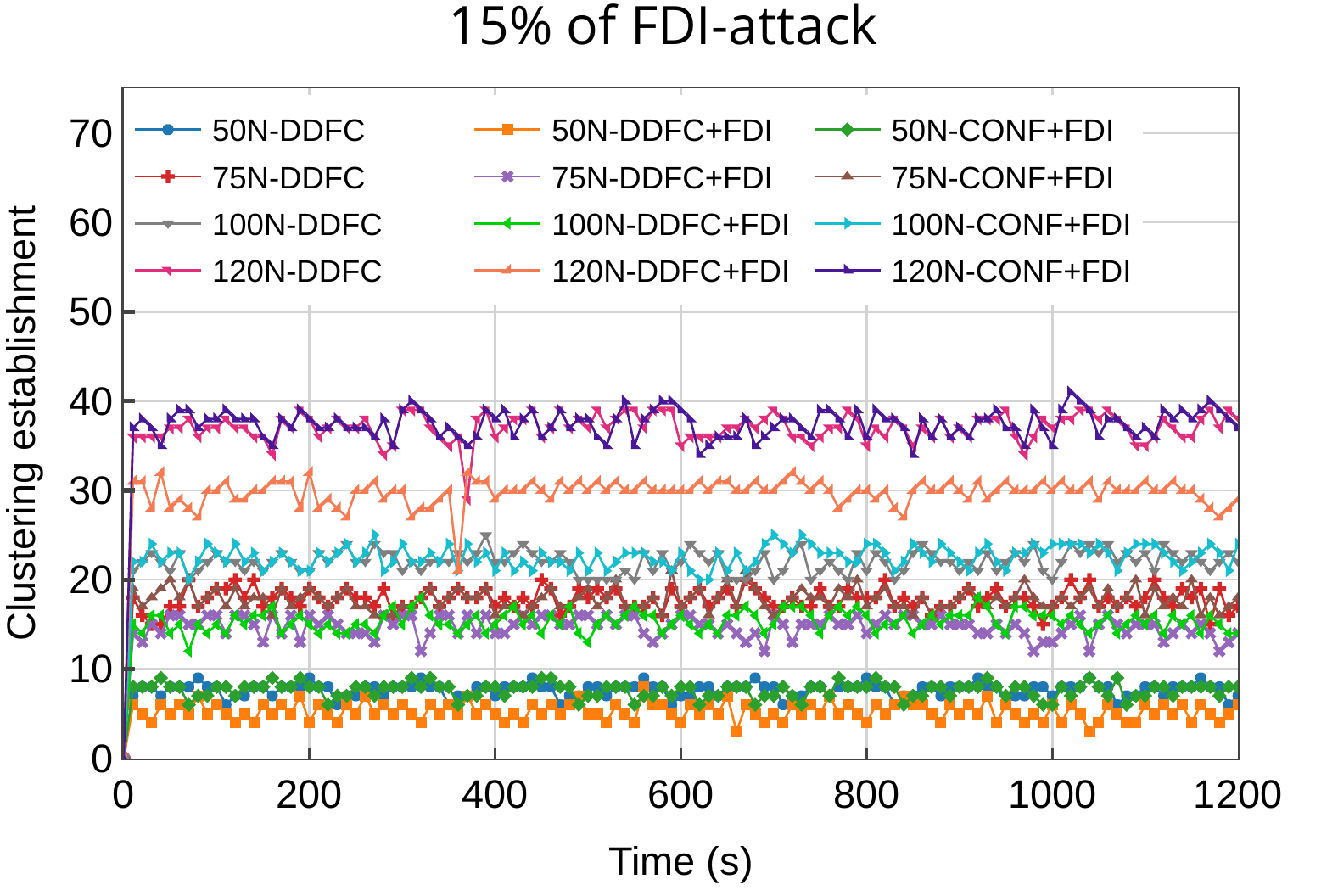}
    \includegraphics[width=65mm]{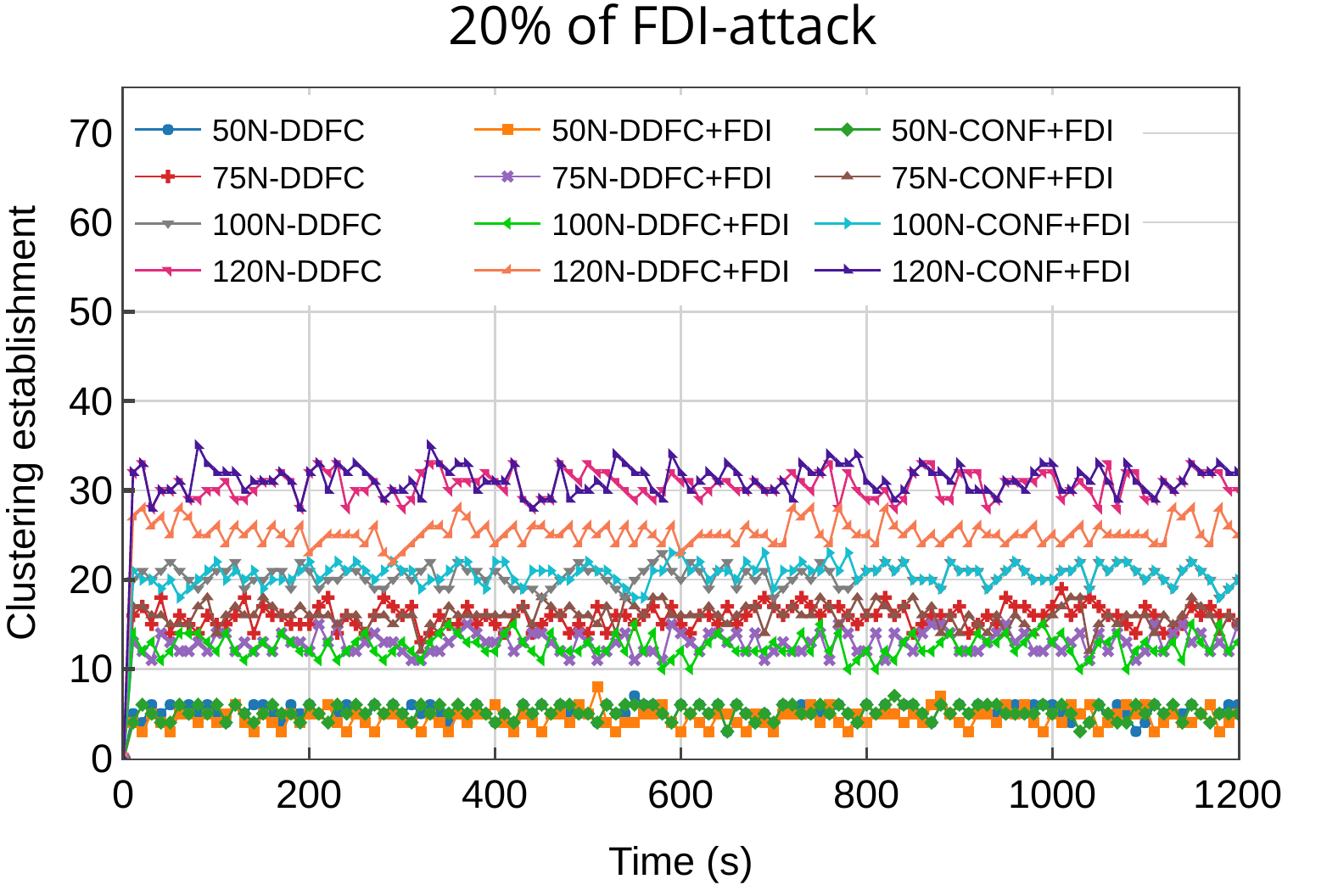}
    \caption{Number of clustering established over time}
    \label{Fig:fun}
\end{figure}

This evaluation takes into account how many clusters are formed over the simulation time. Thus, we analyze the availability of clusters with and without FDI attackers. Firstly, the DDFC performance on the gas pressure IIoT network without FDI attacks. Next, the CONFINIT and DDFC performance by supporting the data availability of the established clusters based on the number of nodes in the network and the percentage of FDI attacks. The graphs in Fig.~\ref{Fig:fun} show the number of cluster established over time in the IIoT network. Note that the number of available clusters varies depending on the influence of the number of nodes in the network and the percentage of FDI attacks. Further, the FDI attack directly impacts the number of clusters formed over time, as the manner it works makes impossible for the captured nodes to join the clusters. We observe that the scenarios with 20\% and 15\% of FDI attackers had a higher decline in the availability of clusters compared to ones with the 10\%, 5\%, and 2\% of FDI attacks. The decline is due to the higher amount of attackers trying to participate in the clusters, which disturbs the similarity calculations by other nodes, hampering the establishment of clusters. We also observed that DDFC under FDI attack showed a drop of 37\% in the average of available clusters, showing how harmful the attack is to the network. In comparison to DDFC, in some cases, CONFINIT increased the number of clusters available by 35\% and in other ones 40\% in reason of the module of fault management, which employs cross-party monitoring to detect FDI Attackers. In general, the attacker behavior affects the amount of available data and interferes negatively with decision-making. As the gas-pressure IIoT network is static, the formation of clusters does not suffer changes in the node position, and only the readings vary in their values. Moreover, clusters contain a variable number of members, i.e., there is no fixed number of members.

\begin{figure}[ht]
    \centering
    \includegraphics[width=40mm]{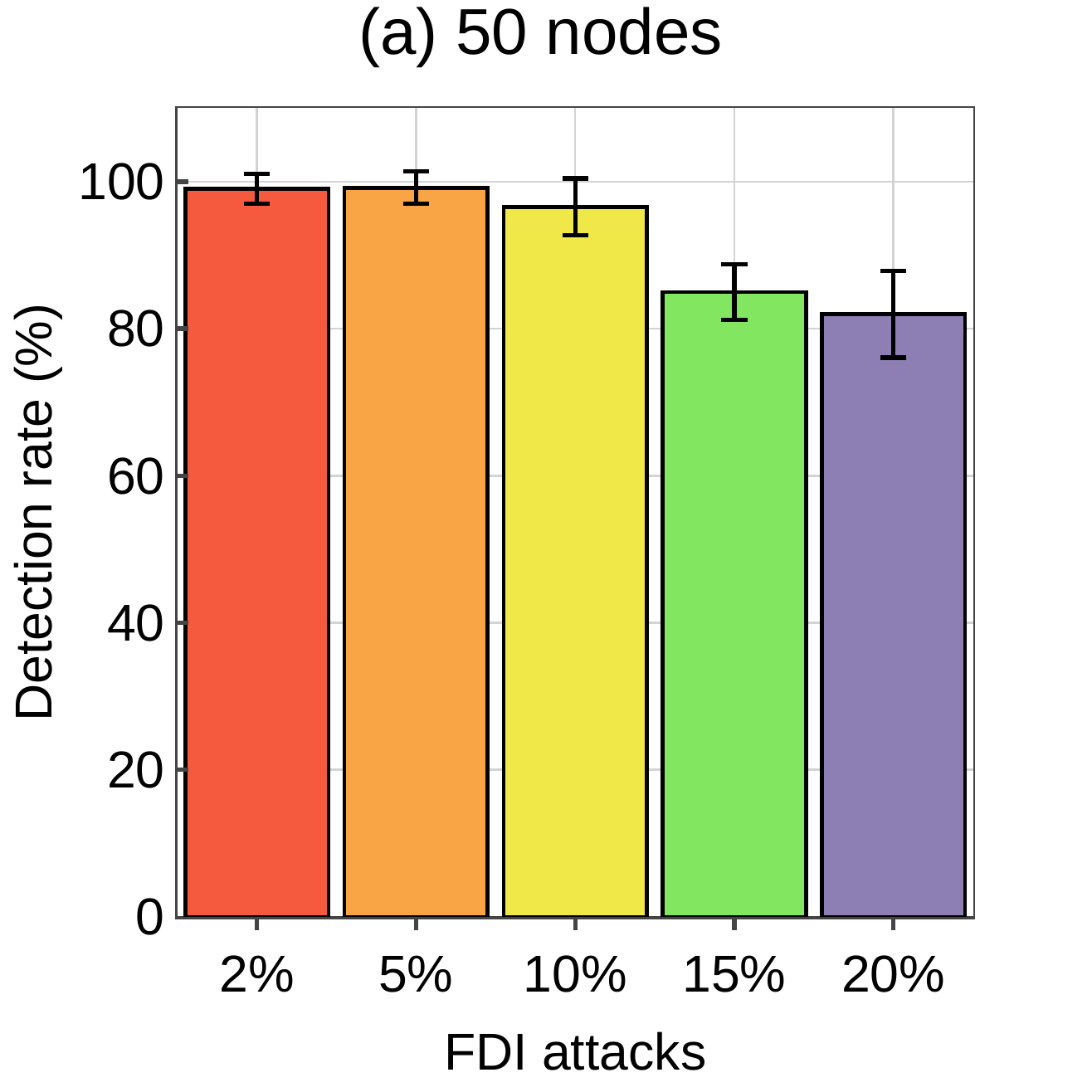}
    \includegraphics[width=40mm]{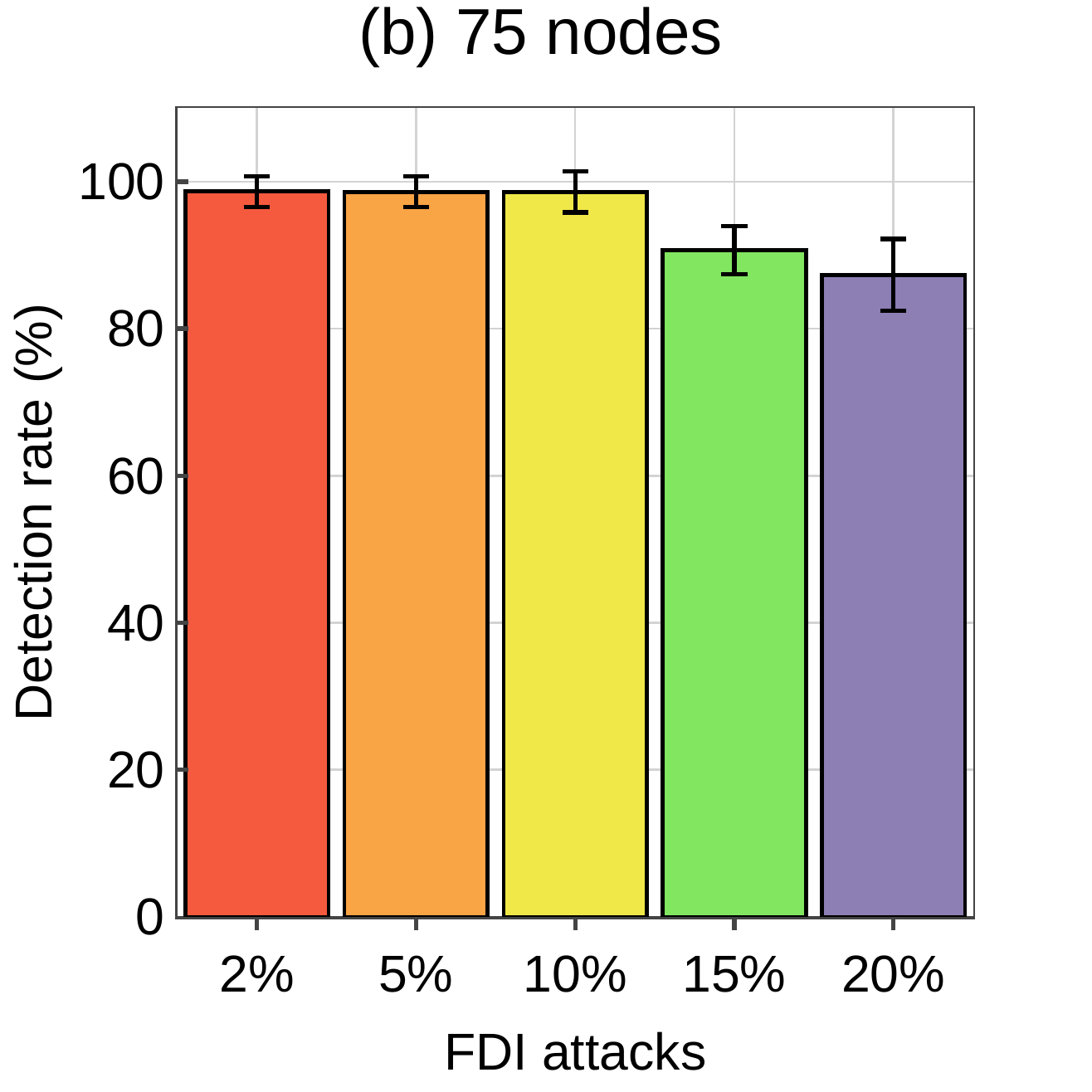}
    \includegraphics[width=40mm]{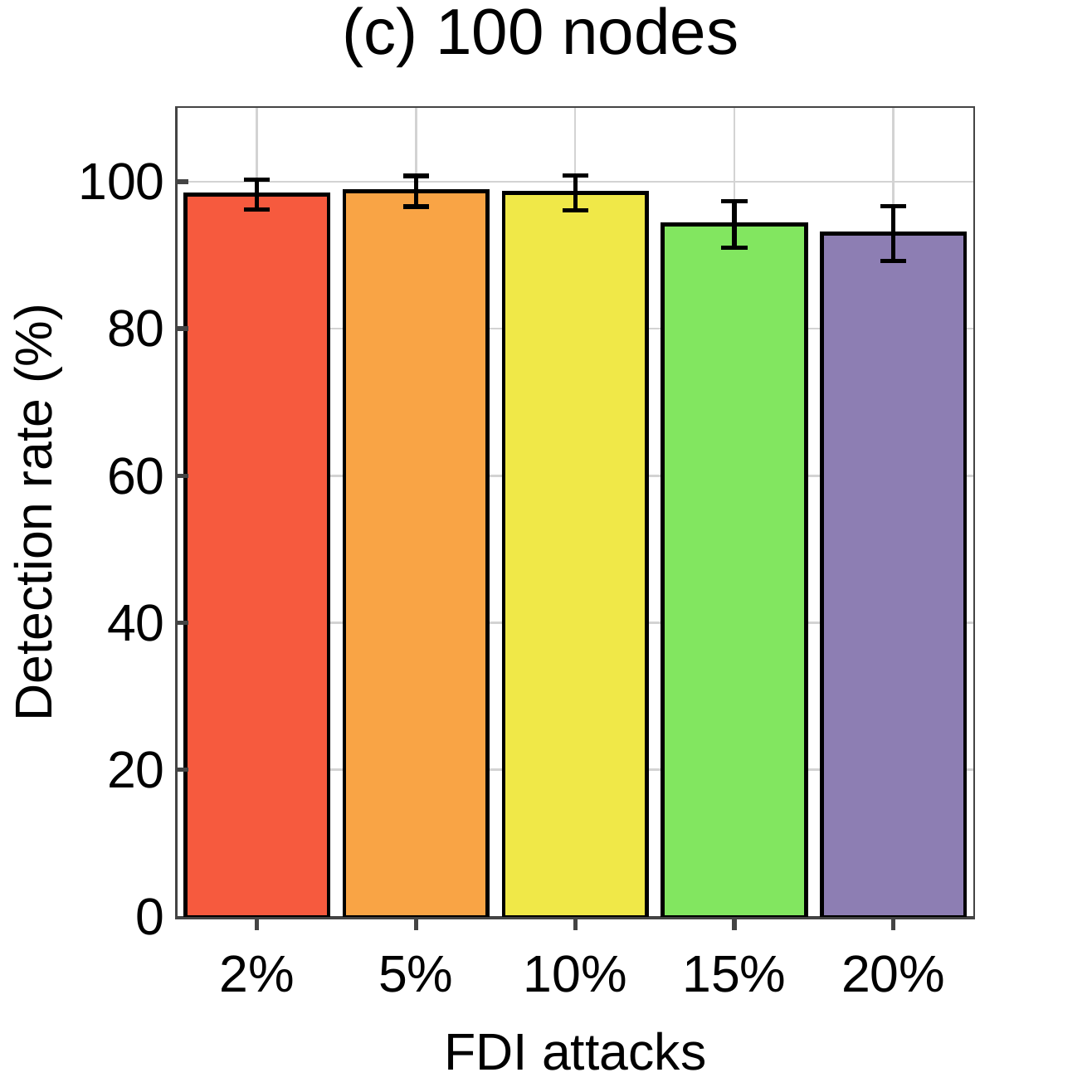}
    \includegraphics[width=40mm]{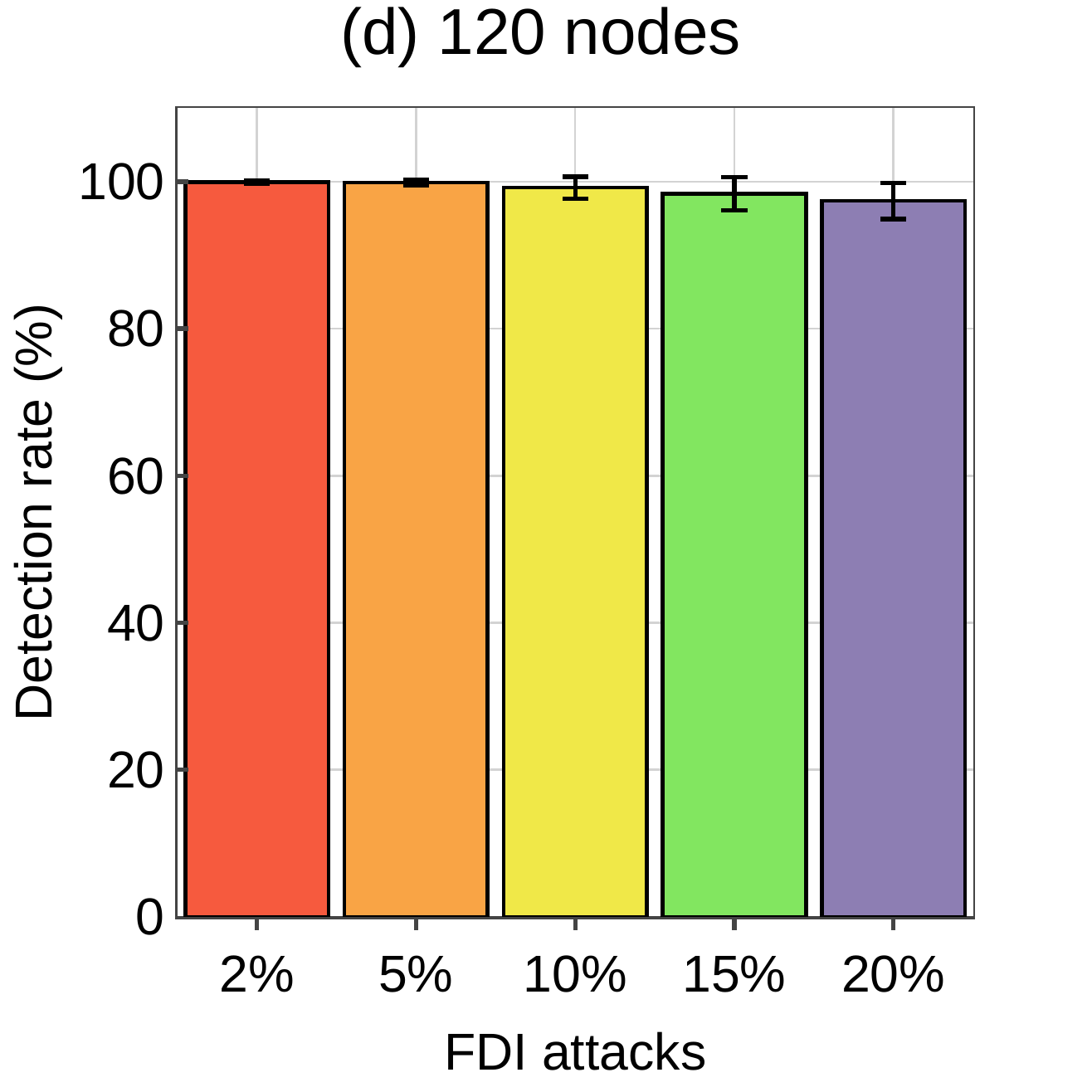}
    \caption{Detection rate of FDI attacks $({T_{det}})$ for 50, 75, 100 and 120 nodes}
    \label{Fig:tx_detec}
\end{figure}

Figure~\ref{Fig:tx_detec} shows the graphs about the FDI attack detection rate~\textbf{(DR)} achieved by CONFINIT taking account the number of nodes on the network and the percentage of FDI attacks. We observed that CONFINIT  in general was able to reach an average DR of 90\%, being effective in detecting FDI intruders, and hence supporting the data availability. Particularly, for instance, with 2\% and 5\% of FDI attackers with 100 and 120 nodes, CONFINIT reached a DR around 100\%.  The \textbf{DR} variation is due to the IIoT density, the denser, more efficient the detection of FDI attacks is. We observed that CONFINIT obtained results less effective in scenarios with 15\% and 20\% attackers for 50 and 75 nodes because the detection is collaborative and with a higher presence of attackers, it had a decline in detection rates. The high DR gotten by CONFINIT is due to the validation of data messages exchanged between participants to identify anomalous behaviors.  This surveillance model helps the IIoT network nodes themselves to keep the network secure. In parallel with this monitoring, the formation of collaborative consensus contributes to decisions about the existence attackers, working in a distributed way among the cluster participants. The application of watchdog and consensus approaches allows maintaining the security of the network distributed among all members by contributing to collaborative filtering and increasing the number of clusters formed, ensuring the availability of only authenticated data and eliminating FDI attacks.

\begin{figure}[ht]
    \centering
    \includegraphics[width=40mm]{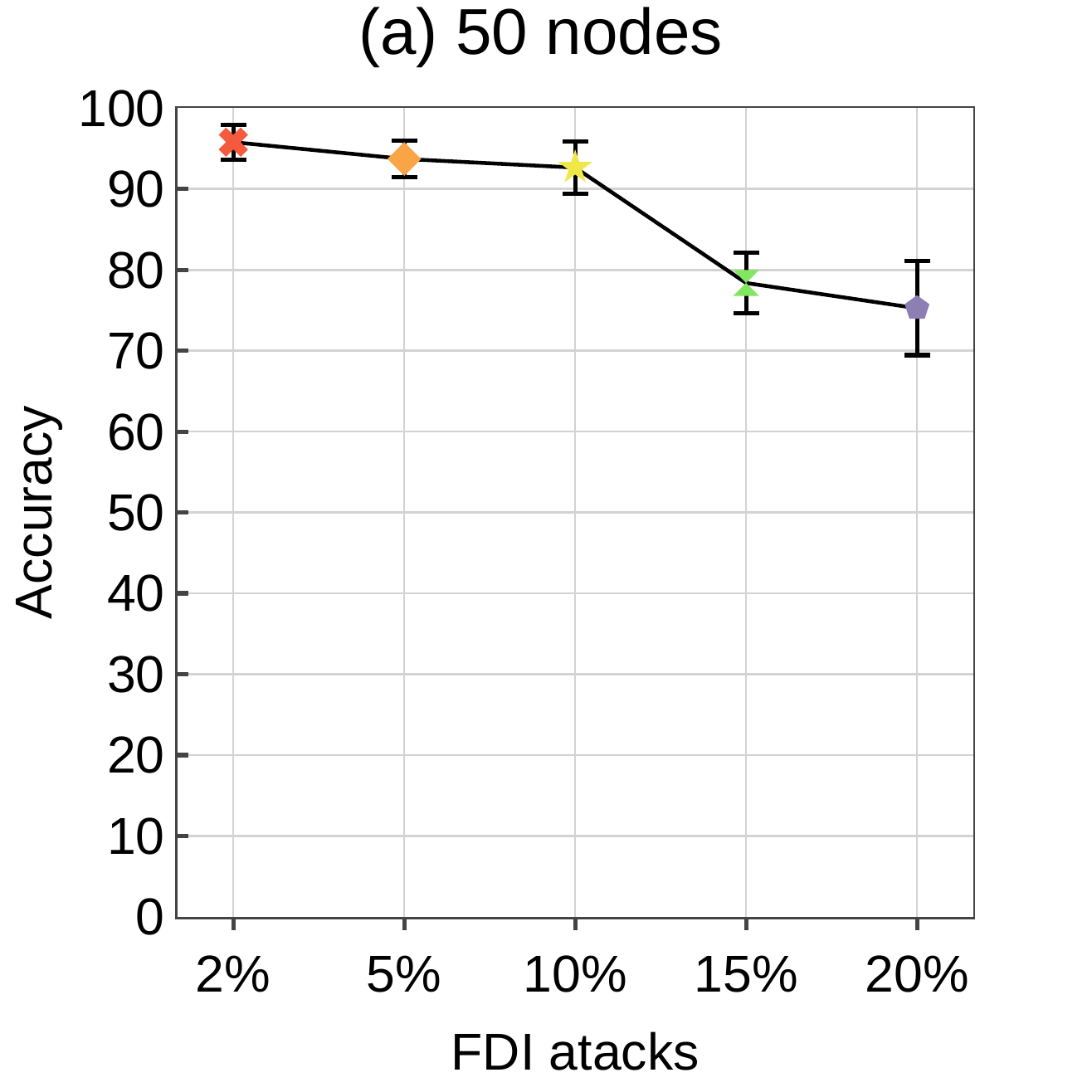}
    \includegraphics[width=40mm]{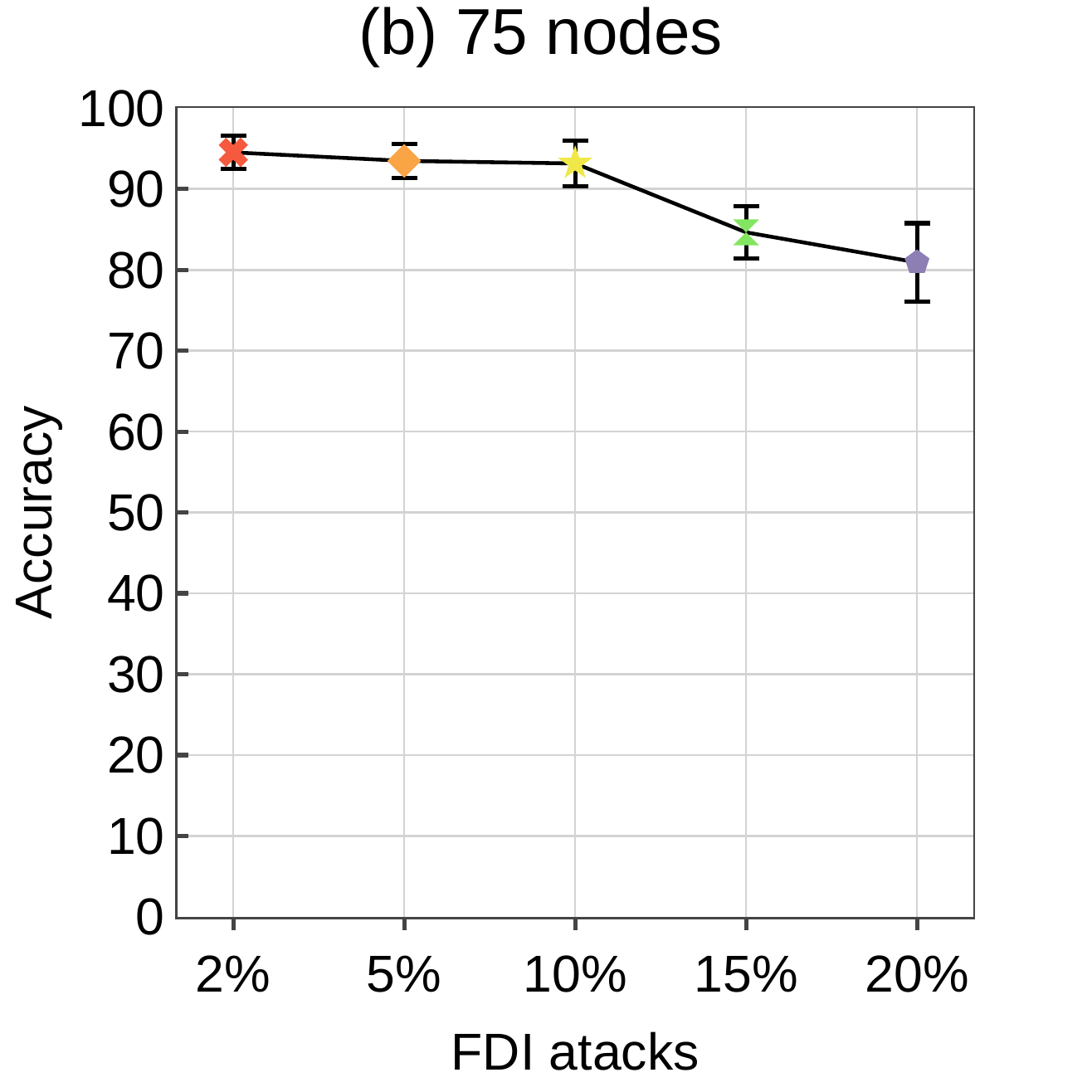}
    \includegraphics[width=40mm]{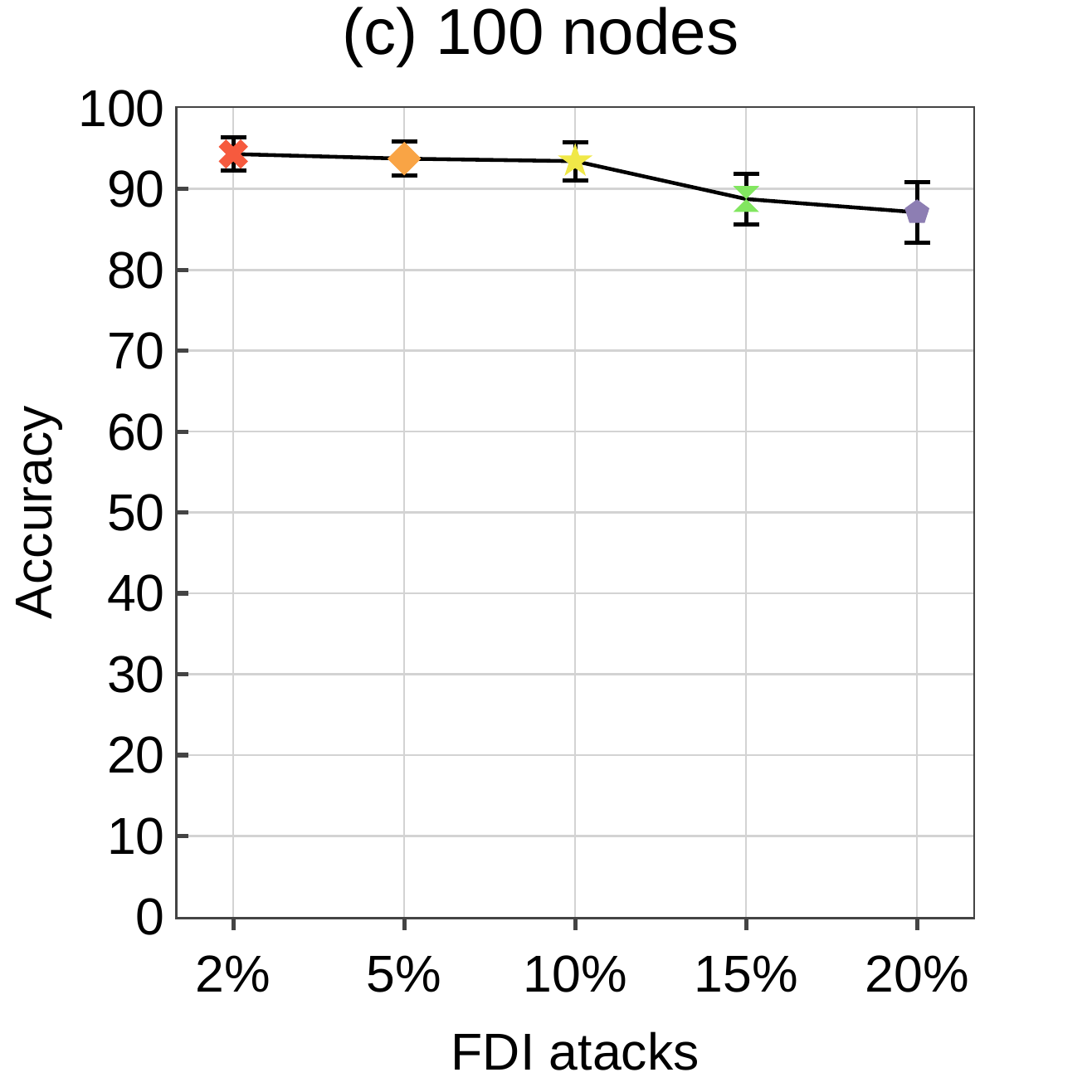}
    \includegraphics[width=40mm]{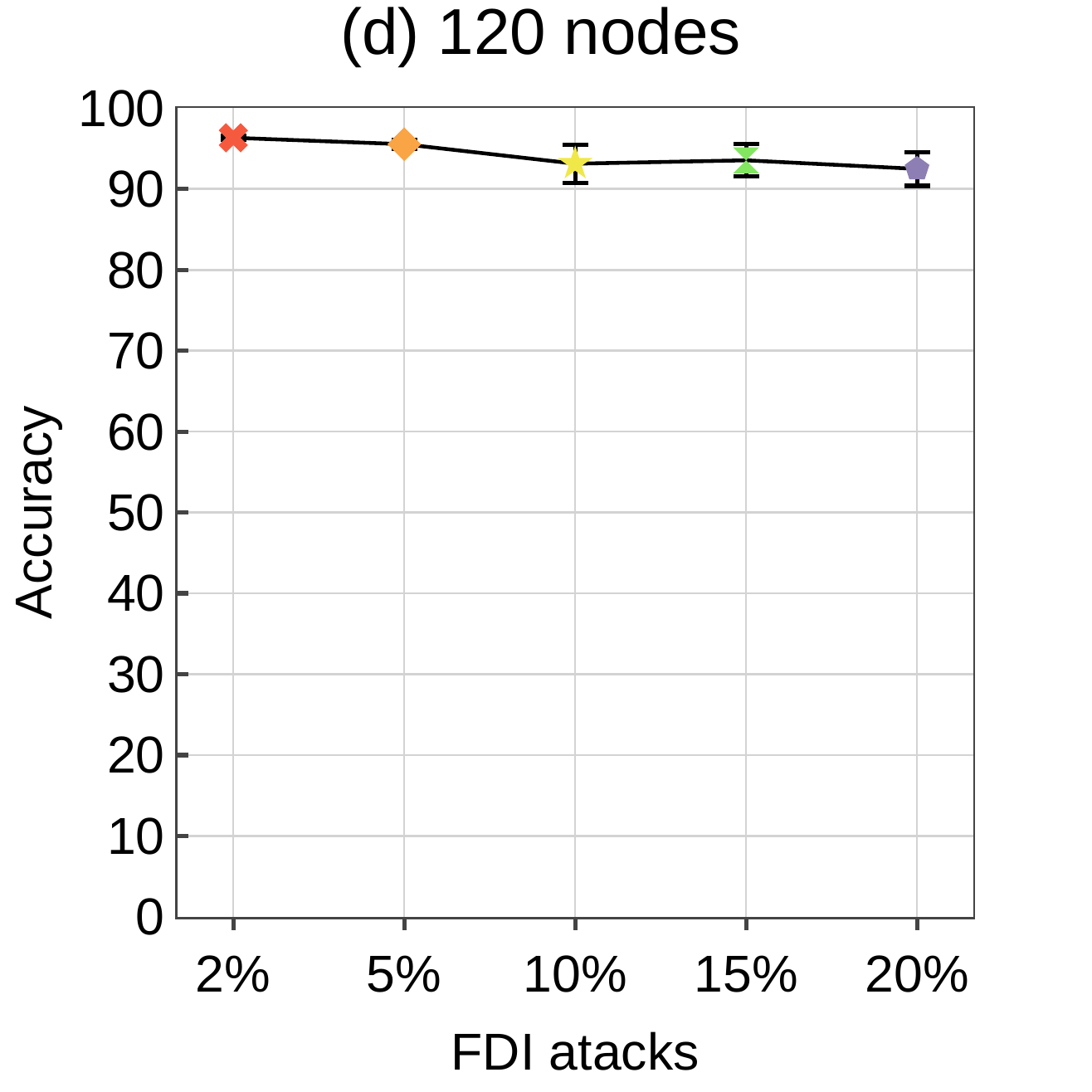}
    \caption{Accuracy $({R_{a}})$ for 50, 75, 100 and 120 nodes}
    \label{Fig:acur_total}
\end{figure}

Figure~\ref{Fig:acur_total} shows the graphs of accuracy (\textbf{AC}) obtained by CONFINIT for supporting the effective detection of FDI attackers. CONFINIT achieved \textbf{AC} values between $71$ and $98$  for all scenarios.  We noticed that with 120 and 100 nodes, \textbf{AC} showed low variation than with 75 and 50 nodes, due to collaborative filtering, being more effective with more nodes in the network. Fixed nodes favor these~\textbf{AC} values, contributing to a high detection rate with low variation about the number of nodes in the network and attackers inserted. We also observed that, regardless of the number of nodes, the~\textbf{AC} variation is minimal. In addition, the CONFINIT was effective against the FDI attack. The high accuracy is due to the watchdog strategy, and hence this accuracy is reflected in the cluster availability, the fewer FDI attacks participate in the clusters, and the less false data is made available to the application.

\begin{figure}[ht]
    \centering
    \includegraphics[width=40mm]{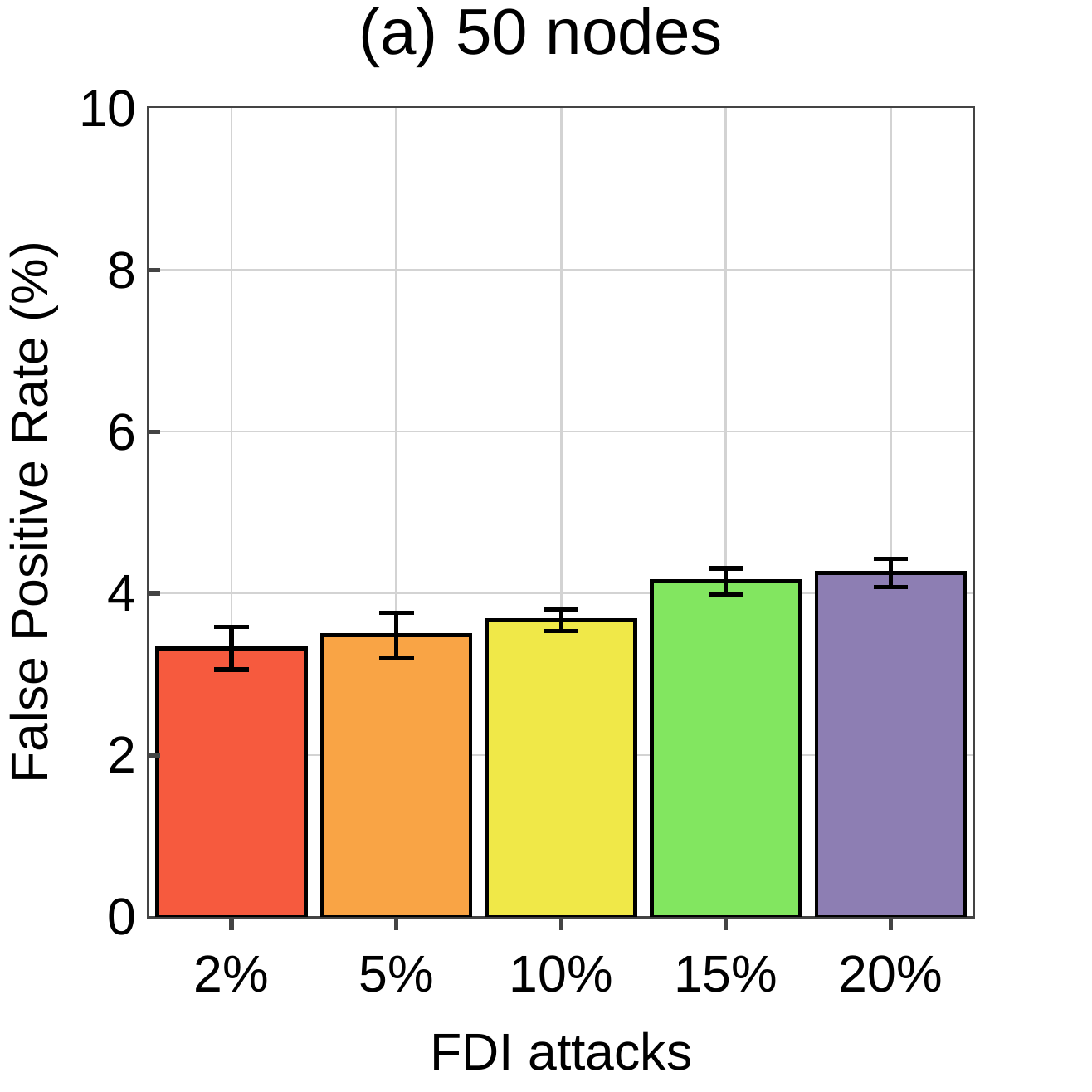}
    \includegraphics[width=40mm]{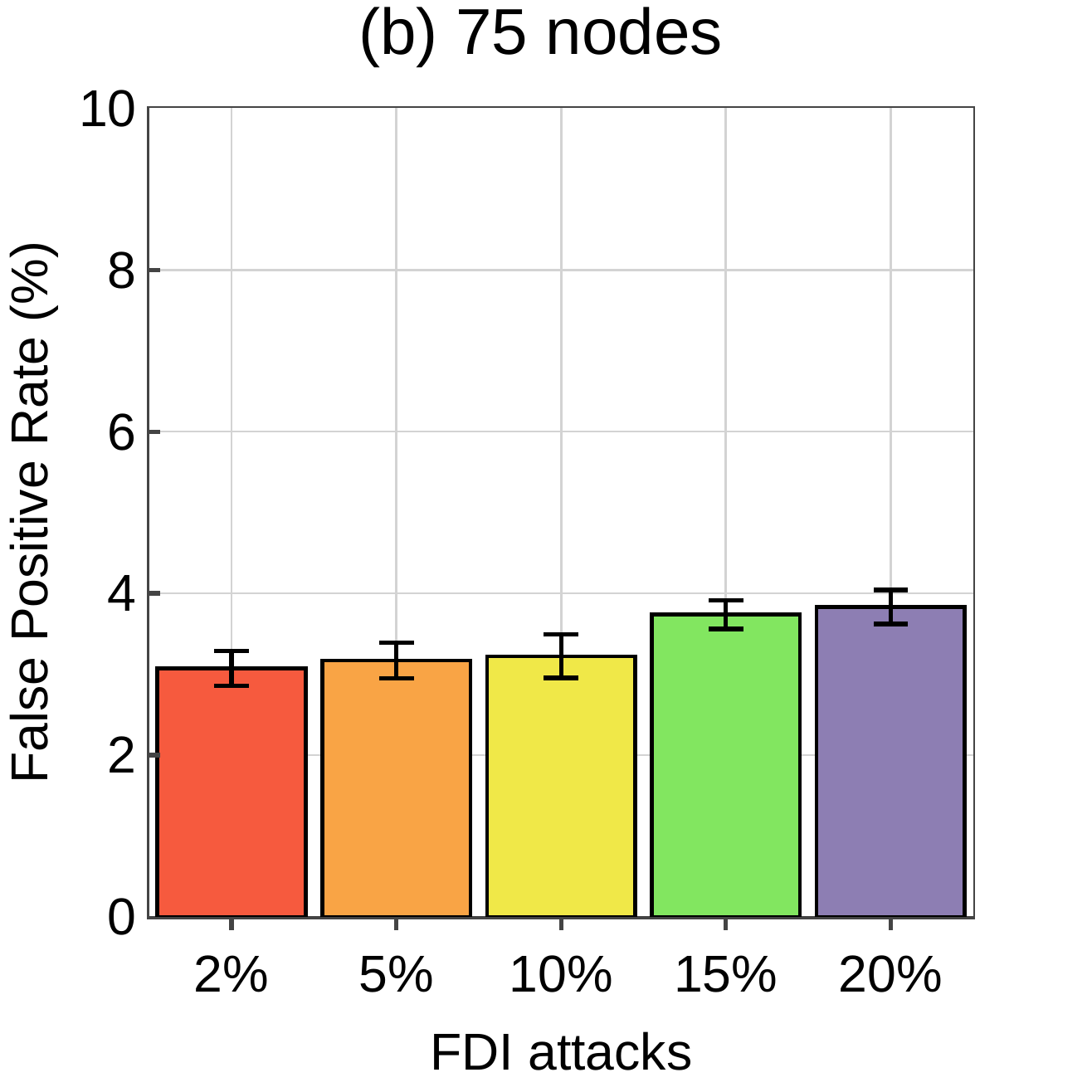}
    \includegraphics[width=40mm]{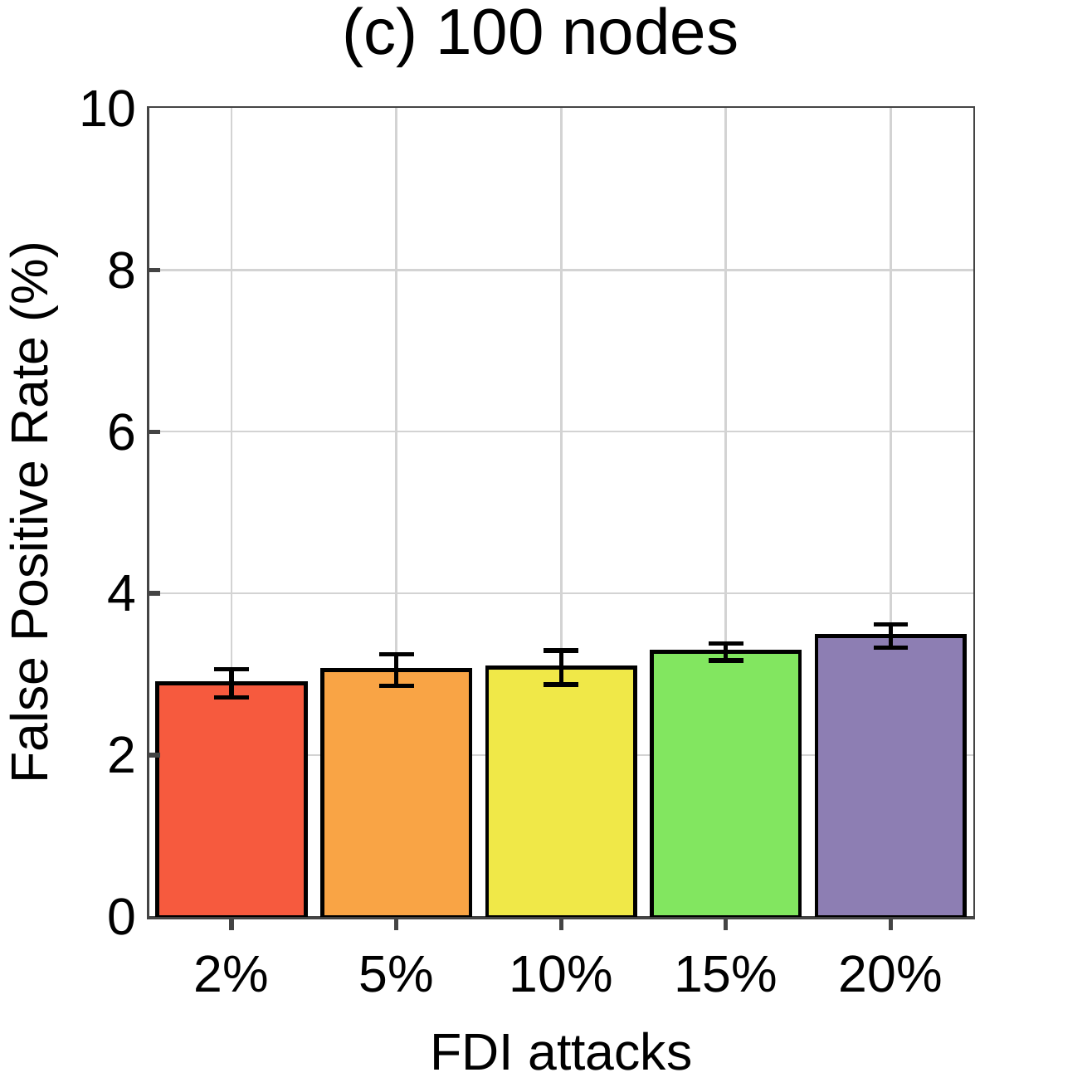}
    \includegraphics[width=40mm]{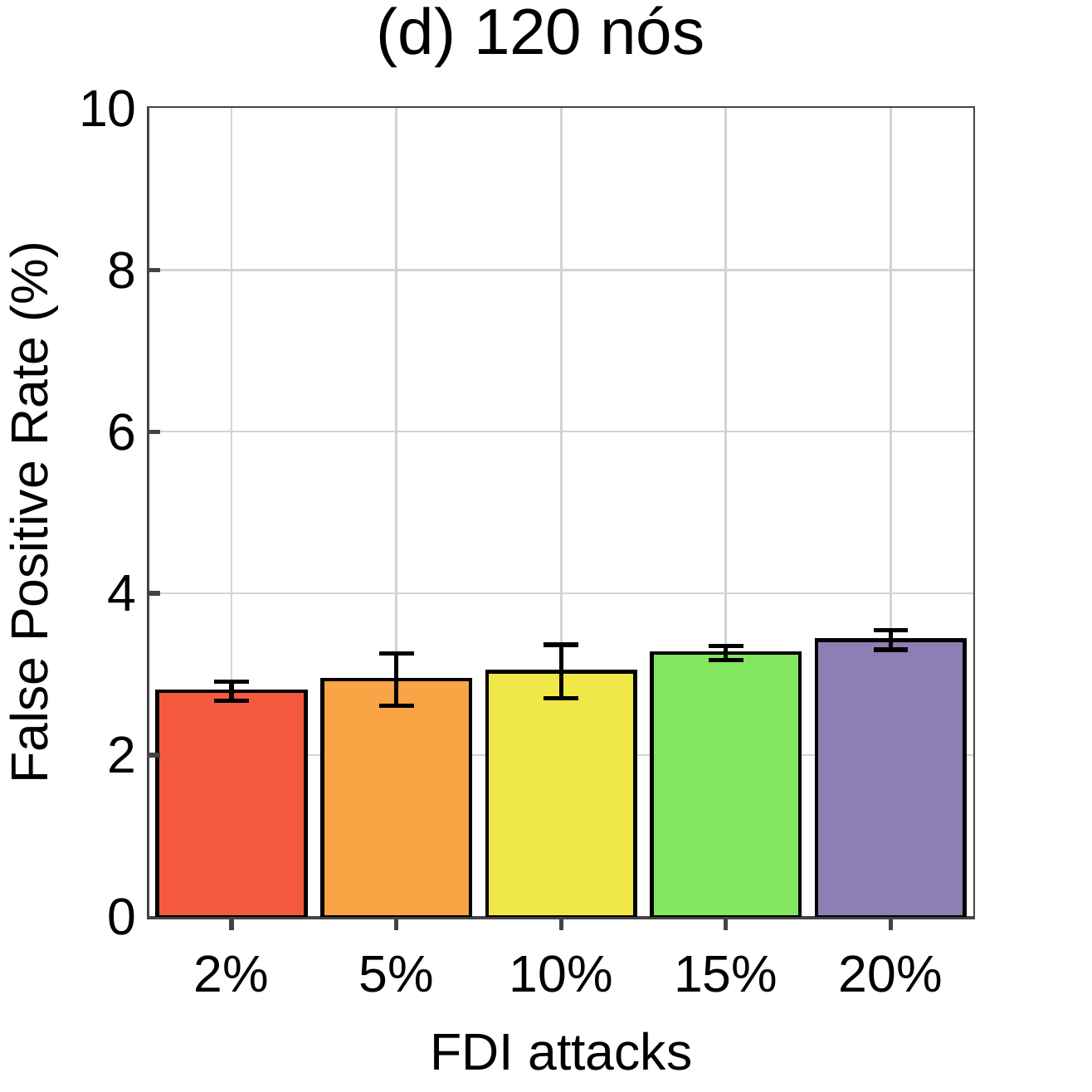}
    \caption{False positive rate $({t_{fp}})$ for 50, 75, 100 and 120 nodes}
    \label{Fig:fPotal}
\end{figure}

Figure~\ref{Fig:fPotal} presents the graphs  around  the false positive rate~(\textbf{FPR}). CONFINIT obtained an average \textbf{FPR} ranging from $2.8\%$ to 2\% of attackers and $4.1\%$ to 20\% of attackers, and 5\%, 10\%, and 15\% of attackers achieved an average \textbf{FPR} between $2\%$ and $4\%$ in all scenarios. The scenarios showed a low percentage difference among them, with an average variation of only $1.8\%$, demonstrating the stability of CONFINIT in operating with different amounts of nodes and attackers in the IIoT network. This low variation is a result of the nodes being static, facilitating the correct identification of attackers.  Although the~\textbf{FPR} values are low, some adjustments need to be made to improve part of the CONFINIT related to the similarity and consensus equation according to the type of application. In this way, those adjustments are related to the types of data collected and how they are made available. The incorrect detection is due to errors in monitoring and the consensus calculation among the participants, it can identify a legitimate node as an FDI attacker based on tiny deviations in the reading values. Thus, nodes seen as suspicious can be labeled as attackers as new interactions and the exchange of data messages take place. We also observed the number of nodes in the network does not influence the results, as the \textbf{FPR} variation from one scenario to another is low. Only the scenario with  50 nodes showed sharp variations in the \textbf{FPR}. Further, the greater the number of devices participating in the network, the greater the resulting FPR. Besides, the smaller number of attackers, the more effective the detection is, which yields the lowest rate of false positives.

\begin{figure}[ht]
  \centering
    \includegraphics[width=40mm]{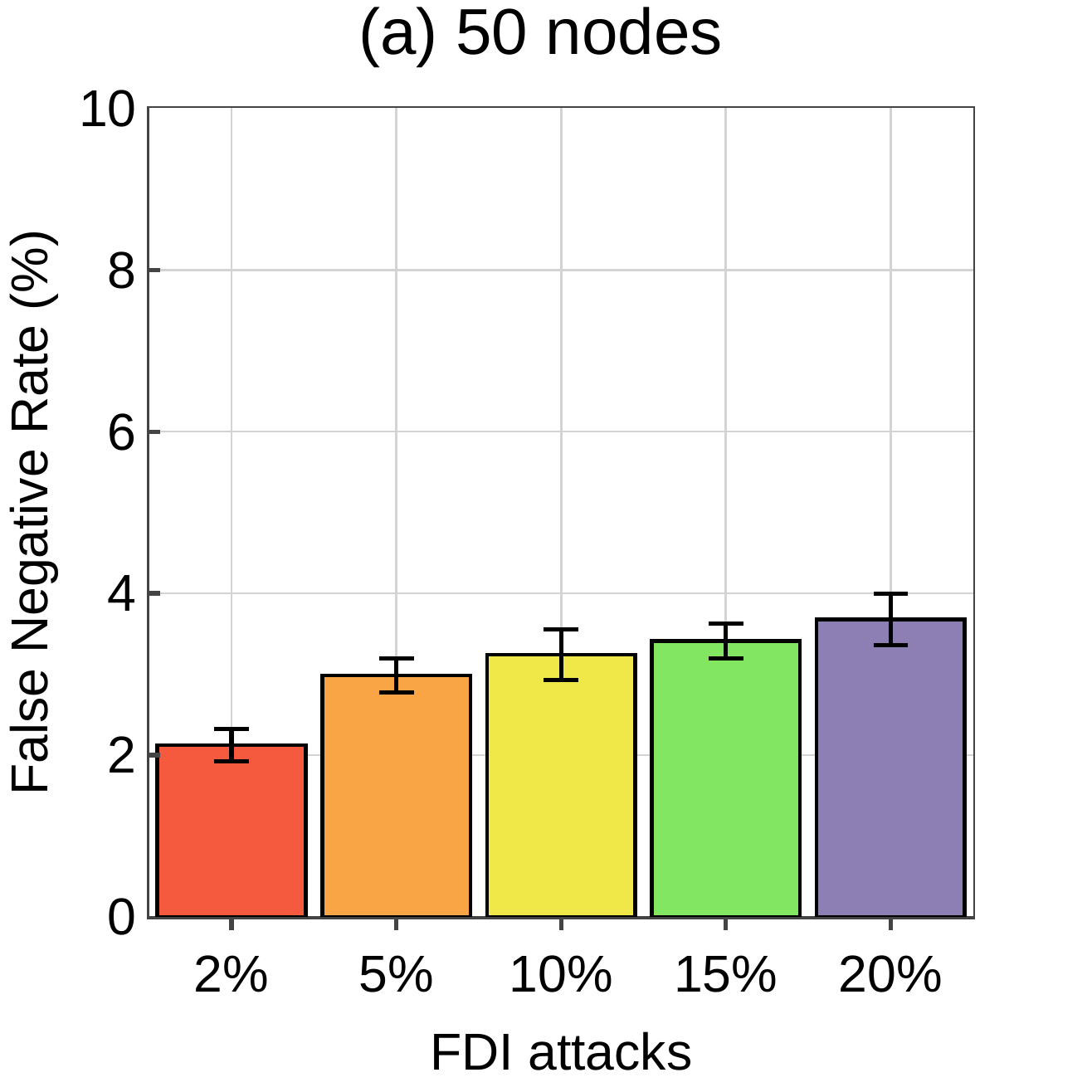}
    \includegraphics[width=40mm]{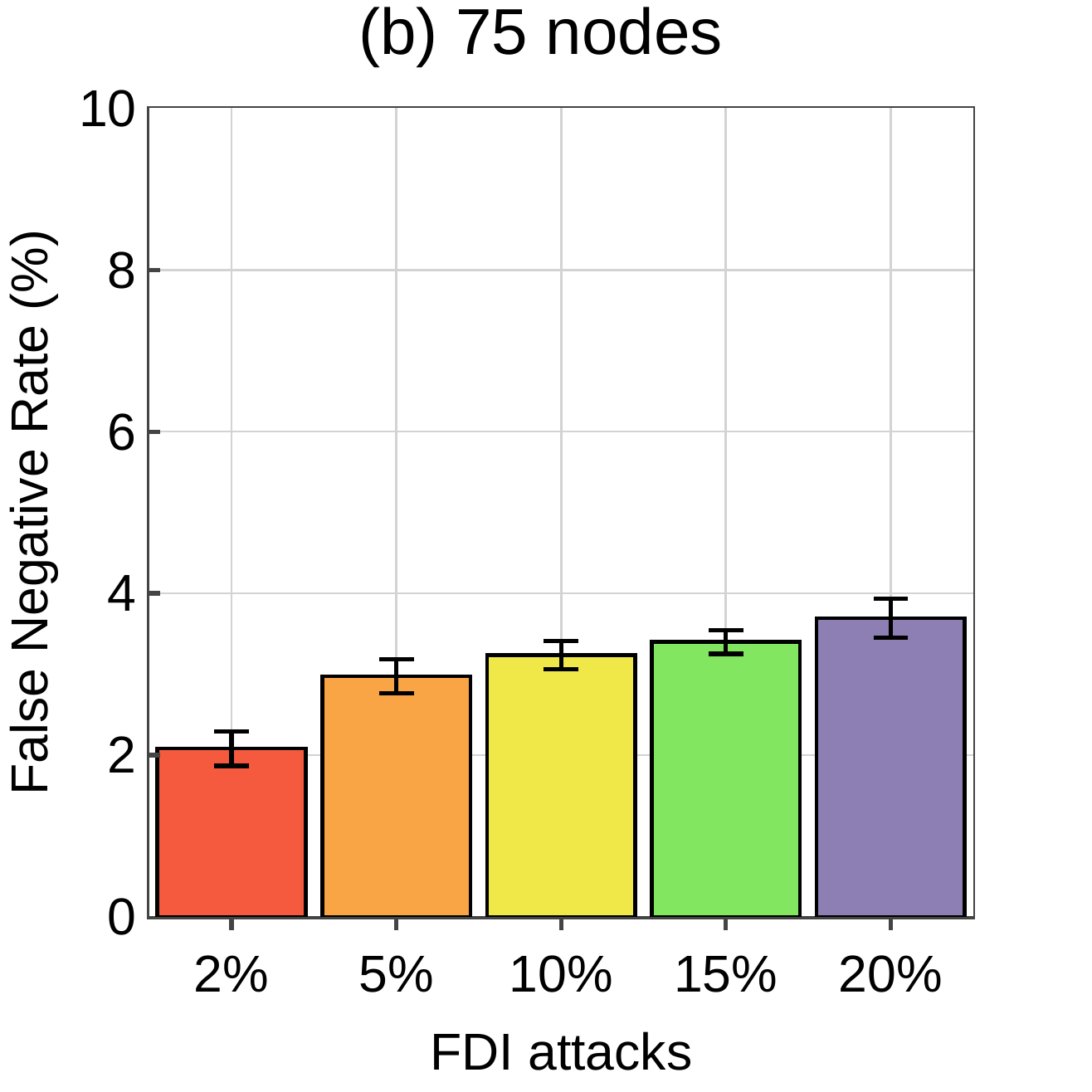}
    \includegraphics[width=40mm]{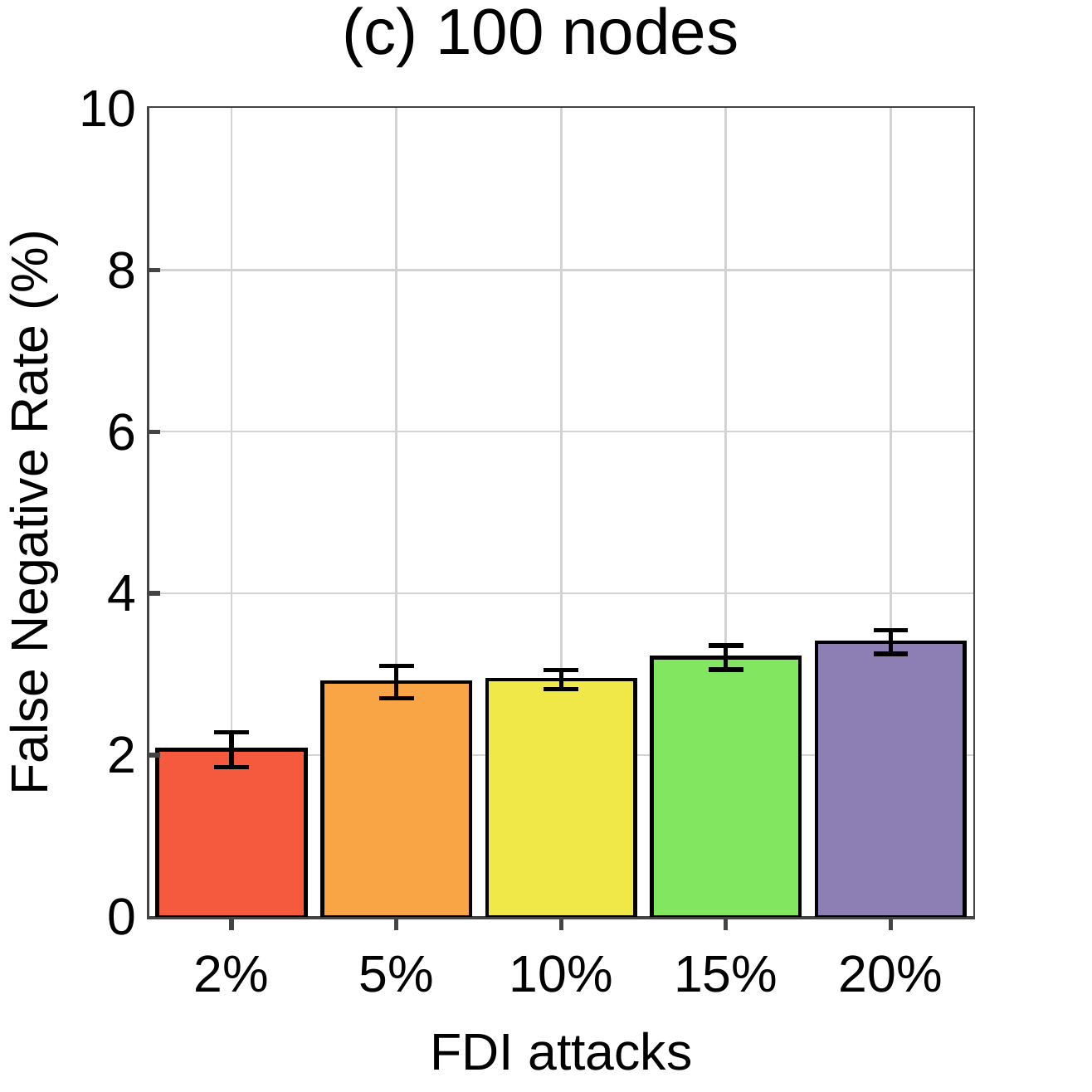}
    \includegraphics[width=40mm]{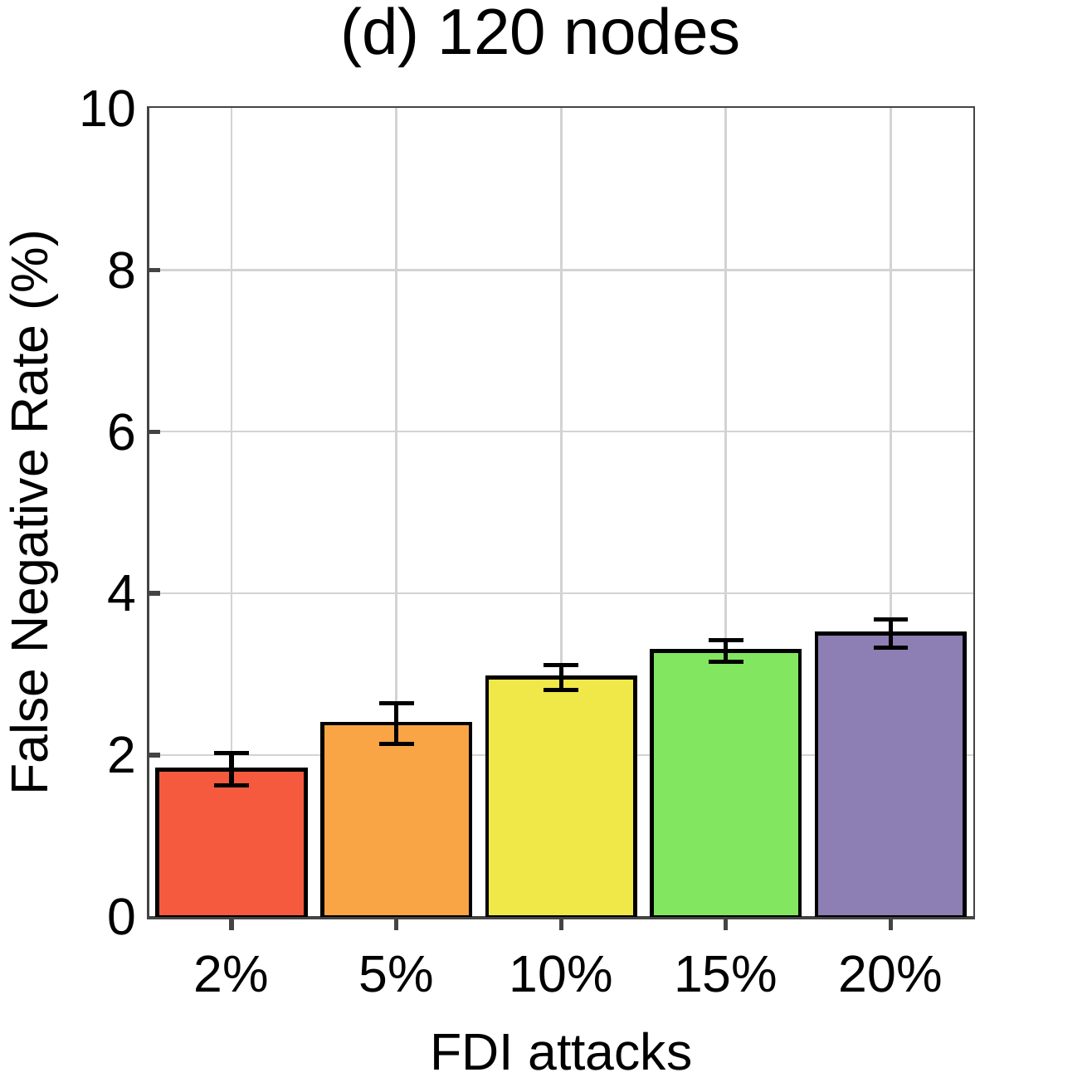}
    \caption{False negative rate $({t_{fn}})$ for 50, 75, 100 and 120 nodes}
    \label{Fig:fNtotal}
\end{figure}

Figure~\ref{Fig:fNtotal} shows the graphs about the false negative rate  \textbf{(FNR)}  against the FDI attack. An average \textbf{FNRs} assessed  by CONFINIT were of $2\%$ and $3.8\%$ to 2\% and  20\% of attackers, respectively, and to 5\%, 10\%, and 15\% attackers they remained stable between $2.9\%$ to $3.6\%$ in all scenarios.  We observe that FNRs presented a variation of only $1.9\%$, demonstrating the effectiveness of CONFINIT under different numbers of nodes and attackers, and hence few FDI attacks were not detected by CONFINIT regardless of the number of nodes in the network. Failure to detect an attacker can appear when there is an error in the similarity calculation, and the node is labeled as suspect, and over the consensus computation phase, that node is seen as a honest node. In this way, IIoT nodes may take a long time to identify an attacker and compromise the detection process.

\begin{figure}[ht]
  \centering
    \includegraphics[width=40mm]{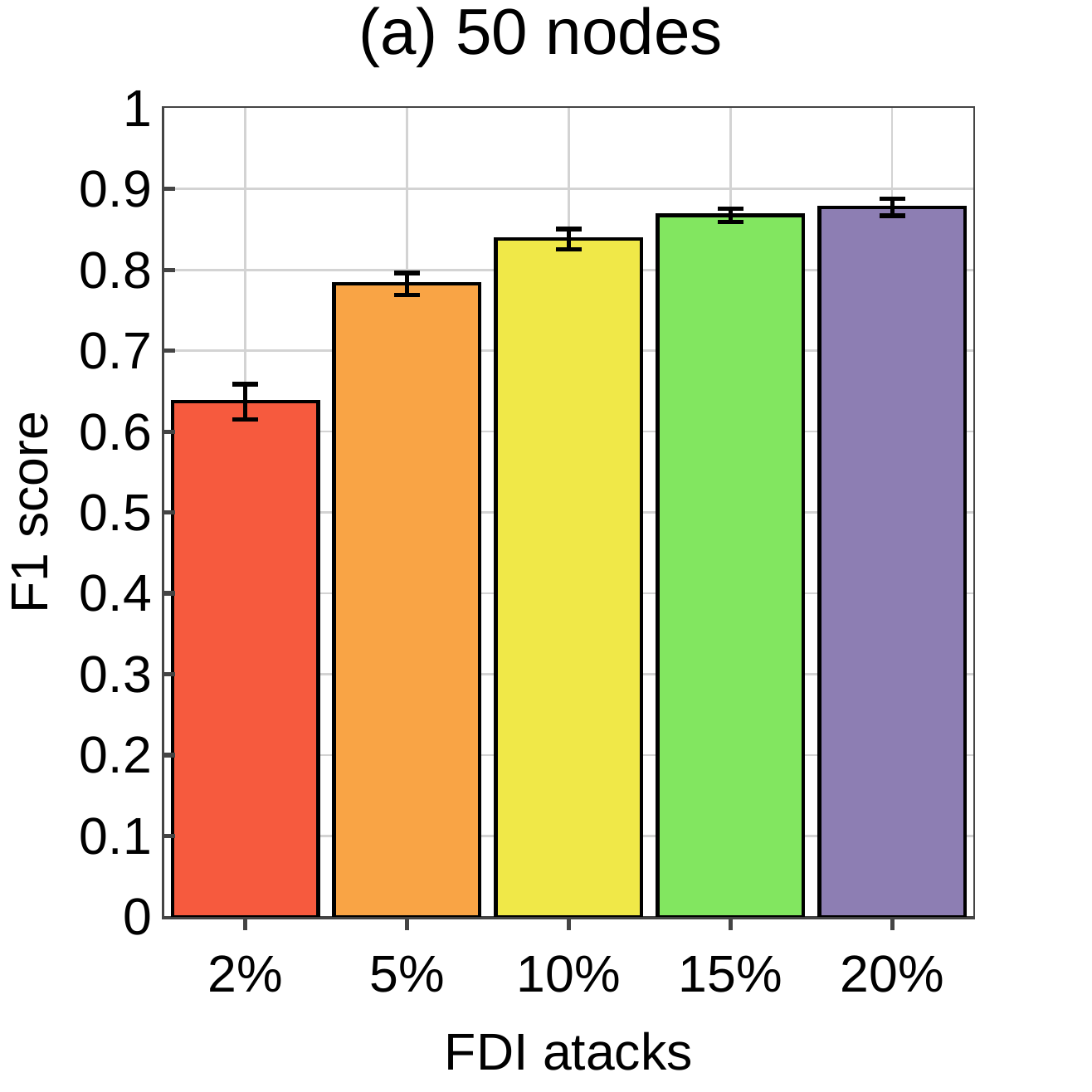}
    \includegraphics[width=40mm]{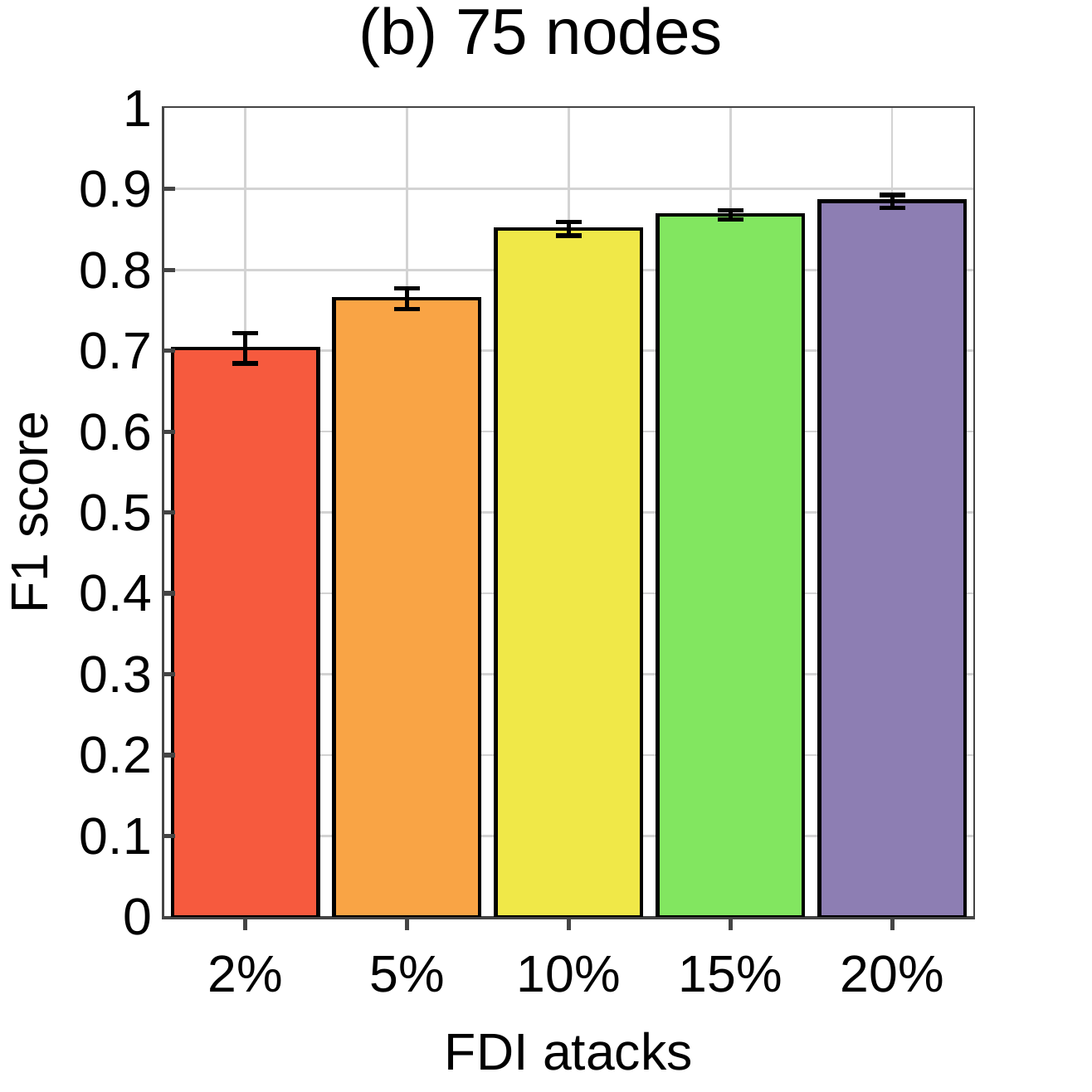}
    \includegraphics[width=40mm]{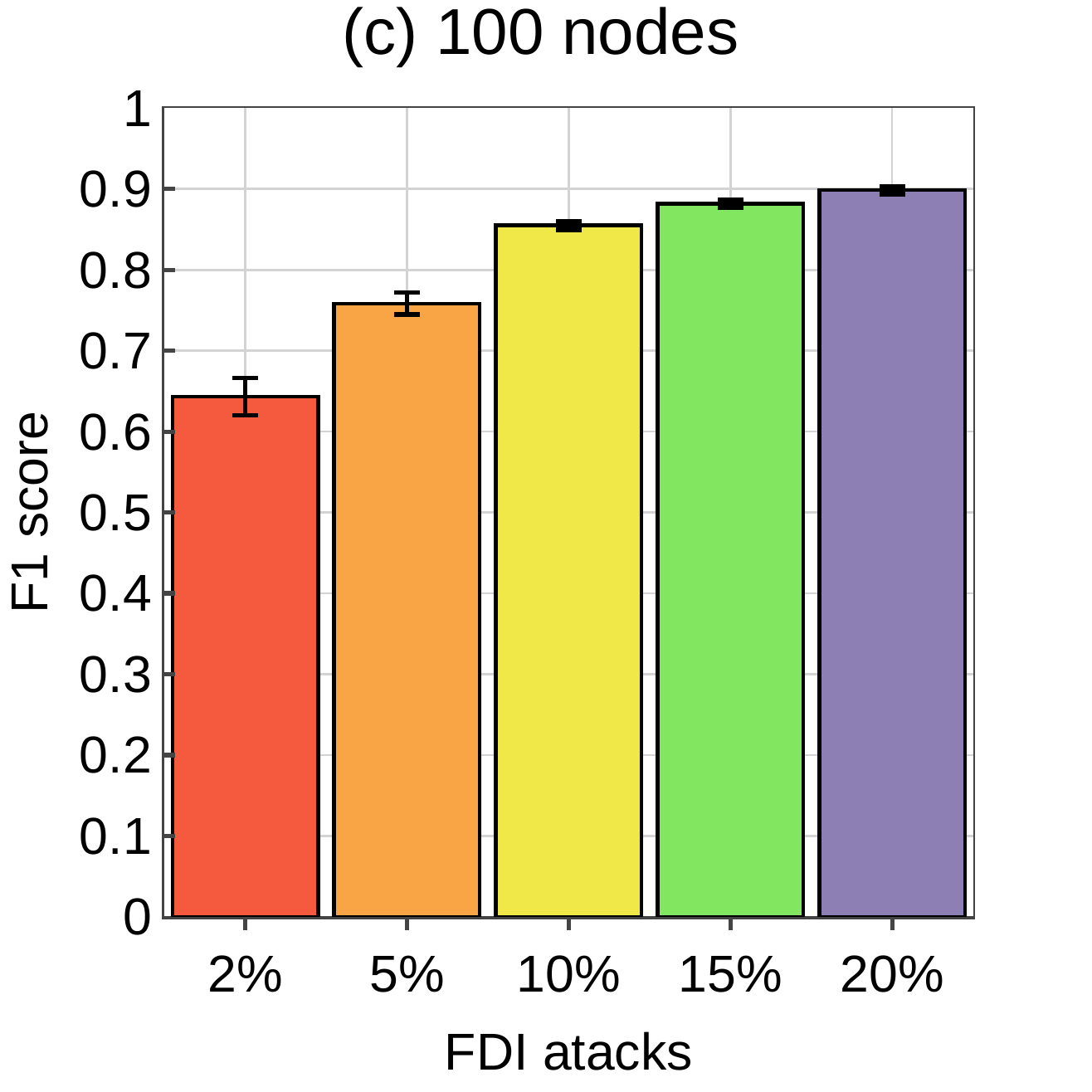}
    \includegraphics[width=40mm]{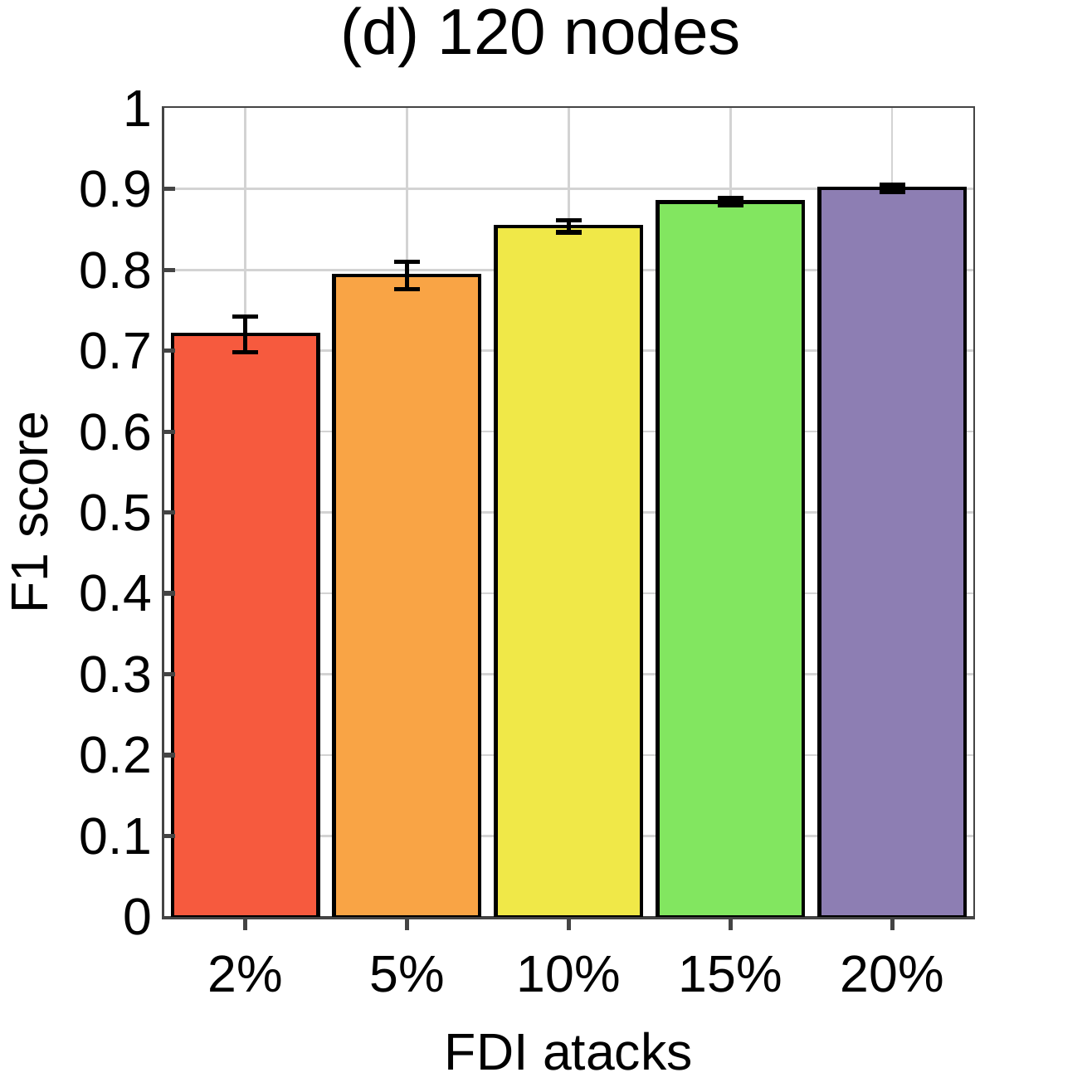}
    \caption{F1 score $({f_{c}})$ for 50, 75, 100 and 120 nodes}
    \label{Fig:fscore}
\end{figure}

Figure~\ref{Fig:fscore} shows the graphs about the F1 score \textbf{(FC)} of the CONFINIT against the FDI attack.  CONFINIT  obtained values ranging $0.63$ and $0.90$ for all scenarios, showing its  assertiveness in detecting  FDI attackers. With 120 and 100 nodes in the network, FC reached better results than with 75 and 50 nodes, since the collaborative filtering works better with more nodes in the network. The FC values are favored as the nodes are fixed, contributing to a high detection rate with low variance according to the number of nodes and intruders in the IIoT network. We also noted the FC variation was bigger with $2\%$ of FDI than others, since the low presence of attackers in the network and errors in the similarity calculation close to the \textit{ThresholdConsensus}. CONFINIT presented FC values among $0.79$ and $0.90$ for the other percentages of attackers, being more effective. Table~\ref{tab:tab3} shows the CONFINIT results for the precision and recall metrics. For all scenarios there was a high precision with values between $0.94$ and $0.96$, indicating the detection capability of CONFINIT. In the recall, the scenarios with 2\% of FDI attackers had the lowest values, ranging from $0.47$ to $0.57$,  the others with values between $0.64$ and $0.84$. That points out the correct identification of attackers and honest nodes by CONFINIT, and hence the capacity to provide the cluster availability and also authentic data to the application.

\begin{table}[ht]
\caption{Precision and recall obtained by CONFINIT}
\label{tab:tab3}
\centering
\scriptsize
\begin{tabular}{|c|c|c|c|c|c|c|c|c|c|c|c|}
\hline
\textbf{Scenarios}   & \textbf{Metrics} & \textbf{2\%} & \textbf{\begin{tabular}[c]{@{}c@{}}standard \\ deviation\end{tabular}} & \textbf{5\%} & \textbf{\begin{tabular}[c]{@{}c@{}}standard \\ deviation\end{tabular}} & \textbf{10\%} & \textbf{\begin{tabular}[c]{@{}c@{}}standard \\ deviation\end{tabular}} & \textbf{15\%} & \textbf{\begin{tabular}[c]{@{}c@{}}standard \\ deviation\end{tabular}} & \textbf{20\%} & \textbf{\begin{tabular}[c]{@{}c@{}}standard \\ deviation\end{tabular}} \\ \hline
\multirow{2}{*}{50}  & Precision        & 0,967        &  0,003                                                                  & 0,965        & 0,003                                                                  & 0,963         & 0,001                                                                  & 0,959         & 0,002                                                                  & 0,957         & 0,002                                                                  \\ \cline{2-12} 
                     & Recall           & 0,475        & 0,023                                                                  & 0,658        & 0,018                                                                  & 0,742         & 0,020                                                                  & 0,793         & 0,013                                                                  & 0,811         & 0,018                                                                  \\ \hline
\multirow{2}{*}{75}  & Precision        & 0,969        & 0,002                                                                  & 0,968        & 0,002                                                                  & 0,968         & 0,003                                                                  & 0,963         & 0,002                                                                  & 0,962         & 0,002                                                                  \\ \cline{2-12} 
                     & Recall           & 0,552        & 0,022                                                                  & 0,632        & 0,018                                                                  & 0,759         & 0,012                                                                  & 0,790         & 0,009                                                                  & 0,819         & 0,013                                                                  \\ \hline
\multirow{2}{*}{100} & Precision        & 0,971        & 0,002                                                                  & 0,969        & 0,002                                                                  & 0,969         & 0,002                                                                  & 0,967         & 0,001                                                                  & 0,965         & 0,001                                                                  \\ \cline{2-12} 
                     & Recall           & 0,481        & 0,025                                                                  & 0,623        & 0,019                                                                  & 0,765         & 0,008                                                                  & 0,810         & 0,009                                                                  & 0,841         & 0,008                                                                  \\ \hline
\multirow{2}{*}{120} & Precision        & 0,972        & 0,001                                                                  & 0,971        & 0,003                                                                  & 0,970         & 0,003                                                                  & 0,967         & 0,001                                                                  & 0,966         & 0,001                                                                  \\ \cline{2-12} 
                     & Recall           & 0,573        & 0,029                                                                  & 0,671        & 0,024                                                                  & 0,763         & 0,011                                                                  & 0,814         & 0,008                                                                  & 0,844         & 0,008                                                                  \\ \hline
\end{tabular}
\end{table}

\subsection{Churn and Sensitive FDI attacks}

We analyzed two variants of FDI attacks, called {\it Churn} and {\it Sensitive}, to better understand other behaviors of false data injection (FDI) attacks. The churn attack continually spreads false data on the network with values similar or not to the values sent by honest nodes. The churn attack differs from the FDI attacks by changing its behavior over time. In this was, after integrating the cluster, at certain times it pretends to be an honest one sending legitimate data to other cluster participants, and other moments it disseminates false data. The FDI-sensitive attack, in contrast, continuously sends false data close to the similarity threshold, making it difficult to be identified. We set up a scenario with 100 nodes being 10\% of a given attack (Churn or Sensitive). We evaluate the {\bf DR},  {\bf FNR},  {\bf FPR},  {\bf AC}, and  {\bf FC} for FDI-Churn and FDI-sensitive attackers.

\begin{figure}[ht]
\centering
    \includegraphics[width=40mm]{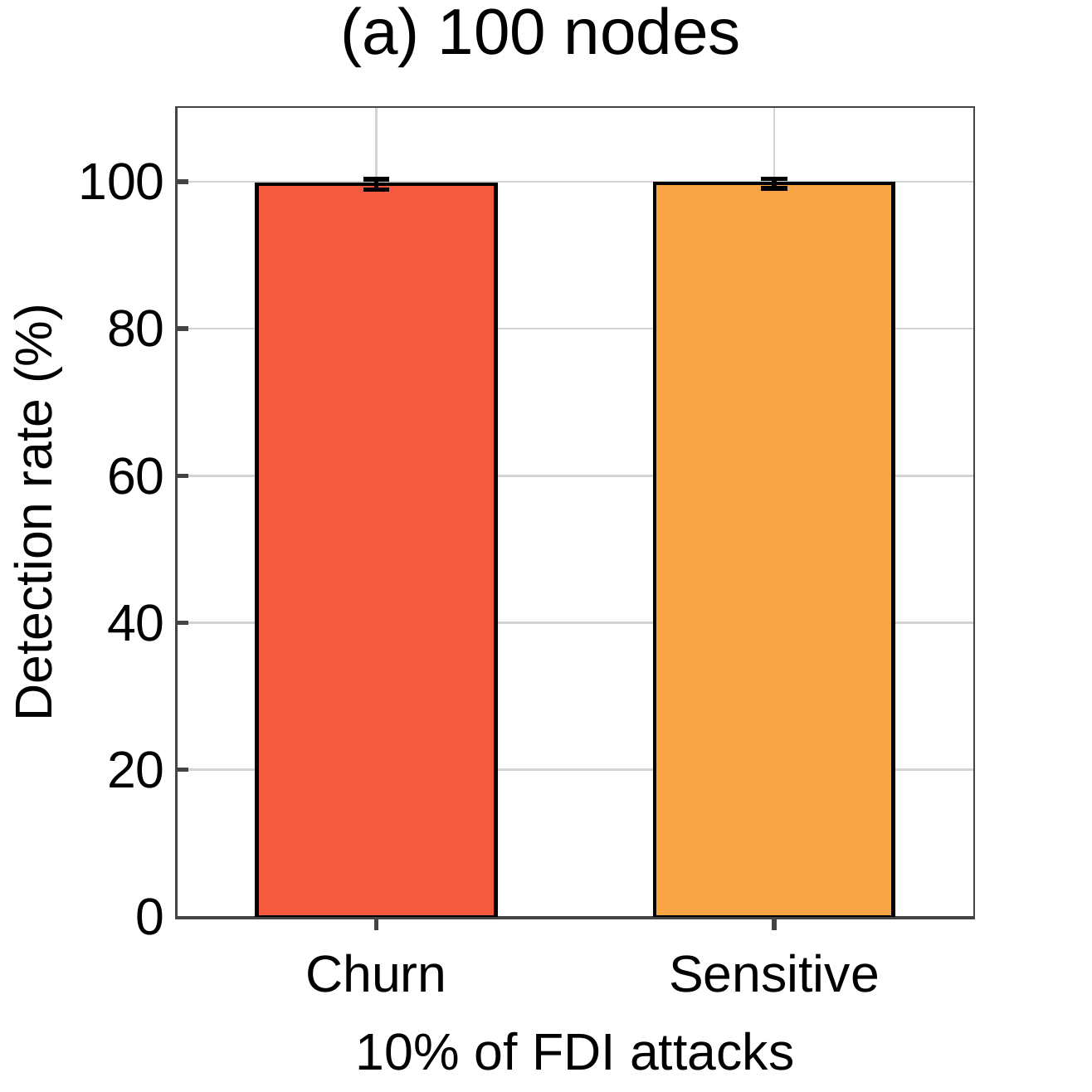}
    \includegraphics[width=40mm]{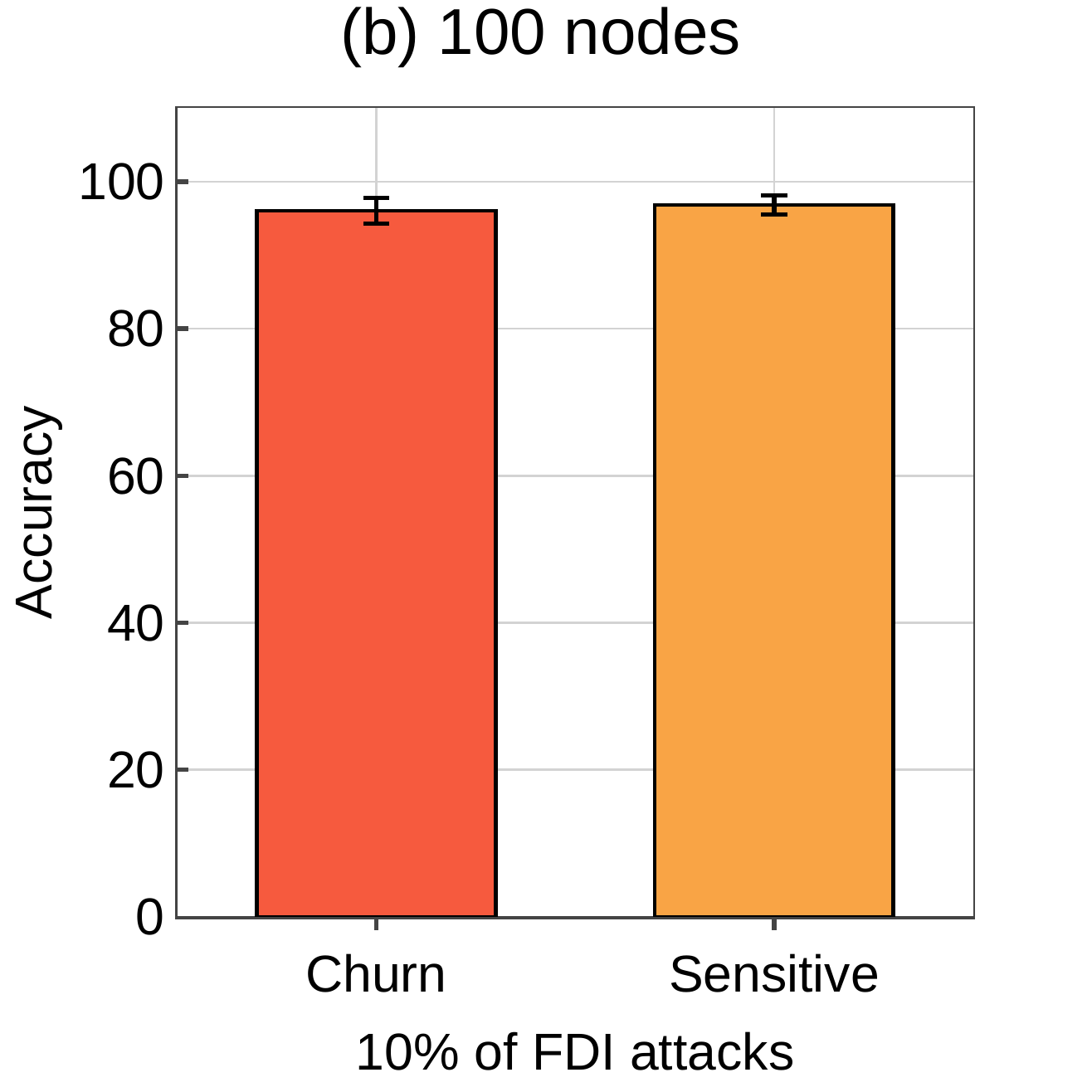}
    \includegraphics[width=40mm]{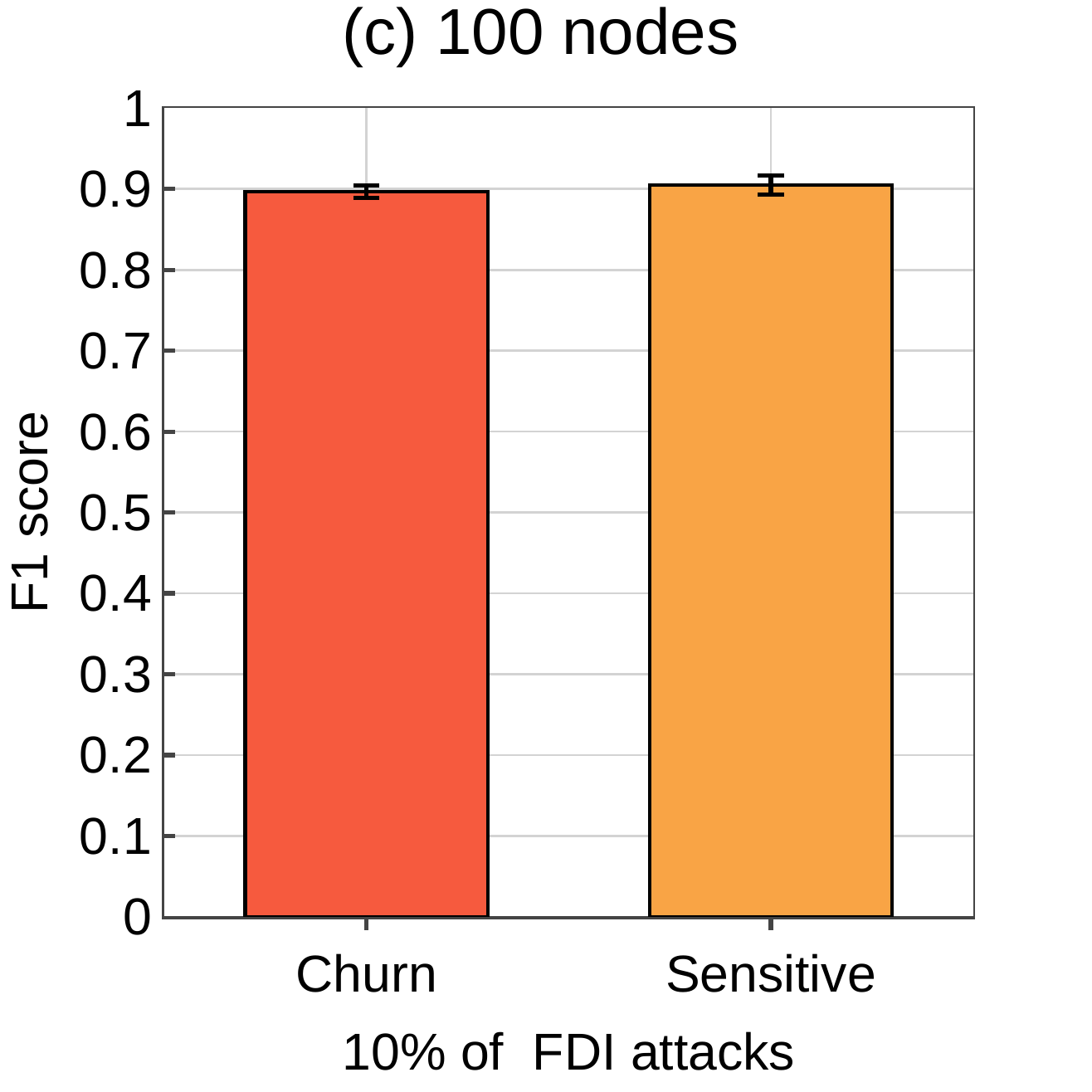}
    \includegraphics[width=40mm]{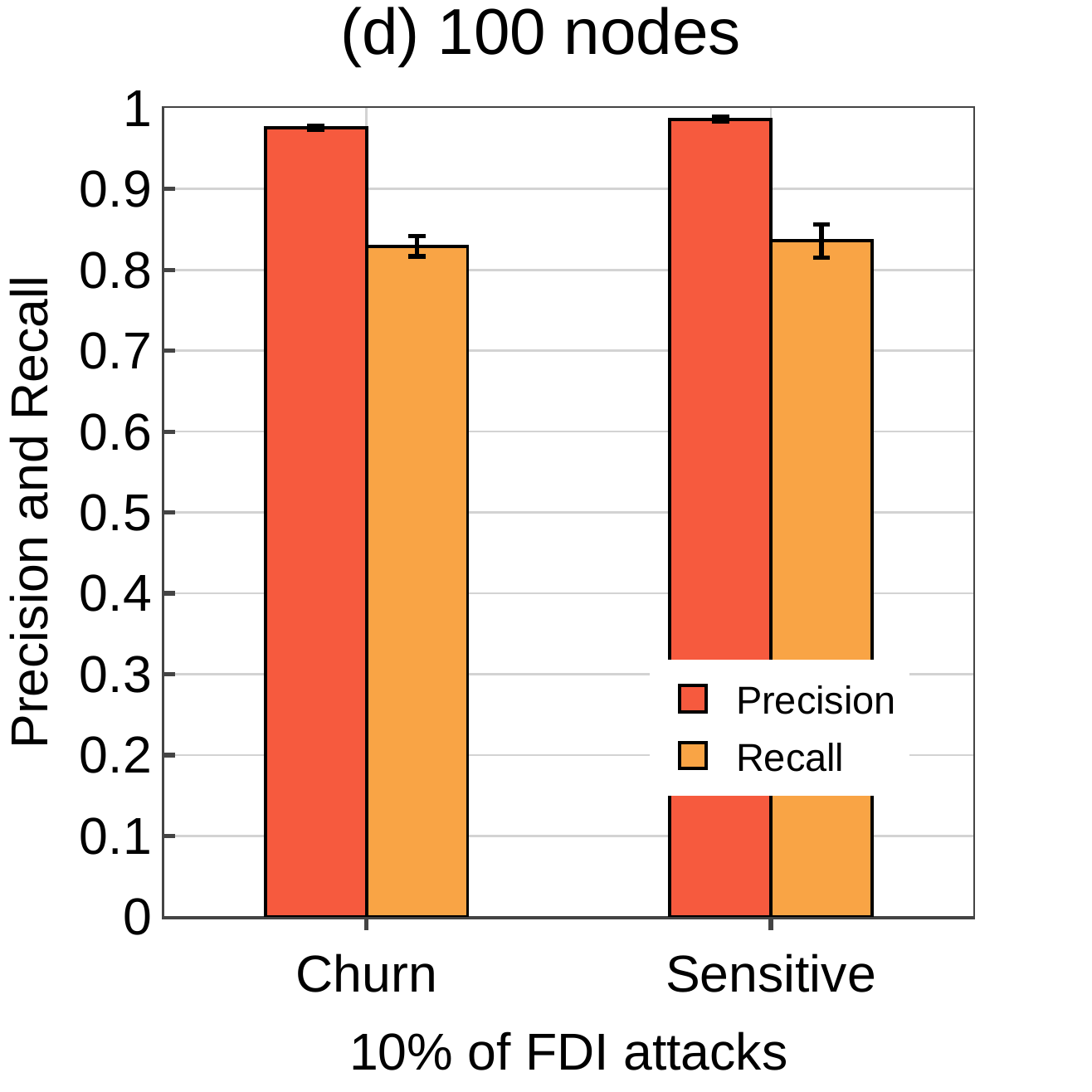}
    \caption{Detection, Accuracy and F1 Score of FDI-Churn and FDI-Sensitive for 100 nodes and 10\% of attacks}
    \label{Fig:chur}
\end{figure}

The detection and mitigation effectiveness of CONFINIT under Churn and Sensitive FDI attacks is shown by~\textbf{(DR)} graph in Fig.~\ref{Fig:chur}~\textbf{(a)}. CONFINIT obtained an average~\textbf{(DR)} of $99\%$, sometimes even reaching $100\%$ for both attacks, showing the CONFINIT's capacity to handle variations of FDI attacks in the IIoT. Under the FDI-churn behavior, the detection was easier than FDI-sensitive, since the variation in the frequency of sending data facilitates the insertion of intruders in the list of suspects and subsequent exclusion from the network. The FDI-sensitive is not easy to be identified, as the data sent by attacker contain values close to the honest nodes. In~\textbf{(AC)} graph, shown in Fig.~\ref{Fig:chur}~\textbf{(b)}, CONFINIT achieved high~\textbf{AC} between $95$ and $98$, supported  the watchdog surveillance among participants by evaluating the exchanged data messages. Further, the collaborative consensus ensured a better validation and a high detection rate among nodes. As the scenarios are statics, we have also noted slight variations in the detection rate, showing CONFINIT high ability to handle Churn and Sensitive-FDI attacks in dense IIoT. In~\textbf{(FC)} graph, shown in Fig.~\ref{Fig:chur}~\textbf{(c)}, CONFINIT reached~\textbf{FC} values of $0.92$ for churn and $0.94$ for sensitive in reason of the adoption of a suspect list that allows labeling honest nodes and attacker ones. In precision and recall graph, shown in Fig.~\ref{Fig:chur}~\textbf{(d)}, we observed that the precision remained stable with values between $0.97$ and $0.99$, and for the recall, CONFINIT achieved values between $0.84$ and $0.87$, showing the correct detection capability of honest nodes and attackers.

\begin{figure}[ht]
\centering
    \includegraphics[width=50mm]{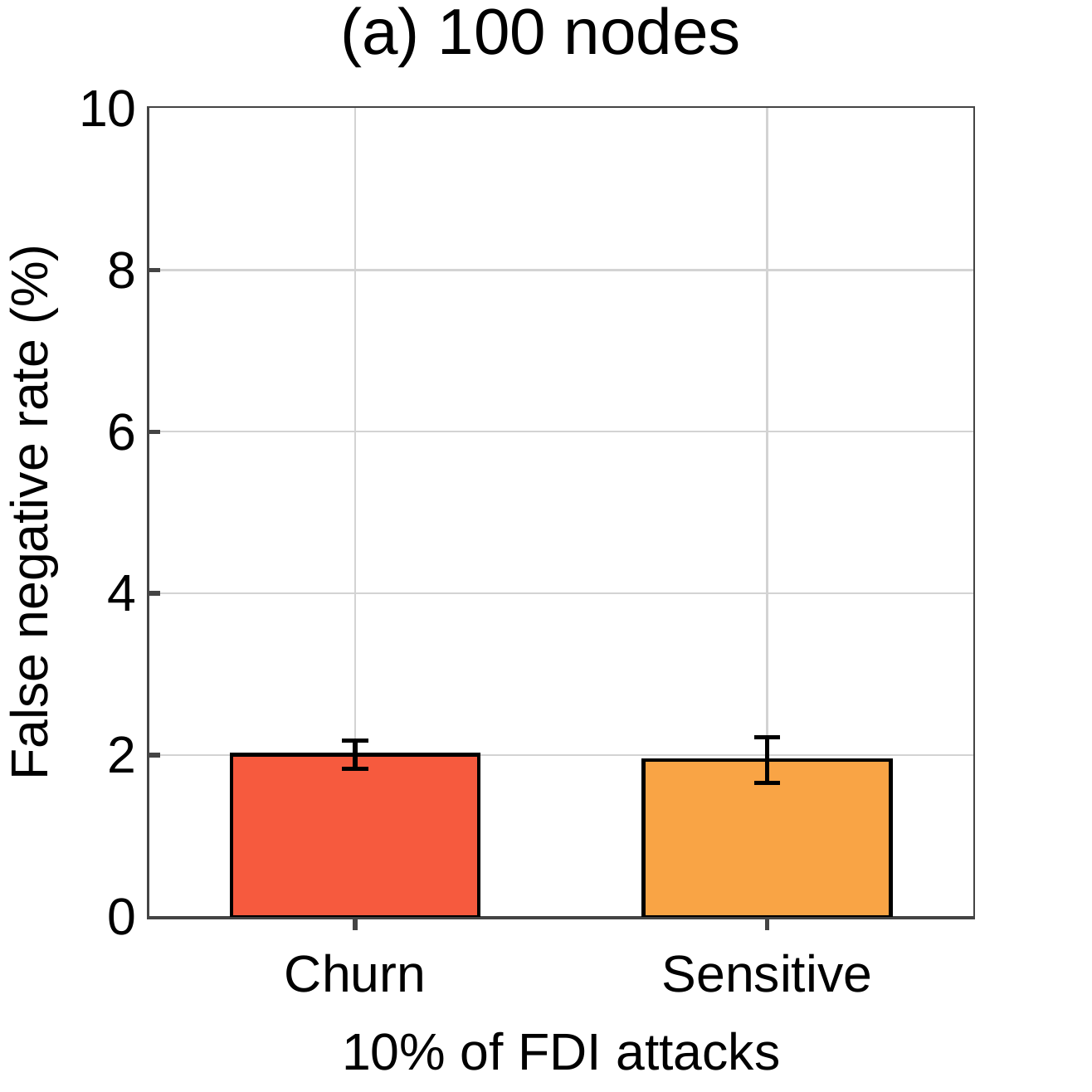}
    \includegraphics[width=50mm]{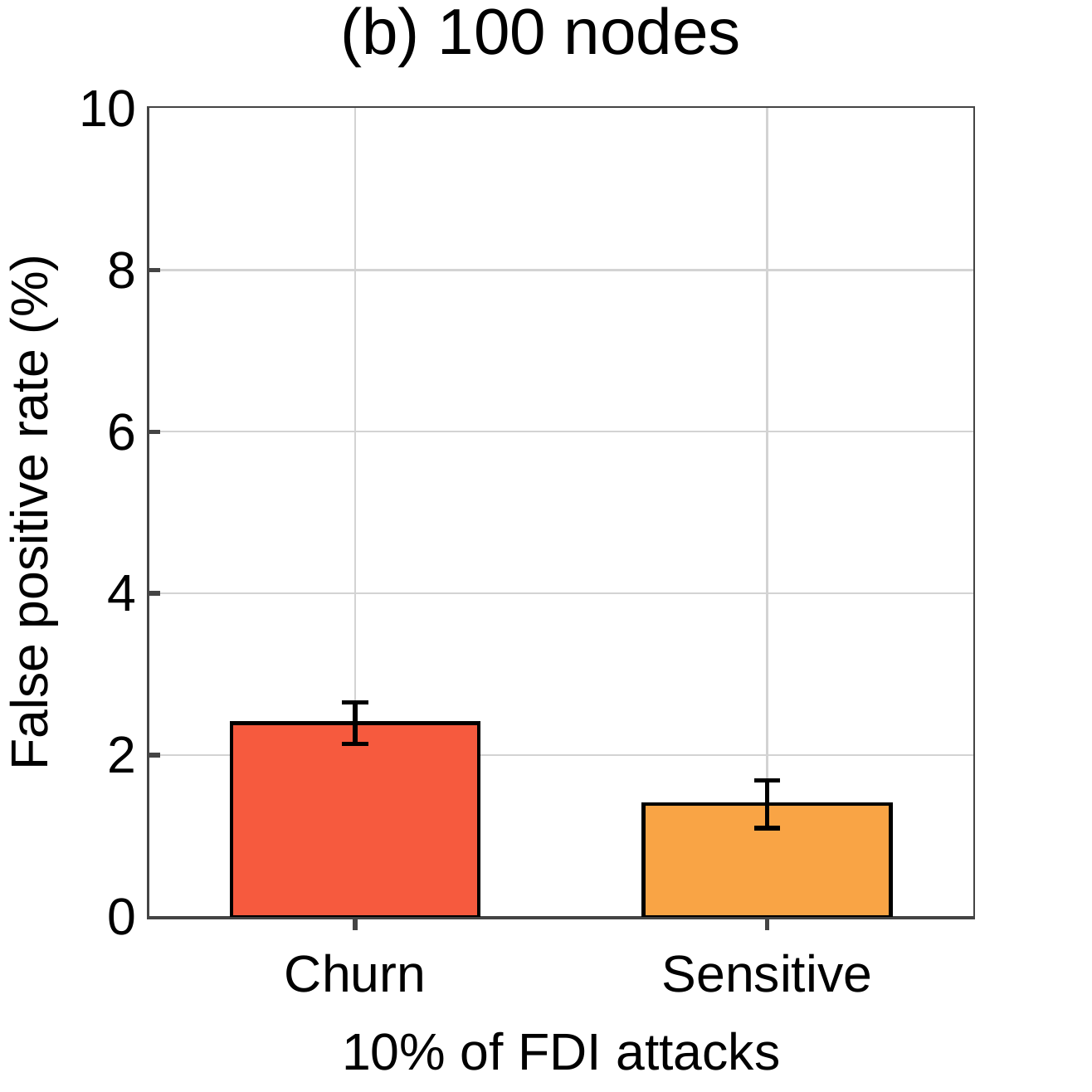}
    \caption{False negative and False positive rates of FDI-Churn FDI-Sensitive for 100 nodes and 10\% of attacks}
    \label{Fig:sensi}
\end{figure}

CONFINIT obtained stable values of FNR across the graph~\textbf{(a)} in Fig.~\ref{Fig:sensi}, being less than $2\%$ for both types of FDI attack. The detection failure sometimes appears when there were errors in the similarity computation stage performed by neighboring nodes responsible for identifying the node labeled as  honest or attacker due to inconsistency of its data reading, so certain nodes become suspicious. This failure is more usual in face of FDI-sensitive attacks since they typically send data with few variations to real data readings, making those attacks difficult to be identified by CONFINIT. Thus, in the collaborative consensus step, these nodes were incorrectly identified as intruders. On the other hand, the detection of FDI-Churn behavior can have errors in both equations, since it can start by spreading correct data and then spreading false data. The \textbf{FPR} remained stable for both attacks as shown in graph~\textbf{(b)} in Fig.~\ref{Fig:sensi}, around $1.8\%$  for Sensitive and $2.2\%$ for Churn attacks. Those detection errors were also due to errors in the consensus computation among watchdogs nodes regarding the node status, which took a low deviation from their readings. We noted this error is more common in FDI-churn due to randomness greater in the frequency of the false data sending, as the churn can send a sequence of data that exceed the similarity threshold and then data that does not, interfering in the identification and generating a higher rate of the wrong detection. Those nodes were added to the individual suspect, but with new data message exchange interactions, new calculations pointed to them as honest nodes.

In summary, based on the results obtained in the evaluations carried out, we concluded that CONFINIT was able to deal with the nodes density of the gas pressure IIoT network, mitigating the action of the FDI attack and preserving the availability and authenticity of the data sent by nodes to the application. CONFINIT achieved up to 40\% more clusters formed than DDFC, showing its effectiveness in dealing with device density and making more data available. In terms of security, CONFINIT obtained an average detection rate of 97\%, and in some cases, it reached 100\% for certain gas pressure IIoT scenarios. Further, for variations of the attack such as FDI-churn and FDI-Sensitive, CONFINIT achieved satisfactory results with a 100\% of detection rate that demonstrates the ability to handle different behaviors of FDI attacks and maintain data authenticity. Notably, accuracy values, F1-Score, false positive, and false negative rates confirm these results, contributing to the mechanism's high rate of hits in detecting attacks. This low variation in results related to changes in scenarios such as the number of nodes, and percentage of attacks, demonstrates the effectiveness of CONFINIT in maintaining security against different types of FDI. The  effectiveness against FDI attacks is due to the benefit of employing multi-level collaborative filtering and watchdog monitoring techniques, suitable for dealing with the network density, the large volume of data, and supporting the authenticity and data availability. Furthermore, we understand that CONFINIT focuses on IoT environments with fixed nodes, a high nodes density, continuous data flow, and clusters based on data similarity. In addition, CONFINIT only deal with attacks that act with data manipulation.

\section{Conclusion and future works}
\label{sec:concl}

This work presented the CONFINIT system for mitigating false data injection attacks in dense IIoT networks. It arranges massive IIoT networks in clusters through the reading similarity among fixed nodes to deal with the density issue. Furthermore, it embraces a watchdog strategy and collaborative consensus to monitor misbehavior nodes concerning their reading information, neighbors, and aggregate readings to determine nodes with malicious behavior about others. It also applies collaborative filtering to separate honest nodes from FDI attackers. The simulation results showed the effectiveness of CONFINIT against the FDI attack, ensuring only legitimate data availability for the IIoT application. Also, we demonstrated its robustness against other FDI behaviors - Churn and Sensitive. As future work, we intend to compare CONFINIT against other systems, evaluate its performance under different contexts of dense IoT networks that demand data clusters, and also analyze the impact of mobility in cluster formation and attack detection.

\section*{Acknowledgments}

We would like to acknowledge the support of the Brazilian Agency CNPq - grant \#309238/2017-0 and grant \#436649/2018-7.

\subsection*{Conflict of interest}

The authors declare no potential conflict of interests.

\bibliography{confinit}

\clearpage

\section*{Author Biography}

\noindent
{\textbf{Carlos Pedroso} is currently a Ph.D. student in the research group on Wireless Networks and Advanced Networks (NR2) at the Federal University of Paraná (UFPR). Graduation in Computer Networks by the Faculty of Technology of São Paulo (FATEC) (2016). Master in Informatics from Federal University of Paraná (UFPR) (2019). Has experience in Computer Science, with emphasis on Hardware, Computer networks, Wireless Sensor Networks and Internet of Things, acting mainly on the following topics: IoT and Security. Member of the Brazilian Computer Society (SBC) and IEEE Communication Society Communication (ComSoc).}

\noindent

\vspace{8.0pt}

\noindent
{\textbf{Aldri L. dos Santos} is professor of the Department of Computer Science  at Federal University of Minas Gerais (UFMG). Aldri is PhD in Computer Science from the Federal University of Minas Gerais, Master in Informatics and Bachelor of Computer Science at UFPR. Aldri is working in the following research areas: network management, fault tolerance, security, data dissemination, wireless ad hoc networks and sensor networks. He is leader of the research group (Wireless and Advanced Networks). Aldri has also acted as reviewer for publications as IEEE ComMag, IEEE ComNet, ComCom, IEEE Communications Surveys and Tutorials, IEEE eTNSM, JNSM, Ad hoc Networks. Aldri has served as member of the technical committee of security information and IEEE Communication Society Communication (ComSoc).}

\end{document}